\documentclass{aa}  
\usepackage{graphicx}
\usepackage{txfonts}
\usepackage{epstopdf}
\usepackage{subcaption}
\usepackage{color, colortbl}
\usepackage{bm}
\usepackage{array}

\newcommand{\CII}{C{\scriptsize\,II}}

\newcommand{\OI}{O{\scriptsize\,I}}
\newcommand{\HII}{H{\scriptsize\,II~}}
\newcommand{\HI}{H{\scriptsize\,I~}}
\newcommand{\Tex}{T_{\mathrm{ex}}}
\newcommand{\kb}{k_{\mathrm{B}}}

\newcommand{\vel}{\mathrm{v}}

\definecolor{Megenta}{rgb}{1,0,1}
\definecolor{LightGray}{gray}{0.9}

\begin{document} 

\title{Self-absorption in [C\,{\sc{ii}}], $^{12}$CO, and H\,{\sc{i}} in RCW120}
    
\subtitle{Building up a geometrical and physical model of the region\thanks{The $^{12}$CO and $^{13}$CO (3$\to$2)  data shown in Fig.~4 are available in fits format form the CDS via anonymous ftp to cdsarc.u-strasbg.fr 
(130.79.128.5) or via http://cdsweb.u-strasbg.fr/cgi-bin/qcat?J/A+A/. The [\CII] data are provided at the NASA/IPAC Infrared science archive at https://irsa.ipac.caltech.edu/Missions/sofia.html.}}

\author{S. Kabanovic \inst{1} \and
N. Schneider\inst{1} \and
V. Ossenkopf-Okada\inst{1} \and
F. Falasca \inst{2} \and 
R. G\"usten \inst{3} \and 
J. Stutzki \inst{1} \and 
R. Simon \inst{1} \and
C. Buchbender \inst{1} \and 
L. Anderson \inst{4,5} \and 
L. Bonne \inst{6} \and 
C. Guevara \inst{1} \and 
R. Higgins \inst{1} \and 
B. Koribalski \inst{7,8} \and 
M. Luisi \inst{5,9} \and 
M. Mertens \inst{1} \and 
Y. Okada \inst{1} \and 
M. R\"ollig \inst{1} \and 
D. Seifried \inst{1} \and 
M. Tiwari \inst{3, 10} \and 
F. Wyrowski \inst{3} \and 
A. Zavagno \inst{11,12} \and 
A.G.G.M. Tielens \inst{10,13}
}
     
\institute{I. Physikalisches Institut, Universität zu Köln, Z\"ulpicher Str. 77, 50937 K\"oln, Germany\\
\email{kabanovic@ph1.uni-koeln.de}
\and 
Courant Institute of Mathematical Sciences, New York University, New York, NY, USA \and 
Max-Planck Institut f\"ur Radioastronomie, Auf dem H\"ugel 69, 53121 Bonn, Germany \and 
Department of Physics and Astronomy, West Virginia University, Morgantown, WV 26506, USA \and
Center for Gravitational Waves and Cosmology, West Virginia University, Chestnut Ridge Research Building, Morgantown, WV 26505, USA \and
SOFIA Science Center, NASA Ames Research Center, Moffett Field, CA 94 045, USA \and
Australia Telescope National Facility, CSIRO Astronomy and Space Science, PO Box 76, Epping, NSW 1710, Australia \and
Western Sydney University, Locked Bag 1797, Penrith, NSW 2751, Australia \and
Department of Physics, Westminster College, New Wilmington PA 16172, USA \and
Department of Astronomy, University of Maryland, College Park, MD 20742, USA \and 
Aix Marseille Universit\'e, CNRS, CNES, LAM, Marseille, France \and 
Institut Universitaire de France (IUF), Paris, France \and
Leiden Observatory, Leiden University, PO Box 9513, 2300 RA Leiden, The Netherlands 
}
	
\date{draft of \today}
\titlerunning{Self-absorption in RCW~120}  
\authorrunning{S. Kabanovic}  

\abstract
{}
{Revealing the 3D dynamics of \HII\ region bubbles and their associated molecular clouds and \HI\ envelopes is important for developing an understanding of the longstanding problem as to how stellar feedback affects the density structure and kinematics of the different phases of the interstellar medium.} 
%
{We employed observations of the \HII\ region RCW~120 in the [\CII] 158 $\mu$m line, observed within the Stratospheric Observatory for Infrared Astronomy (SOFIA) legacy program FEEDBACK, and in the $^{12}$CO and $^{13}$CO (3$\to$2) lines, obtained with the Atacama Pathfinder Experiment (APEX) to derive the physical properties of the gas in the photodissociation region (PDR) and in the molecular cloud. We used high angular resolution \HI\ data from the Southern Galactic Plane Survey to quantify the physical properties of the cold atomic gas through \HI\ self-absorption.
The high spectral resolution of the heterodyne observations turns out to be essential in order to analyze the physical conditions, geometry, and overall structure of the sources. 
Two types of radiative transfer models were used to fit the observed [\CII] and CO spectra. A line profile analysis 
with the 1D non-LTE radiative transfer code SimLine proves that the CO emission cannot stem from a spherically  symmetric molecular cloud configuration. With a two-layer multicomponent model, we then quantified the amount of warm background and cold foreground gas. To fully exploit the spectral-spatial information in the CO spectra, a Gaussian mixture model was introduced that allows for grouping spectra into clusters with similar properties.} 
%
{The CO emission arises mostly from a limb-brightened, warm molecular ring, or more specifically a torus when extrapolated in 3D. 
There is a deficit of CO emission along the line-of-sight toward the center of the \HII\ region which indicates that the \HII\ region is associated with a flattened molecular cloud. Self-absorption in the CO line may hide signatures of infalling and expanding molecular gas. 
The [\CII] emission arises from an expanding [\CII] bubble and from the PDRs in the ring/torus. 
A  significant part of [\CII] emission is absorbed in a cool ($\sim$60-100 K), low-density ($<$500 cm$^{-3}$) atomic foreground layer with a thickness of a few parsec. 
}
%
{We propose that the RCW~120 \HII\ region formed in a flattened, filamentary, or sheet-like, molecular cloud and is now bursting out of its parental cloud. The compressed surrounding molecular layer formed a torus around the spherically expanding \HII\ bubble. This scenario can possibly be generalized for other \HII\ bubbles and would explain the observed "flat" structure of molecular clouds associated with \HII\ bubbles. 
We suggest that the [\CII] absorption observed in many star-forming regions is at least partly caused by low-density, cool, \HI-envelopes surrounding the molecular clouds.}
	
	
\maketitle
	

\section{Introduction} \label{sec:intro} 

    \begin{figure*}
        \centering
    	\begin{subfigure}[c]{0.40\textwidth}
        \includegraphics[width=1.\textwidth]{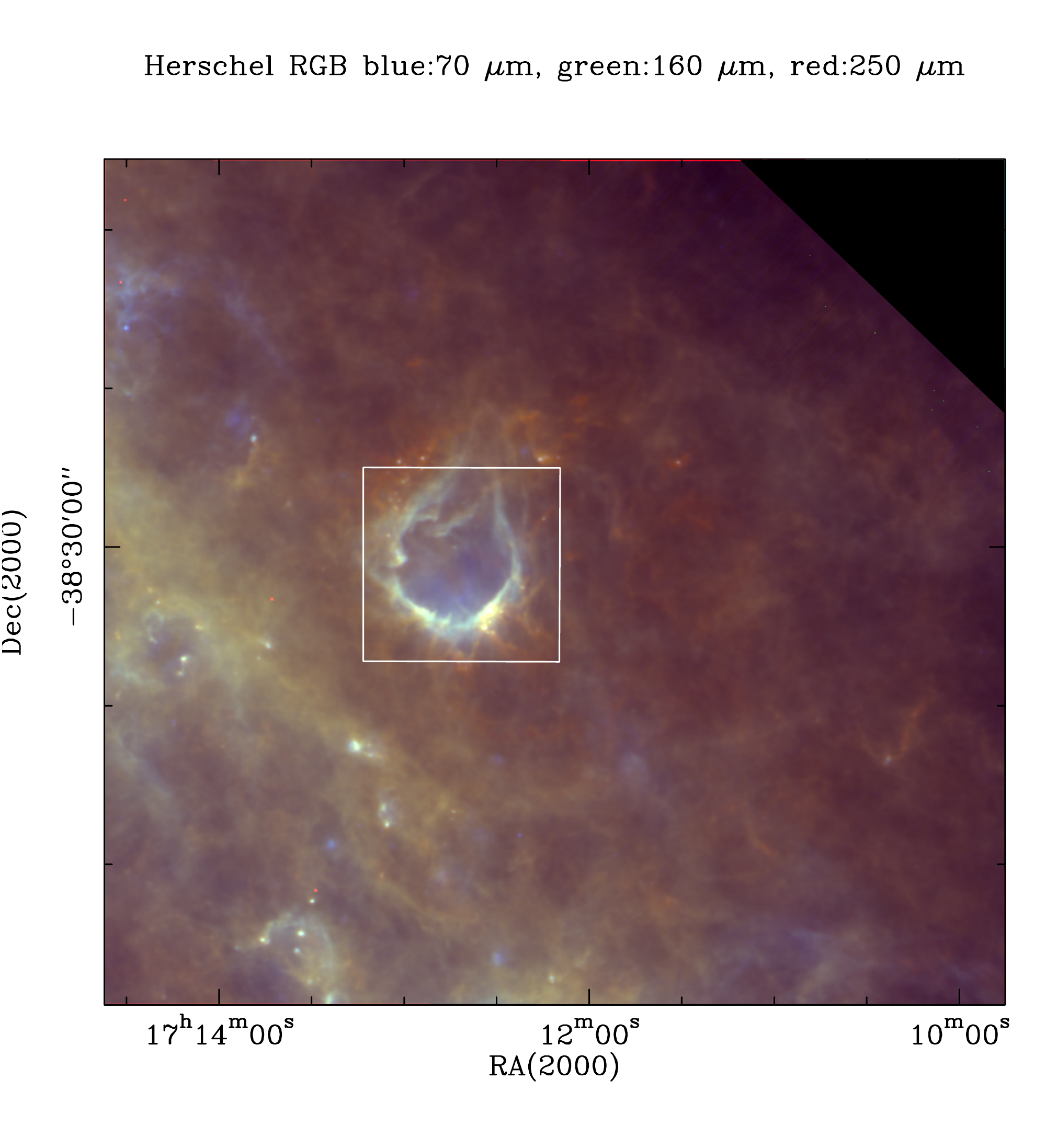}
    	\end{subfigure}
     	\begin{subfigure}[c]{0.40\textwidth}
        \includegraphics[width=1.\textwidth]{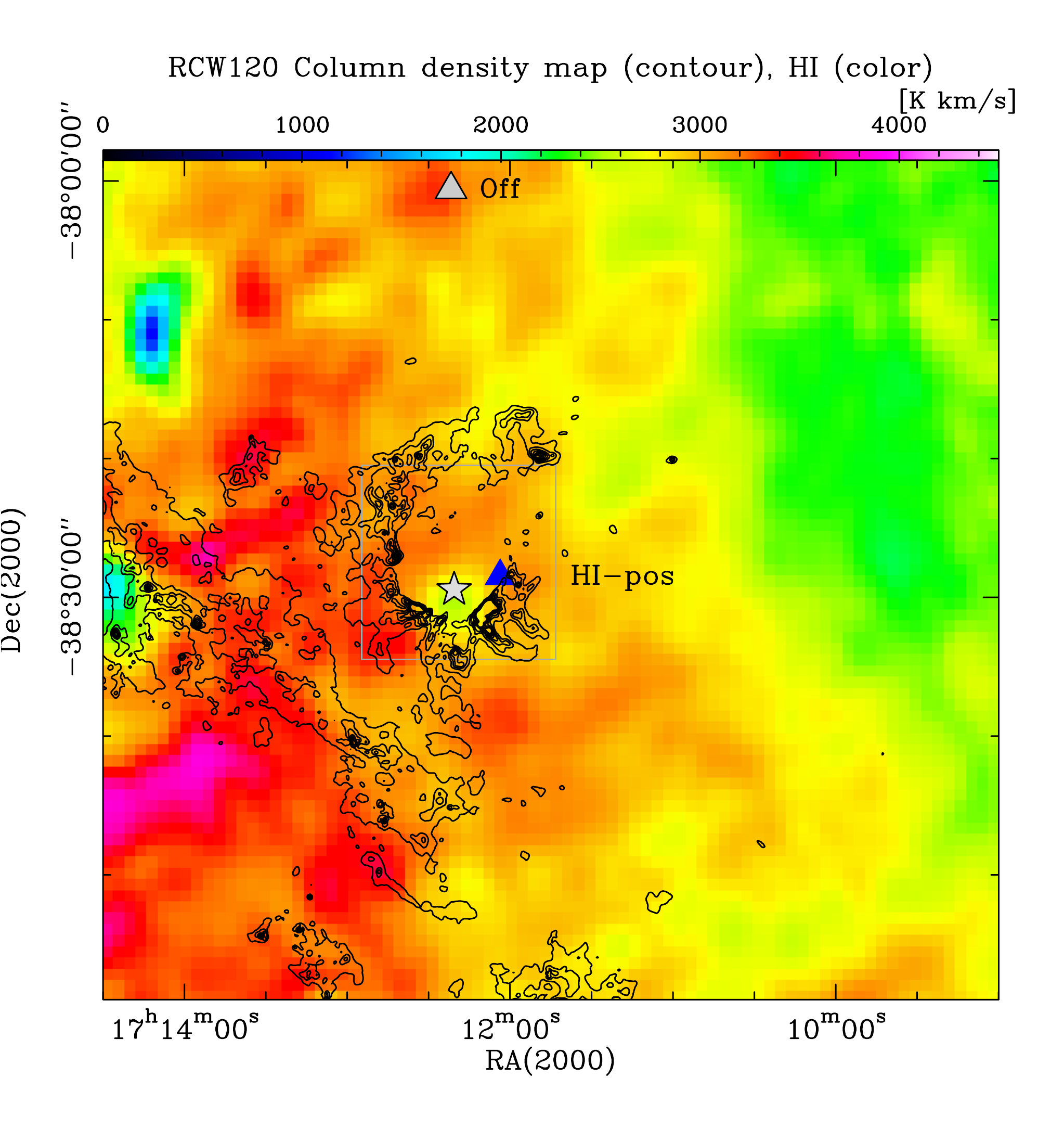}
    	\end{subfigure}
    	
    	\vspace{-.3cm}  
    	
    	\begin{subfigure}[c]{0.40\textwidth}   
        \includegraphics[width=1.\textwidth]{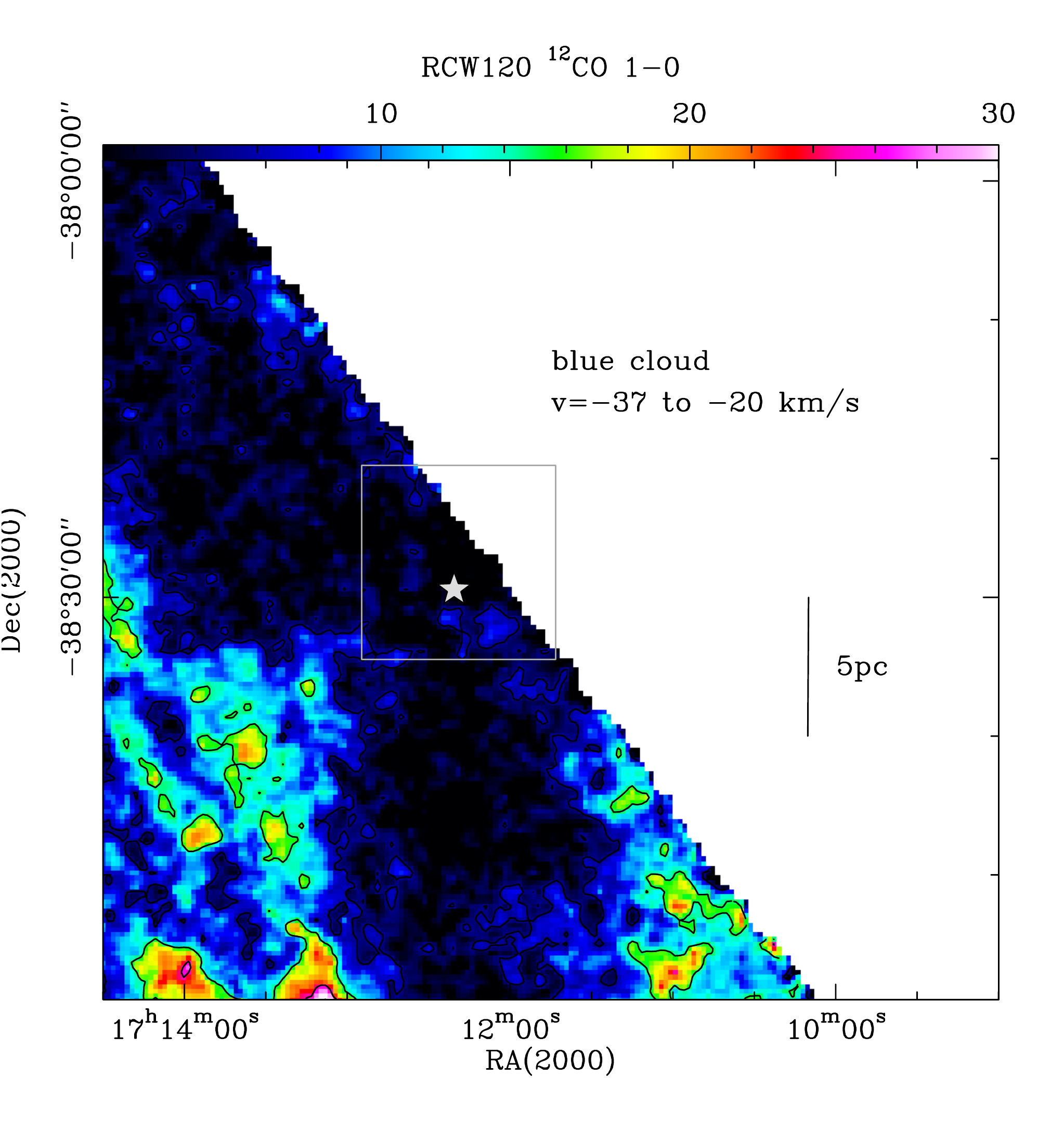}
    	\end{subfigure}
        \begin{subfigure}[c]{0.40\textwidth}  
        \includegraphics[width=1.\textwidth]{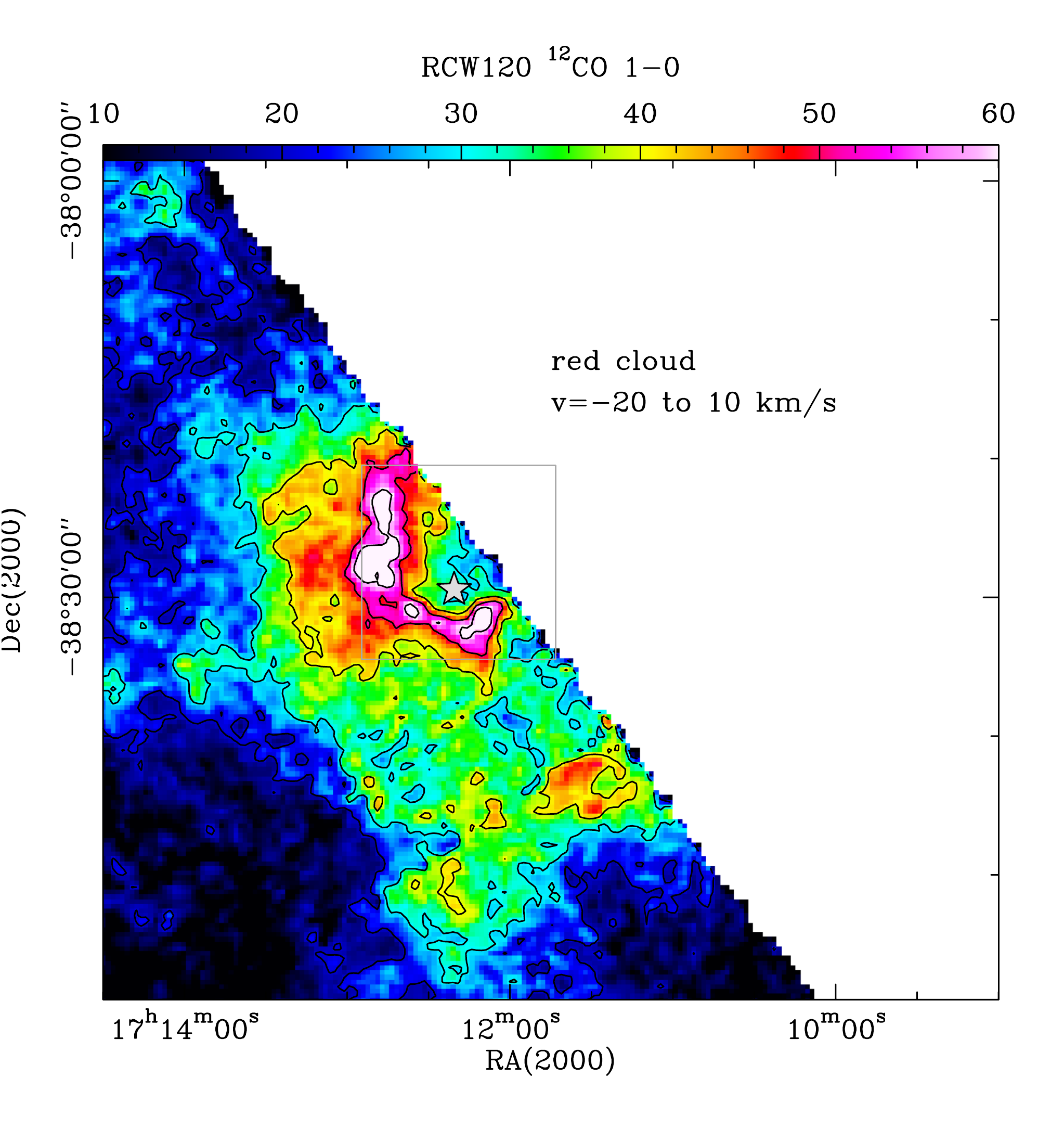}
    	\end{subfigure}
    	\caption{An overview of RCW~120. Top left: False color image of RCW~120 in {\sl Herschel} fluxes at $70\,\mathrm{\mu m}$ (blue), $160\,\mathrm{\mu m}$ (green), and $250\,\mathrm{\mu m}$ (red). The \HII\ bubble of RCW~120 stands out prominently. Top right:  
    	\HI\ line integrated (v = $-$20 to 10\,km\,s$^{-1}$) emission from the SGPS survey \citep{McClure2005} with contours of {\sl Herschel} dust column density overlaid (levels A$_{{\rm v}}$ = 15 to 50 in increments of five). The {\sl Herschel} data were observed within the HiGAL program \citep{Molinari2010} and the column density map was produced using the PPMAP method \citep{Marsh2017}. The position where we show an individual spectrum (Fig.~\ref{fig:HI-spectra}) is indicated by a blue triangle and the off-position we used for the HISA analysis is marked with a gray triangle. The white and gray squares outline the spectra map displayed in Fig.~\ref{fig:HI-map} and the star symbol indicates the exciting star of RCW~120. Bottom left and right: Line integrated $^{12}$CO (1$\to$0) intensity maps of RCW~120 between $-37$ to $-20$\,km\,s$^{-1}$ and $-20$ to 10 km s$^{-1}$, respectively. The CO data were taken with the MOPRA telescope and have a resolution of 30$''$ \citep{Torii2015}.}
    	\label{fig:intensity_maps_co_j10}
	\end{figure*}

    \HII\ regions are created by the ionizing radiation of massive stars. Extreme ultra-violet (EUV) photons with energies $>$13.6$\,$eV photoionize hydrogen in the surrounding medium, and a "bubble" of hot ionized gas (T$>$8000\,K) expands around the star \citep{Stroemgren1939}, at a supersonic sound speed, into the cool (T$\sim$10-20$\,$K) molecular cloud. During this expansion, a layer of dust and gas is formed between the ionization front and the preceding shock. Though the propagation of the expanding \HII\ region is initially spherical and many \HII\ regions indeed look circular or bipolar  \citep{Churchwell2006,Deharveng2010,Anderson2014}, suggesting a 3D bubble structure, more evolved \HII\ regions can have more irregular shapes. This depends on the distribution of the ionizing sources, the initial structures in the embedding gas, and the evolutionary state. 
    Similarly, the structure of the parental molecular cloud can also be nonuniform. \cite{Beaumont2010} found, based on $^{12}$CO (3$\to$2) maps, that the molecular gas distribution of many \HII\ regions with a circular shape forms ring-like structures around \HII\ region bubbles with a deficit of CO emission along the line-of-sight (LOS) toward the bubble center. This suggests that the clouds are "flat", that is sheet-like, with one dimension thinner than the other two. \citet{Anderson2010} and \citet{Kirsanova2019} support this scenario for the prototypical \HII\ region bubble RCW~120, based on dust continuum and CO line observations. However, special attention has to be given to self-absorption effects in optically thick low- to mid-J CO lines because concentrations of cold dense material in front of a warm emitting cloud can mimic a CO deficit.

    It has been discovered recently that many circular \HII\ regions are associated with expanding shells seen in ionized carbon \citep{Pabst2019,Pabst2020,Tiwari2021,Luisi2021}, based on [\CII] $158\,\mathrm{\mu m}$ fine-structure line observations with the Stratospheric Observatory for Infrared Astronomy (SOFIA). 
    This important cooling line arises mostly in the atomic-to-molecular transition layer between the \HII\ region and the dense molecular cloud, where a warm ($T\gtrsim 90\,\mathrm{K}$) photodissociation region \citep[PDR,][]{ Hollenbach1999} is formed. 
    For RCW~120, it was shown that the 3D C$^+$ bubble is driven by the stellar wind of the central star and has a significant impact on the star-formation efficiency in the embedding cloud \citep{Luisi2021}. A spherical distribution of warm dust inside the \HII\ region, but a more clumpy structure in the associated molecular cloud, was reported by \citet{Marsh2019} based on {\sl Herschel} data.   
    
    We here present a study of [\CII], CO, and \HI\ line emission in the RCW~120 region, complementary to the study presented in \citet{Luisi2021}. The [CII] 158 $\mu$m line was observed in the context of the SOFIA Legacy Program
    FEEDBACK\footnote{https://feedback.astro.umd.edu} \citep{Schneider2020}. Maps of $^{12}$CO and $^{13}$CO (3$\to$2) emission were obtained with the Atacama Pathfinder Experiment (APEX) telescope. In addition, there is a wealth of archival data on this source at various wavelengths we employ to support the interpretation of our data. Particularly important are high angular resolution ($\sim$2$'$) \HI\ 21$\,$cm line  emission data from the Southern Galactic Plane Survey \citep[SGPS,][]{McClure2005}.
     
    The objectives of this paper are to determine the physical properties of the different gas components that are responsible for emission and absorption in the CO, [\CII], and \HI\ lines, and to develop a geometrical model for the RCW~120 region. We show that self-absorption features in [\CII] and \HI\ are observed at velocities of the bulk emission of the cloud which suggests a physical connection. We argue that the observed large quantities of cold C$^+$ arise from a cool \HI\ halo around the molecular cloud associated with RCW~120.
    This representation may then serve as a template for explaining the emission properties in other bubble \HII\ regions.
    
    The \HII\ region RCW~120 has a diameter of $4.5\,\mathrm{pc}$ and is excited by a single O6-8V/III star \citep{Martins2010} at a distance of 1.7\,kpc \citep{Kuhn2019}. 
    Figure~\ref{fig:intensity_maps_co_j10} displays the larger environment of RCW~120 in various tracers (dust, \HI, and CO). 
    We observe that there is less atomic and molecular material in the northwestern part of the region and that the \HI\ emission envelops the dense molecular cloud, which is outlined by the dust column density contours. The lower panels of Figure~\ref{fig:intensity_maps_co_j10} show the molecular cloud velocity components that are associated with RCW~120, a blue-shifted velocity cloud (v = $-37$ to $-20\,\mathrm{km\,s^{-1}}$) and a red-shifted one (v = $-20$ to $10\,\mathrm{km\,s^{-1}}$) that are proposed to be in collision \citep{Torii2015}. 
    The most prominent feature in RCW~120 is a ring of dense dust and molecular gas from the red-shifted velocity cloud. Many massive condensations along the ring structure were identified in (sub)-millimeter continuum emission by \cite{Zavagno2007,Deharveng2009, Anderson2012}. Notable is "condensation 1" \citep{Zavagno2010b} in the southern arc of CO emission, that hosts a Class 0 young stellar object (YSO), driving a CO outflow \citep{Figueira2020}. 

    The paper is organized as follows. Section~\ref{sec:obs} gives an overview of the observational data we make use of. Section~\ref{sec:results} briefly discusses the [\CII], CO, and \HI\ emission distributions and Sect.~\ref{sec:analysis} presents an analysis of the self-absorption effects in [\CII], CO (3$\to$2) and \HI. This analysis uses the two-layer multicomponent model for [\CII] and CO, first introduced by \cite{Guevara2020} and applied in \citet{Bonne2020}. To run the model for an entire spectral cube in CO, we employ a Gaussian Mixture Model, which is an unsupervised machine learning approach to cluster spectra along common properties \citep{Brunton2019}. We then constrain the geometry of RCW~120 using SimLine \citep{Ossenkopf2001} on the CO and [\CII] data. \HI\ self-absorption (HISA) is quantified with improved methods for HISA studies.  In Sect.~\ref{sec:discussion} we discuss our results and draw a general picture for the possible formation and evolution of \HII\ regions associated with flat molecular clouds. Section~\ref{sec:summary} summarizes our findings. 
    	
\section{Observations} \label{sec:obs} 
    \subsection{SOFIA} \label{sec:sofia} 
    
        \begin{figure*}[!ht]
        	\centering
        	\includegraphics[width=0.32\textwidth]{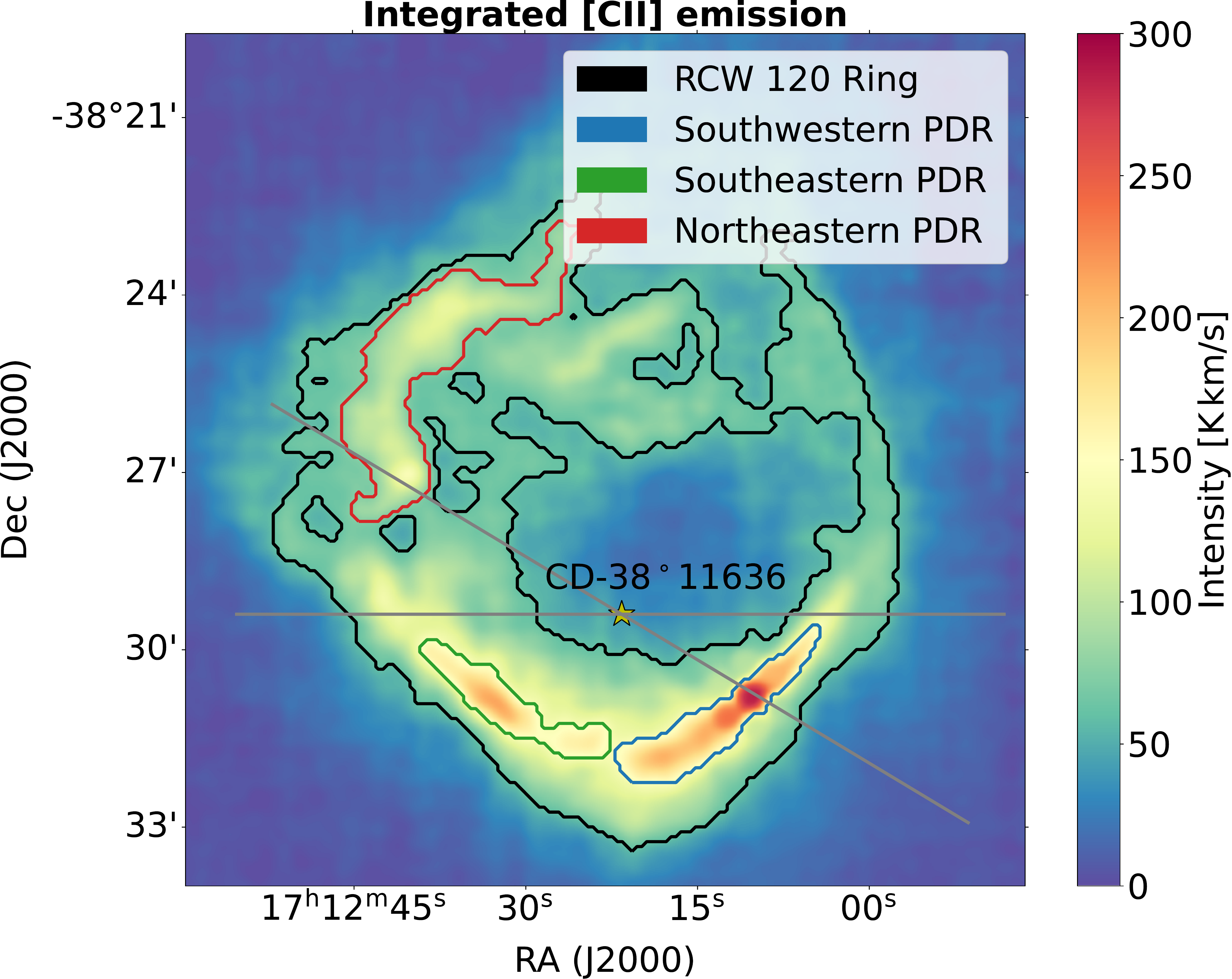}
        	\includegraphics[width=0.32\textwidth]{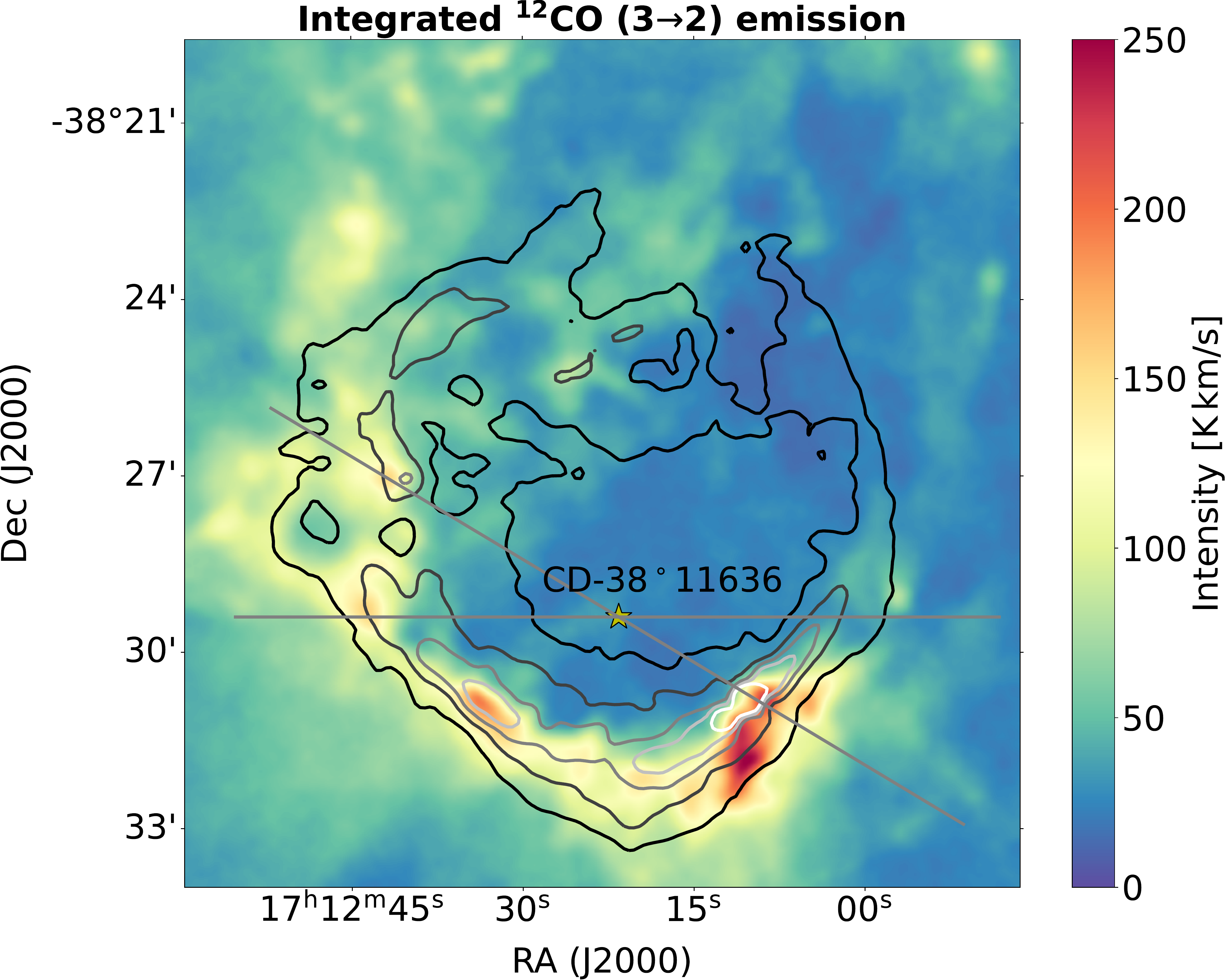}
        	\includegraphics[width=0.32\textwidth]{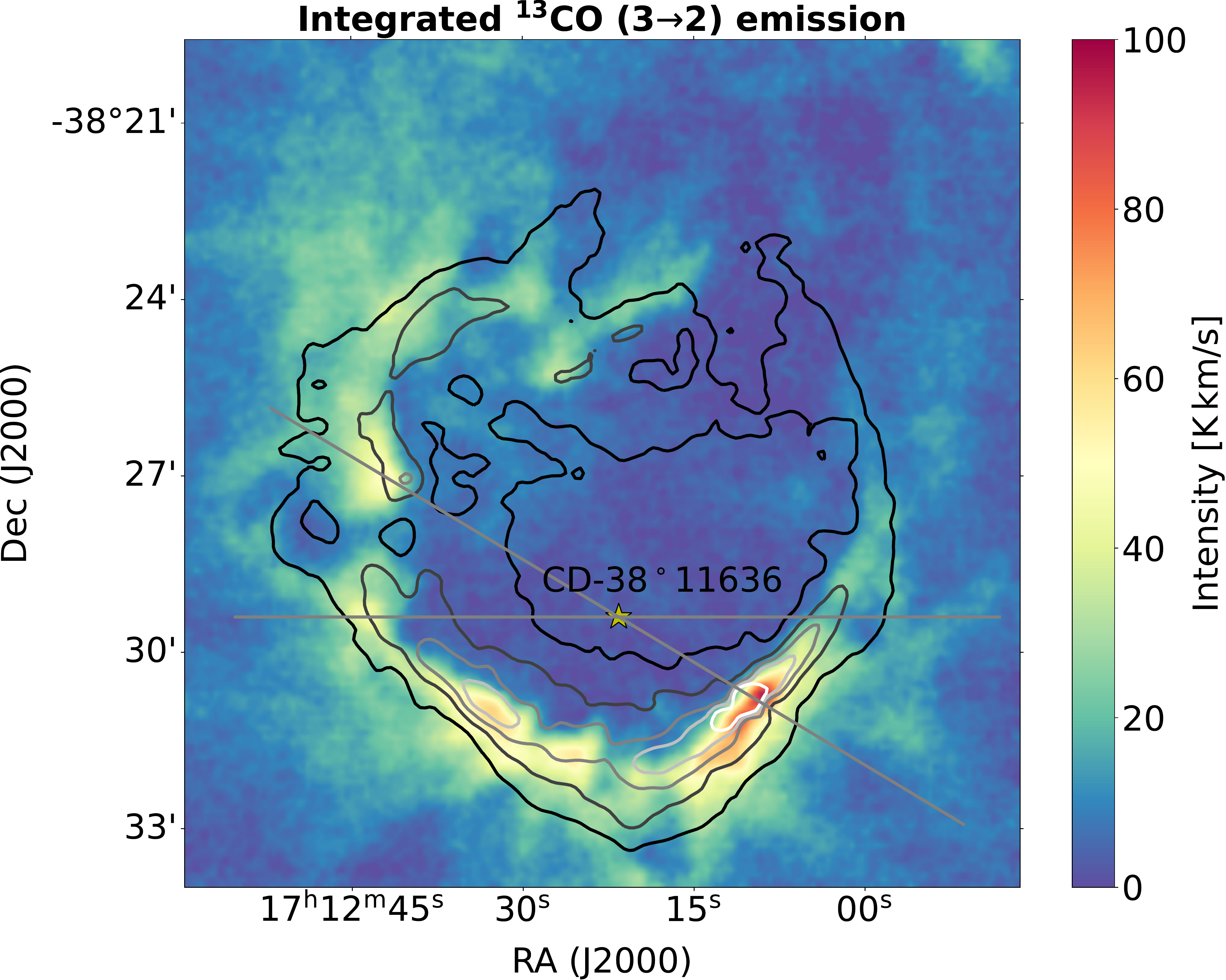}
    		\caption{Velocity integrated maps of [\CII] and CO (3$\to$2) emission in RCW~120 between $-20$ and 10 $\mathrm{km s^{-1}}$. Left panel: Line integrated [\CII] intensity map. The contours outline the regions extracted with a dendrogram based approach \citep{Rosolowsky2008,Robitaille2019}, explained in Sec.~\ref{sec:tau}. Middle and right panel:  Line integrated $^{12}$CO (3$\to$2) and $^{13}$CO (3$\to$2) intensity maps, respectively. The contours give the line integrated [\CII] intensity levels $60, 100, 140, 180, 220 \,\mathrm{K\,km s^{-1}}$, ranging from black to light gray. The two gray lines indicate cuts displaying the line intensities in Fig.~\ref{fig:pi_diagram}. The diagonal line has an angle of $25^{\circ}$ with respect to the horizontal line. The exciting O6-8V/III type star  CD-38 11636 is marked with a star symbol.} \label{fig:intensity_maps}
    	\end{figure*}
	
    	\begin{figure}[!h]
    	\centering
    		\includegraphics[width=0.4\textwidth]{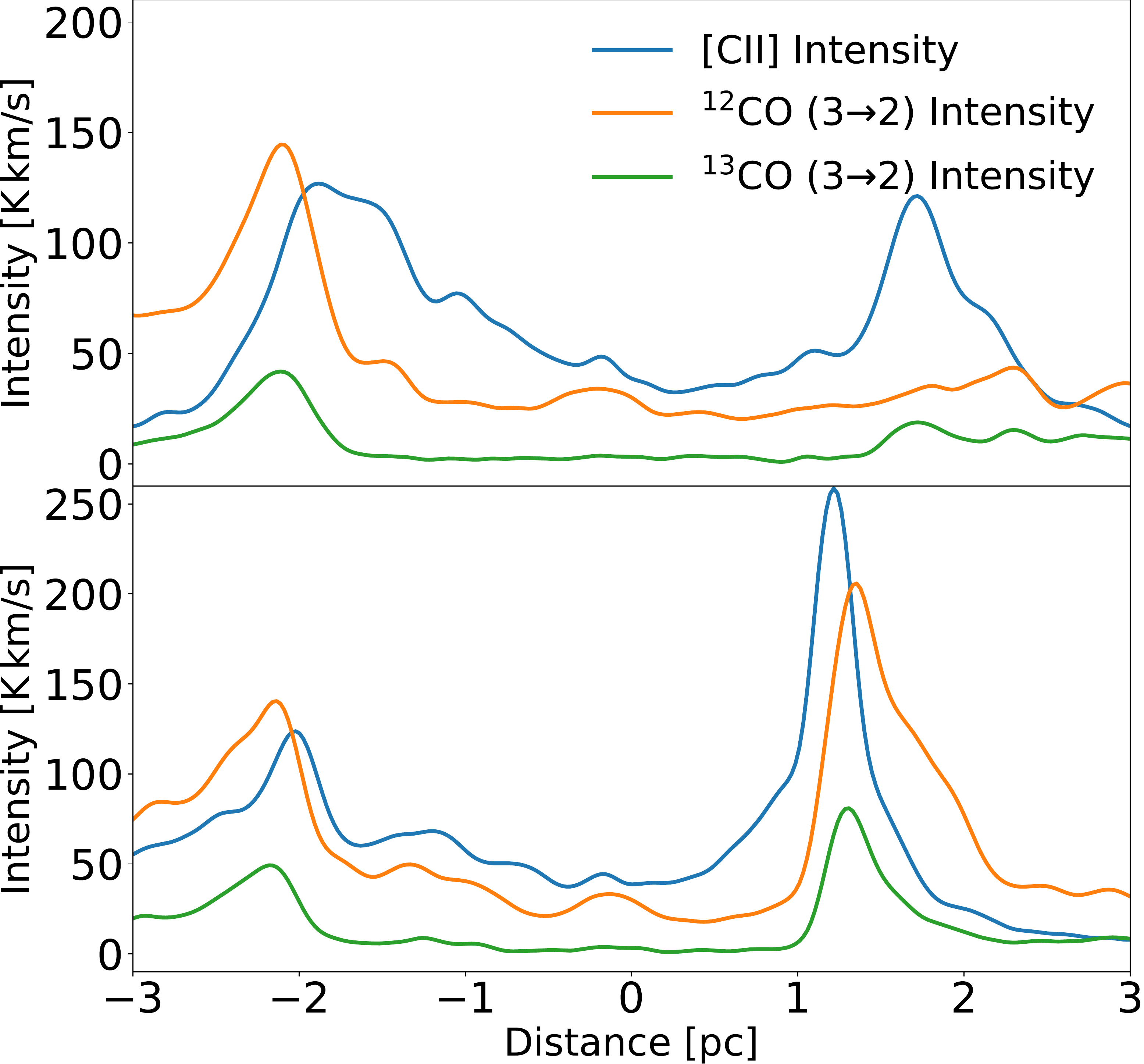}
    		\caption{Position-intensity diagram along two cuts. The positions of both cuts are indicated by the gray lines crossing the ionizing star (central position) in Fig. \ref{fig:intensity_maps}, negative distances are to the east and northeast. The upper panel shows the intensity along the horizontal line and the lower one the intensity along the diagonal cut ($25^{\circ}$). The [\CII] and $^{12}$CO and $^{13}$CO (3$\to$2) intensities are indicated by a blue, orange, and green curve, respectively.} \label{fig:pi_diagram}
    	\end{figure} 
	
        The [\CII] line at $1.9\,\mathrm{THz}$ was observed during one flight from Christchurch, New Zealand, on 10 June 2019, and during one flight from Tahiti, French Polynesia, on 12 August 2021. The map from the 2019 run covers 75\% of the total planned area and is presented in \citet{Luisi2021}, the missing part was obtained in the 2021 southern deployment in Tahiti, though with slightly less integration time (see below). For both runs, the heterodyne receiver upGREAT\footnote{German Receiver for Astronomy at Terahertz. (up)GREAT \citep{Risacher2018} is a development by the MPI für Radioastronomie and the KOSMA Universität zu Köln, in cooperation with the MPI für Sonnensystemforschung and the DLR Institut für Planetenforschung.} onboard SOFIA was used. The target area was split into four squared "tiles" in which each tile has a length of 435.6$''$ = 7.26$'$ and is covered four times, two times in RA-direction and two times in Dec-direction. The first two coverages are performed with the array rotated 19$^\circ$ on the sky. The second two coverages are then shifted by 36$''$ to fill up the gaps of the [\OI] map, which was observed in parallel but will not be discussed here. All tiles were covered in the array-on-the-fly mapping mode. During the Tahiti campaign, the last missing northwestern tile was covered two times in RA-direction but only one time in Dec-direction. This results in a slightly lower signal-to-noise ratio.
    
        The half-power beam width at $1.9\,\mathrm{THz}$ is 14.1$''$. As backend, a Fast Fourier Transform Spectrometer (FFTS) with $4\,\mathrm{GHz}$ instantaneous bandwidth was employed \citep{Klein2012}. The velocity resolved spectra are resampled to a velocity resolution of $0.2\,\mathrm{km s^{-1}}$. We convolved the [\CII] spectra with a Gaussian function to $20''$ resolution to increase the S/N and to better compare to the APEX CO (3$\to$2) data at an original resolution of 18$''$ (and convolved to 20$''$). The [\CII] spectra are presented on a main beam brightness temperature scale $T_{\mathrm{mb}}$, using an average main beam efficiency of $\eta_{\mathrm{mb}}$ = 0.65.  The forward efficiency is $\eta_{f}$ = 0.97. From the spectra, a first order baseline and a set of principle components were removed that were determined from a Principal Component Analysis (PCA) of the spectra from the OFF-position measurements (emission free background). The components originate from the instrument and/or atmosphere (Buchbender et al., in prep.). Systematic variations in the spectra, originating from these instrumental effects, are subtracted in this way from the ON-position spectra (see \citet{Tiwari2021} for more information). 
        The central position of the final map is located at RA(2000) = 17$^h$12$^m$03.65$^s$ and Dec(2000) = -38$^\circ$30$'$28.43$''$. As emission-free reference position, we used RA(2000) = 17$^h$10$^m$41$^s$, Dec(2000) = -37$^\circ$44$'$04$''$.

    \subsection{APEX} \label{sec:apex} 
        RCW~120 was mapped on September 21, 2019 in good weather conditions (precipitable water vapor, $\mathrm{pwv} = 0.5-0.7\,\mathrm{mm}$). The lines observed were $^{12}$CO (3$\to$2) at $345.796\,\mathrm{GHz}$ and $^{13}$CO~(3$\to$2) at $330.588\,\mathrm{GHz}$, using the LAsMA array on the APEX\footnote{APEX, the Atacama Pathfinder Experiment is a collaboration between the Max-Planck-Institut für Radioastronomie, Onsala Space Observatory (OSO), and the European Southern Observatory (ESO).} telescope \citep{Guesten2006}. LAsMA is a 7-pixel single polarization heterodyne array receiver that allows simultaneous observations of the two isotopomers in the upper ($^{12}$CO) and lower ($^{13}$CO) side-band of the receiver, respectively. The array is arranged in a hexagonal configuration around a central pixel with a spacing of about two beam widths ($\theta_{mb}$=18.2$''$ at $345.8\,\mathrm{GHz}$) between the pixels. It uses a K mirror as de-rotator. The backends are advanced Fast Fourier Transform Spectrometers \citep{Klein2012} with a bandwidth of 2 $\times$ $4\,\mathrm{GHz}$ and a native spectral resolution of $61\,\mathrm{kHz}$. The mapping was done in total power on-the-fly mode using a clean reference position at RA(2000) = 17$^h$10$^m$41$^s$, Dec(2000) = -37$^\circ$44$'$04$''$ (same as for [\CII]). The mapped region of 15$'\times$15$'$ was split into 2$\times$2 tiles. Each tile was scanned in both RA and Dec with a spacing of 9$''$ (oversampling to 6$''$ in scanning direction), resulting in a uniformly sampled map with high fidelity. All spectra are calibrated in main beam brightness temperatures $T_{mb}$ with a main-beam efficiency $\eta_{mb}$ = 0.68 at $345.8\,\mathrm{GHz}$.  The observed spectra are convolved with a Gaussian function to $20''$ resolution on a grid with a pixel size of $5''$. The spectra are resampled to a velocity resolution of $0.2\,\mathrm{km s^{-1}}$. 
	
    \subsection{\HI\ data} \label{sec:HI-data}
        We make use of \HI\ 21\,cm hyperfine emission line data and continuum at 1.4\,GHz from the Southern Galactic Plane Survey (SGPS), described in detail in \citet{McClure2005}. In particular, we use the combined images from the Australia Telescope Compact Array and the Parkes Radio Telescope with an angular resolution of $\sim$2$'$. The spectral cube at an angular resolution of $\sim$150$''$ was resampled to the same grid as the [\CII] and CO~(3$\to$2) data with a pixel size of 5$''$. The velocity resolution is 0.8\,km\,s$^{-1}$ and the total velocity range covered is -200 to 150\,km\,s$^{-1}$.

	\begin{figure*}[!ht]
	    \centering
		\includegraphics[width=.8\textwidth]{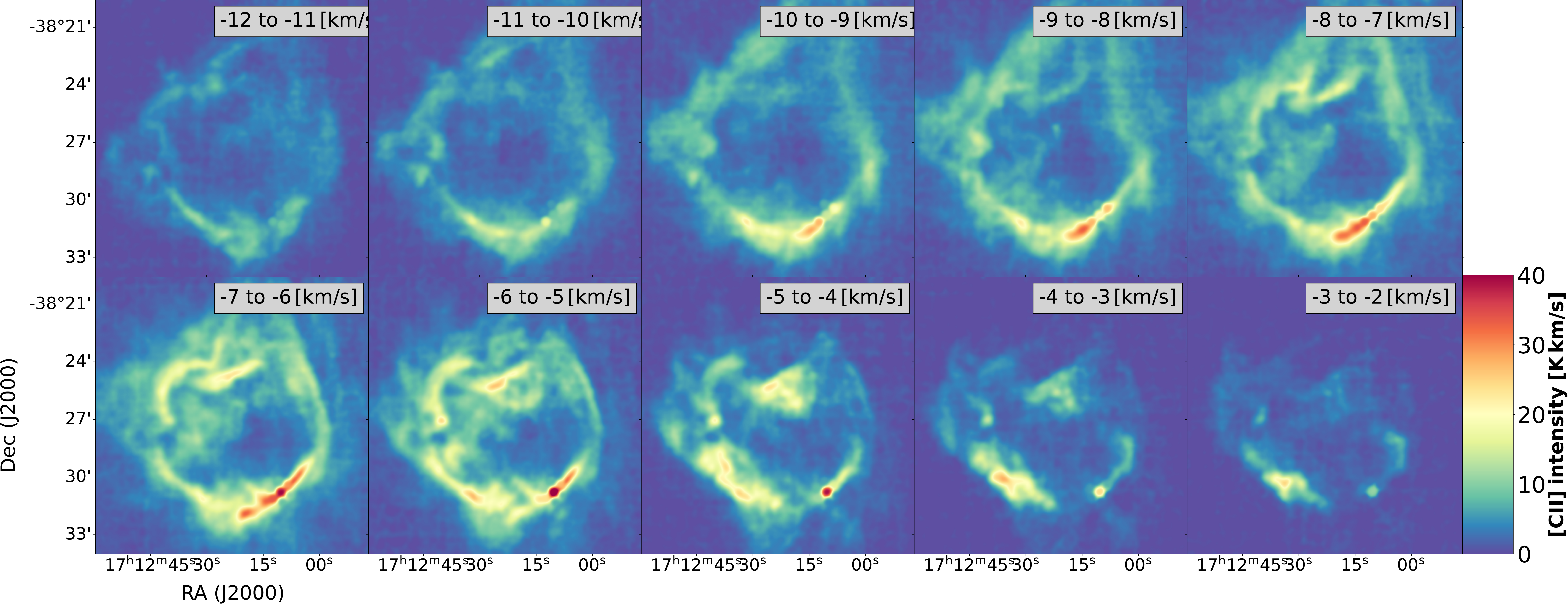}
        \includegraphics[width=.8\textwidth]{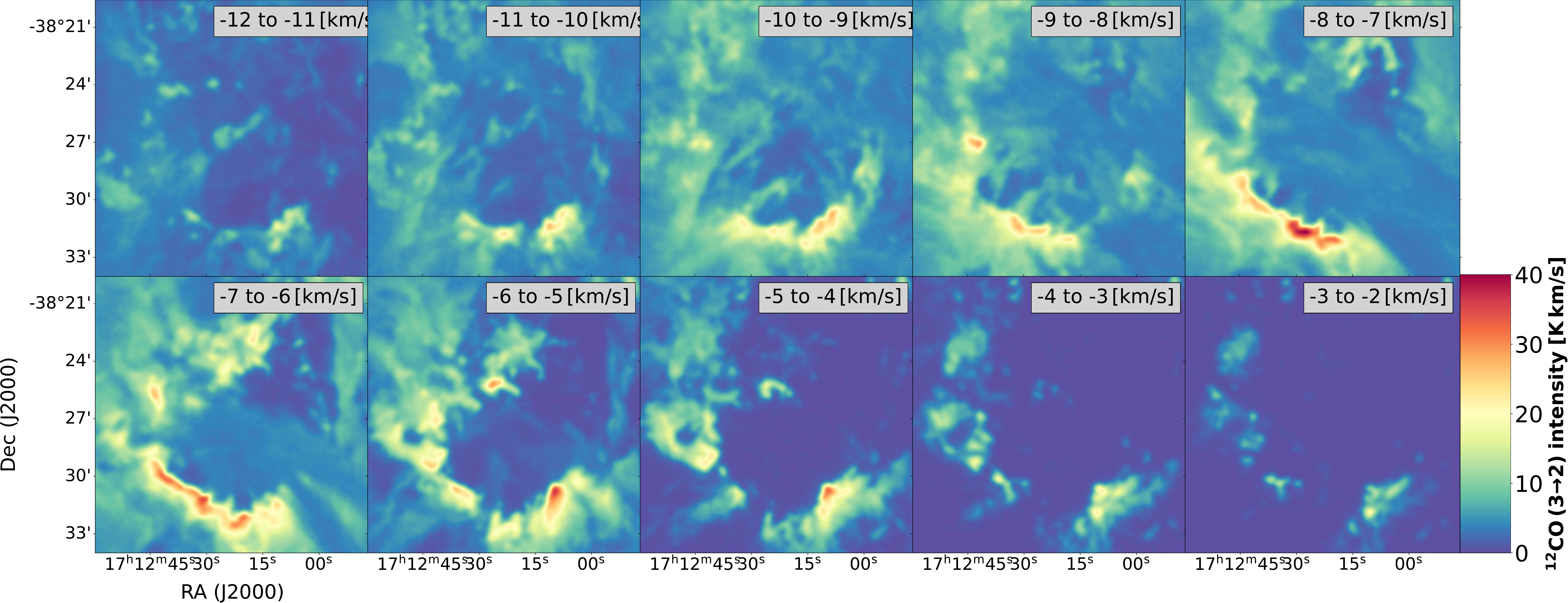}
		\includegraphics[width=.8\textwidth]{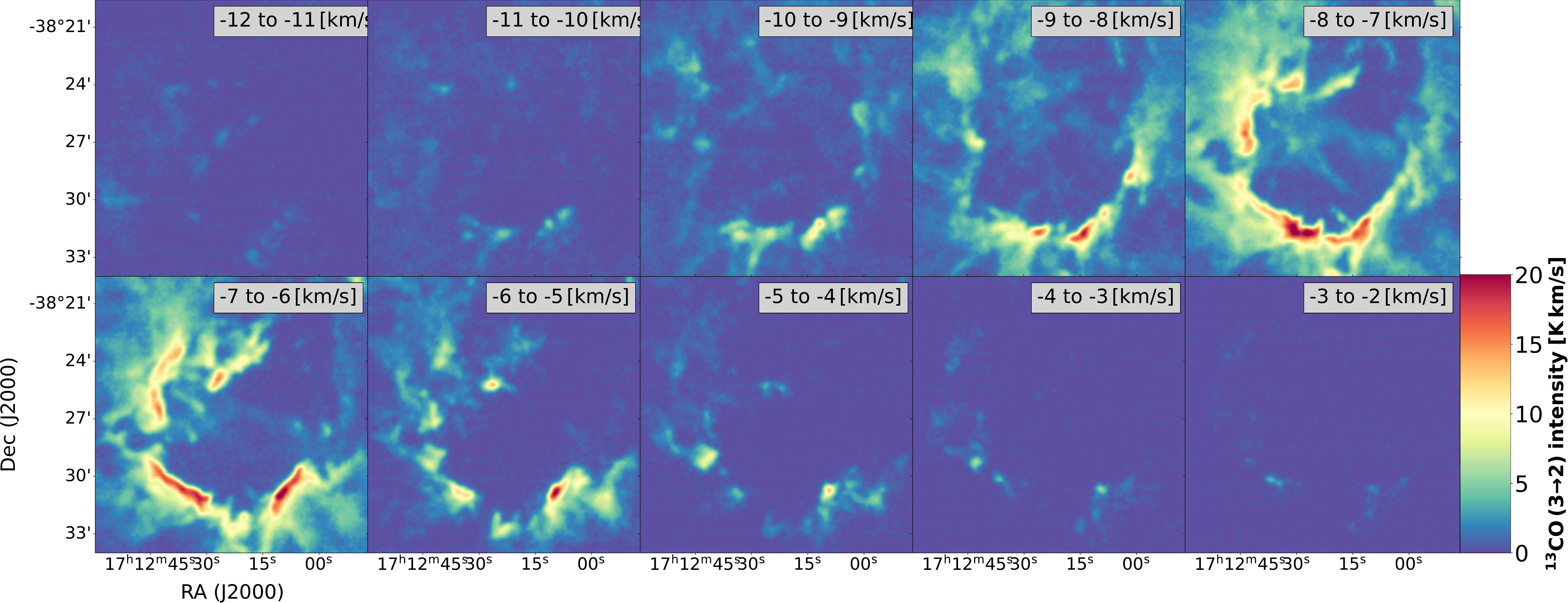}
		\caption{Channel maps of [\CII], $^{12}$CO~(3$\to$2) and  $^{13}$CO (3$\to$2) emission between $-$12 and $-$2\,km\,s$^{-1}$ integrated in steps of 1\,km\,s$^{-1}$.  Top panel: [\CII] line emission. Middle panel: $^{12}$CO~(3$\to$2) line emission. Bottom panel: $^{13}$CO (3$\to$2) line emission.} \label{fig:channel_maps}
	\end{figure*} 
	
\section{Results} \label{sec:results} 
	
    \begin{figure*}
        \centering
    	\includegraphics[width=13cm, angle=0]{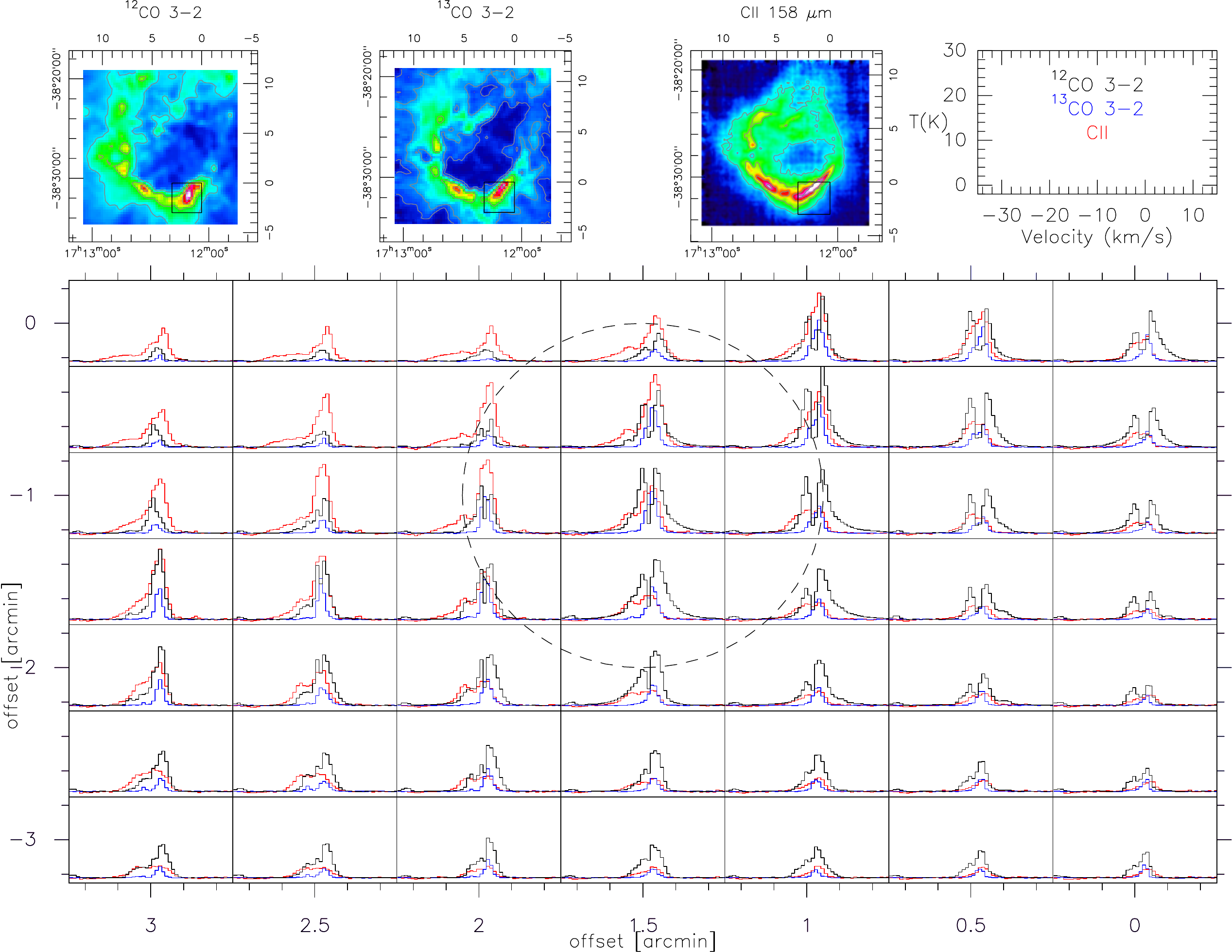}
    	\caption{The three top panels show the line integrated $^{12}$CO, $^{13}$CO (3$\to$2) and [\CII] emission (from left to right, v=-20 to 10\,km\,s$^{-1}$). The black square in each panel indicates the area where spectra of these lines are shown in the box below. $^{12}$CO and $^{13}$CO are indicated in black and blue, respectively, the [\CII] spectra in red. We note that there is no scaling for any of the lines. The x- and y-axis for the spectra are shown in the top right panel. The spatial offsets are given in arcmin in all panels, the (0,0) position refers to RA(2000)=17$^h$12$^m$03.65$^s$, DEC(2000)=-38$^\circ$30$'$28.4$''$ (see upper panels). All spectra were smoothed to an angular resolution of 30$''$ and are displayed in a grid of 30$''$. 
    	This cutout includes the clump with condensation 1 \citep{Zavagno2007}, indicated by a dashed circle, with its prominent outflow source at offsets $\sim$1.5$'$,$-$1$'$.} \label{fig:spectra}
    \end{figure*}
	
    \begin{figure}
        \centering
    	\includegraphics[width=8cm, angle=0]{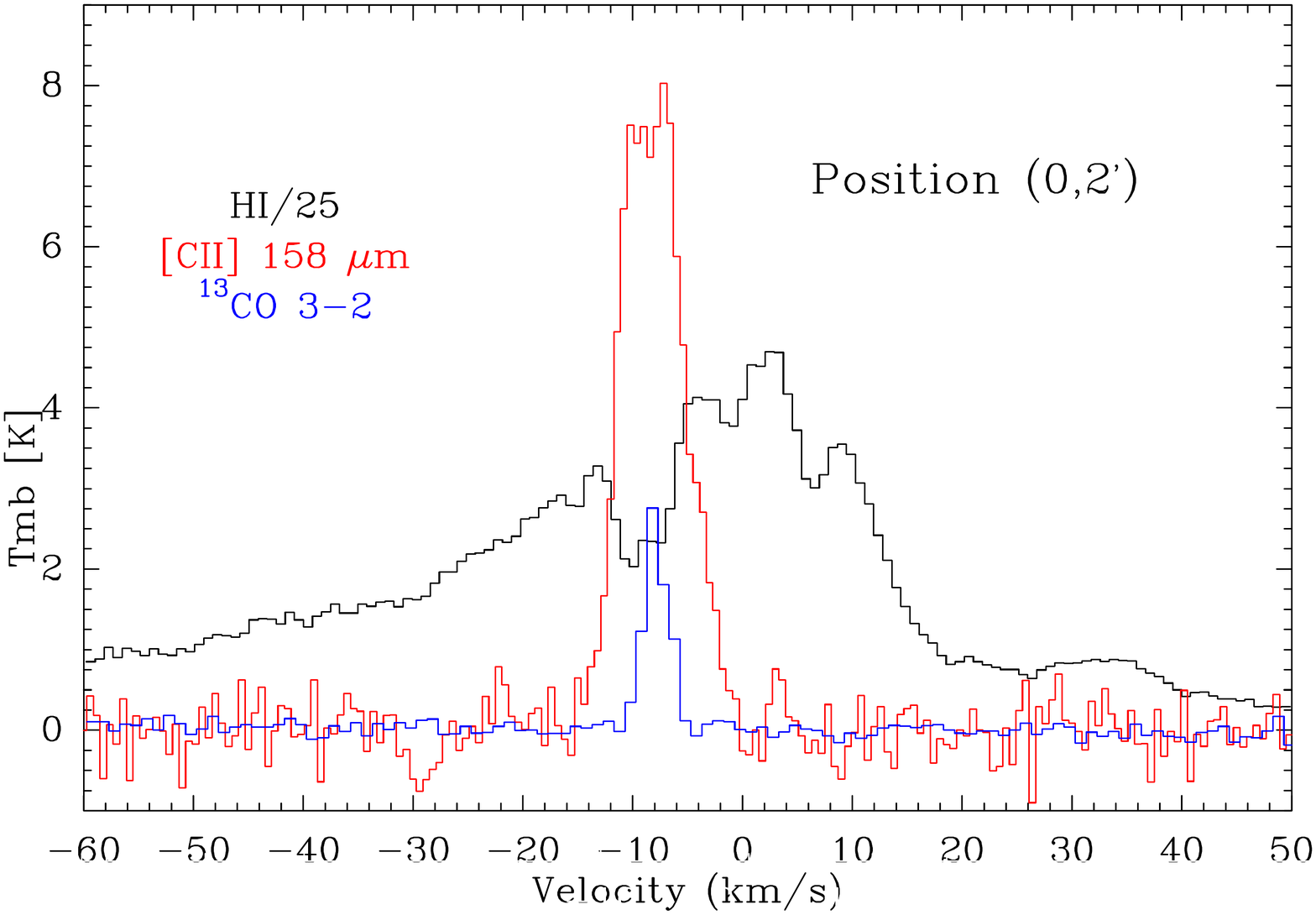}
    	\caption{Overlay between [\CII] (red), \HI\ (black), and $^{13}$CO~(3$\to$2) (blue) spectra at position (0,2$'$), indicated in Fig.~\ref{fig:intensity_maps_co_j10}. The \HI\ data have an angular resolution of $\sim$2$'$, the other spectra have a resolution of 30$''$. This position is a typical example of the line shapes in RCW~120; larger spectral line maps are given in Fig.~\ref{fig:spectra} and ~\ref{fig:HI-map}. } \label{fig:HI-spectra}
    \end{figure}

    \subsection{[\CII] and CO emission distributions} \label{sec:emission}

        Maps of line integrated emission of [\CII], $^{12}$CO and $^{13}$CO (3$\to$2) over a velocity range of $-20$ to 10\,km\,s$^{-1}$ are displayed in Fig.~\ref{fig:intensity_maps}. This is the velocity range of the red-shifted cloud that is clearly associated with the RCW~120 \HII\ region (Fig.~\ref{fig:intensity_maps_co_j10}). Comparing the spatial distribution of [\CII] and CO, we find that the molecular ring, traced by CO, is somewhat larger and envelopes the [\CII] structure. This is demonstrated in Fig. \ref{fig:pi_diagram}, which shows the integrated intensity (between $-20$ to 10\,km\,s$^{-1}$) of [\CII] and CO along a horizontal and a diagonal cut through the star (indicated by the two gray lines in Fig. \ref{fig:intensity_maps}).
        There is a strong drop in emission towards the center of \HII\ region compared to the ring. In $^{12}$CO the intensity ratio is $\sim 5-10$, in $^{13}$CO $\gtrsim10$ and in [\CII] $\sim$3-5. In addition, the  [\CII] emission peaks at slightly smaller radii than CO. This indicates a typical PDR layer structure with the [\CII] emitting layer closer to the exciting source than the molecular gas. 
    
        Figure~\ref{fig:channel_maps} shows channel maps of [\CII] and CO emission between $-12$ and $-2$\,km\,s$^{-1}$. We intentionally leave out the higher-velocity ranges ($-30$ to $-12$\,km\,s$^{-1}$ and $-2$ to $6$\,km\,s$^{-1}$, respectively) that reveal an expanding bubble in C$^+$ (but not in CO), since this feature is discussed in \citet{Luisi2021}. We here focus on the bulk emission of the cloud. All tracers show a ring-like structure around the central \HII\ region in all channels between $-$12 and $-$2\,km\,s$^{-1}$. The ring stands out particularly at the systemic velocity of RCW~120 at $-$7.5\,km\,s$^{-1}$. However, there are subtle differences between the CO and [\CII] emission distributions. The [\CII] ring is rather smooth, with a narrow opening to the east and a larger one to the northwest, but the emission drops sharply going outwards. The CO emission is more fragmented, clearly visible in the middle panel of Fig.~\ref{fig:channel_maps}, and no coherent ring-like structure is visible, but rather a succession of dense clumps along the ring. Similar to the [\CII] emitting gas, characterizing mainly the PDR, we also find the same opening to the northwest, but it extends further out than observed in the [\CII] line. Close to the systemic velocity of the cloud around $-7.5\,\mathrm{km\,s^{-1}}$ we observe a lack of CO emission following a tube-like structure extending diagonally from the southwest to the northeast. As we show in Sect.~\ref{sec:result_co}, this feature is due to self-absorption and indicates that a substantial fraction of emission  is absorbed by cold foreground material. The $^{13}$CO~(3$\to$2) emission mostly reflects the ring structure and is tracing the dense clumps discovered in dust continuum \citep{Zavagno2007,Deharveng2009,Figueira2017}. In addition, in the $^{13}$CO channel maps (Fig.\ref{fig:channel_maps}, bottom panel) we detect two lanes of emission at velocity $\sim -$7.5 km s$^{-1}$ across the \HII\ region. These lanes are clearly in the foreground because they are visible in H$\alpha$ emission as dark features in front of RCW~120 \citep{Figueira2017,Luisi2021} and they are also present in $^{12}$CO where they are optically thick and extend beyond the southwestern rim of RCW~120. All CO (3$\to$2) emission features are similar to the ones seen in lower-J CO lines, reported in \citet{Anderson2015} and \citet{Kirsanova2019}.  
    	
        We note that the ionizing star $\mathrm{CD}-38^{\circ}11636$ is shifted $\sim$110$''$ southward with respect to the geometrical center of RCW~120. As discussed in \citet{Luisi2021}, the stellar wind of the star drives a fast ($\sim$15\,km\,s$^{-1}$) expanding bubble in [\CII]. \citet{Luisi2021} report that there is no sign of this expanding shell in the CO~(3$\to$2) data. Likewise, this expansion is also not present in CO~(1$\to$0) and (2$\to$1) observations \citep{Anderson2015,Kirsanova2019}. 
	
    \subsection{[\CII], CO, and \HI\ spectra} \label{sec:spectra} 
        \subsubsection{[\CII] and CO spectra} \label{sec:CII-and-CO-spectra}
            Figure~\ref{fig:spectra} shows as an example CO and [\CII] spectra from the area around 
            condensation 1, the brightest emission region in RCW~120 in these tracers. The most prominent feature in the [\CII] spectra (red in Fig.~\ref{fig:spectra}) is the very extended, high-velocity blue wing in the northwestern area of the box (RA-offset 2$'$ to 3$'$, Dec-offsets 0 to -2$'$) which is missing in $^{12}$CO (black spectra). This is the signature of the expanding [\CII] shell discussed in \citet{Luisi2021}. On the other hand, high-velocity red- and blue-shifted $^{12}$CO emission is found at spectra around 1.5$'$,$-$1.5$'$, the peak position of $^{12}$CO emission in condensation 1. This is a protostellar CO outflow, recently reported in \citet{Figueira2020}. The $^{12}$CO spectra show a deep dip in the line shape over several velocity channels around $-$8\,km\,s$^{-1}$ where $^{13}$CO (blue spectra) has its peak emission. This is the classical picture of an optically thick $^{12}$CO line that is self-absorbed. The [\CII] line is less affected by optical depth effects in this area of RCW~120, but shows clear self-absorption features, mostly in the form of "flat-top" spectra, i.e. broad absorption over several velocity channels ($\sim-$7 km s$^{-1}$ to $-$9 km s$^{-1}$) at many positions in- and outside of the ring. The complicated line shapes in CO and [\CII] indicate a complex spatial structure of the emitting gas components with cooler material in front of the bulk emission of the cloud along the LOS. 

        \subsubsection{[\CII] and \HI\ spectra} \label{sec:HI-spectra} 
            The phase transition from warm neutral gas to cold neutral and finally molecular gas creates typical correlations between  $^{12}$CO emission and \HI\ self-absorption \citep{Wang2020}. Figure~\ref{fig:HI-spectra} displays an overlay of [\CII], $^{13}$CO (3$\to$2) and \HI\ emission at a representative position, where a prominent dip in emission in the \HI\ line is visible over a large velocity range of $-$6 to $-$11 km s$^{-1}$, centering at the bulk emission of the cloud at $\sim -7.5$\,km\,s$^{-1}$. The bulk emission is well-traced by the optically thin $^{13}$CO (3$\to$2) line that has a Gaussian shape. In contrast, the [\CII] line is not Gaussian but also shows self-absorption in the form of a "flat-top" spectrum in the same velocity range as the \HI\ self-absorption. 
            We note that the [\CII] and $^{13}$CO spectra for this plot are on a resolution of 30$''$ while the resolution of the \HI\ line is $\sim$2$'$. In 
            Appendix~\ref{HI-more}, we display overlays of [\CII] and \HI\ spectra both at $\sim$2$'$ resolution in order to demonstrate that the \HI\ broad self-absorption around  $-$7.5\,km\,s$^{-1}$ persist across the whole area observed in [\CII] and that there is little variation in the line profile. 
            In the next section, we perform a  quantitative analysis of the line emission in order to assess the amount of self-absorption in [\CII], CO, and \HI.
	
\section{Analysis} \label{sec:analysis}
    In the following, we start with a determination of the optical depths of the $^{12}$CO and [\CII] lines along the PDR ring of emission in RCW~120 (Sec.~\ref{sec:tau}). We then apply (Sec.~\ref{sec:two-layer}) a two-layer multicomponent model \citep{Guevara2020} and present the results for the [\CII] emission.
    To solve the radiative transfer equation for the CO data, we need to find a set of initial conditions that satisfies each spectrum of the spectral cube.  This is done by clustering the spectral cube via a Gaussian Mixture Model (Sec.~\ref{sec:gmm}), which allows us to focus on groups of spectra (i.e., clusters) and initializes the radiative transfer equation with fewer, but physically consistent, initial conditions.   
    In Sect.~\ref{sec:result_co}, we show and discuss the resulting CO synthetic background emission and foreground absorption maps. Next, in Sect.~\ref{sec:simline}, we perform SimLine simulations in order to test if RCW~120 is embedded in a spherical molecular cloud.
    In Sect.~\ref{hisa} we present a study of \HI\ self-absorption. 
	
    \subsection{[\CII] and CO optical depth from a one-layer model} \label{sec:tau}
        The velocity resolved optical depth of $^{12}$CO or [$^{12}$\CII] can be determined from the isotopic $^{12}$C/$^{13}$C-brightness temperature ratio with 
	
    	\begin{equation}
    		\frac{T_{\mathrm{mb, ^{12}C}}(\vel)}{T_{\mathrm{mb, ^{13}C}}(\vel)}= \frac{1-e^{-\tau(\vel)}}{\tau(\vel)} \alpha~, 
    		\label{eq:optical_depth}
    	\end{equation}
	
        \noindent
        assuming that the $^{13}$C-isotopologue is optically thin. 
        Here, $\vel$ is the respective velocity (or channel) and  $\alpha = 59\pm 10$ the local $^{12}$C/$^{13}$C abundance ratio in RCW~120, which is derived from a linear fit, $\alpha = 6.21(\pm1.00)\cdot d_{\mathrm{GC}} + 18.71(\pm7.37)$, over the carbon isotopic rate of multiple sources in the Galactic plane as a function of distance $d_{\mathrm{GC}}$ (in kpc) to the Galactic center \citep{Milam2005}. We assume the same abundance value for ionized and molecular carbon and identical excitation temperatures for all isotopologues (since they have the same collision partners) in the single-layer model. 
    
        The large abundance of CO in interstellar molecular clouds results in large optical depths of the main isotope, and in contrast to the less abundant  [$^{13}$\CII] hyper-fine emission, the $^{13}$CO isotope can also be affected by optical depth effects. In this case, the determined optical depth values from the comparison of both isotopes must be interpreted as a lower limit.

        While the [$^{12}$\CII] line is bright in PDR regions such as the one in RCW~120, the [$^{13}$\CII] line is typically faint and requires deep integration at single points \citep{Guevara2020} or averaging over larger areas.  In addition, the [$^{13}$\CII] transition splits into three hyperfine components, with a relative strength of $s_{\mathrm{2\to 1}} = 0.625$, $s_{\mathrm{1\to 0}} = 0.25$ and $s_{\mathrm{1\to 1}} = 0.125$.
        The three satellites are velocity shifted by  $\Delta {\rm v}_{\mathrm{2\to 1}} = 11.2 \,\mathrm{km s^{-1}}$, $\Delta {\rm v}_{\mathrm{1\to 0}} = -65.2\,\mathrm{km s^{-1}}$ and $\Delta {\rm v}_{\mathrm{1\to 1}} = 63.2\,\mathrm{km s^{-1}}$ with respect to the [\CII] fine structure line \citep{Ossenkopf2013, Guevara2020}. The strongest hyper-fine structure line lies closest to the main line and is therefore often affected by the wing emission of the [$^{12}$\CII] line. In such a case it is required to fit the red-shifted wing and the satellite simultaneously by a Gaussian profile in order to disentangle the wing component of the [\CII] emission from the isotope. Here, we use the conservative assumption that the total intensity in the superposition of both transition lines is dominated by the wing of the main isotope, not the [$^{13}$\CII] line. In this way we may rather underestimate the [$^{13}$\CII] intensity and thus the optical depth, also providing a lower limit.

    	\begin{figure*}[!h]
    	    \centering
    		\includegraphics[width=0.4\textwidth]{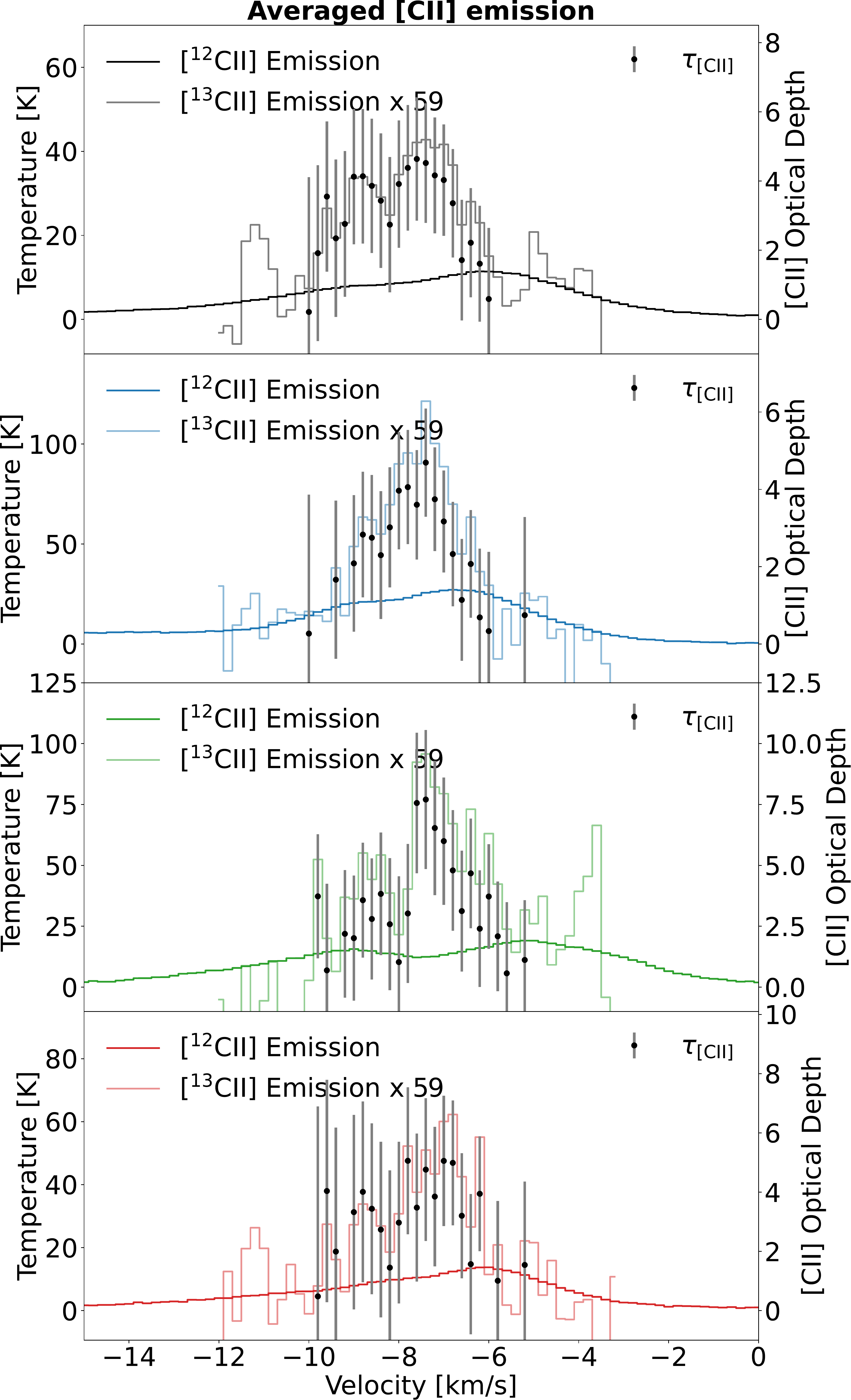}
    		\includegraphics[width=.475\textwidth]{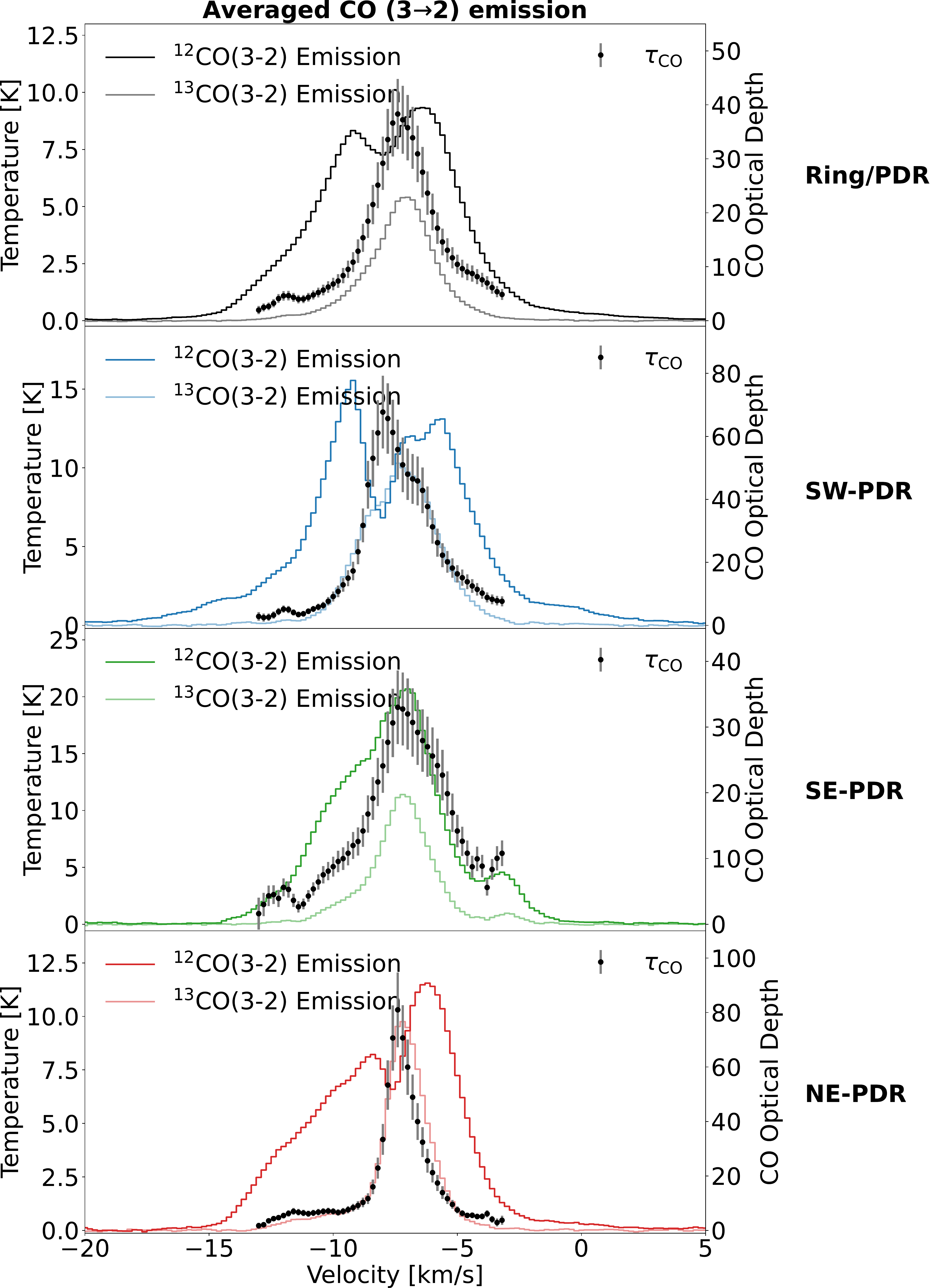}
    	    \caption{Averaged spectra of the structures along the ring of RCW~120, sorted by the regions indicated in Figure~\ref{fig:intensity_maps} (ring, southwestern PDR, southeastern PDR and northeastern PDR). Left panels: averaged [$^{12}$\CII] and scaled up [$^{13}$\CII] emission by the local abundance $^{12}$C/$^{13}$C-ratio of 59. Right panels: averaged $^{12}$CO and $^{13}$CO (3$\to$2) emission. The black data points with error bars indicate the velocity resolved optical depth. \label{fig:optical_depth}}
    	\end{figure*} 
	
        The weak intensity and splitting of the [$^{13}$\CII] hyper-fine transition into three lines result in a small signal-to-noise ratio in each pixel and thus do not allow one to spatially resolve the isotope emission.
        Thus, we average emission over arcmin$^2$ large regions with a dendogram based approach \citep{Rosolowsky2008,Robitaille2019}. Dendrograms allow for a hierarchical structure analysis of spectral data cubes and thus provide a quantitative way to determine the most prominent extended structures in RCW~120 and local bright structure within those. The determined areas are indicated by contours in the [\CII] map (left panel) in Fig. \ref{fig:intensity_maps}. The black contour outlines the entire PDR and bulk of the [\CII] emission from RCW~120. The blue, green and red contours split the emission into the three brightest regions: the southwestern PDR with condensation 1, 
        that includes a protostellar outflow, the southeastern PDR with condensation 2 \citep{Zavagno2007}, and the northeastern PDR, respectively. The corresponding averaged spectra of [$^{12}$\CII] and [$^{13}$\CII] and $^{12}$CO and $^{13}$CO (3$\to$2) emission originating from these regions are shown in Fig. \ref{fig:optical_depth} (left and right, respectively). The dark colored spectra represent the main $^{12}$C line and the light colors show the $^{13}$C isotope emission. The [$^{13}$\CII] line is averaged over all three hyper-fine transitions (weighted by the relative strength of each satellite) and multiplied by the local carbon abundance ratio of 59. The CO averaged spectra are shown on the original intensity scale without multiplying by the abundance ratio. The velocity dependent optical depth is indicated by the black data points with error bars, which are determined by 
	
    	\begin{equation}
        	\Delta \tau = \left| \frac{\tau^2 }{(1+\tau)e^{-\tau}-1} \right| \sqrt{\frac{ \alpha^2 T_{\mathrm{^{13}C}}^2 \sigma^2_{\mathrm{^{12}C}} + \alpha^2T_{\mathrm{^{12}C}}^2 \sigma^2_{\mathrm{^{13}C}} +  \Delta\alpha^2 T_{\mathrm{^{13}C}}^2 T_{\mathrm{^{12}C}}^2  }{\alpha^4 T_{\mathrm{^{13}C}}^4}}~,
        	\label{eq:optical_depth_error}
        \end{equation}

        \noindent
        with $\sigma$ the root mean square (rms) of the brightness temperature of each  respective isotope (see Sect.~\ref{app:tau-error} in the appendix for the derivation of this equation). 
        	 	
        All [\CII] spectra show a double peak structure followed by a blue-shifted tail but the scaled up [$^{13}$\CII] line overshoots the [\CII] emission and peaks where we observe the dip in the [\CII] spectrum, indicating strong self-absorption. We emphasize that the velocity range of the overshoot is not restricted to the bulk emission of the cloud at $-7.5$ km s$^{-1}$ but covers the whole velocity range between $\sim -10$ to $-$6~km s$^{-1}$. The optical depth effects in [\CII] thus cover a broad distribution in velocity, resulting in "flat-top" spectra and a pronounced dip at $-7.5$ km s$^{-1}$. Both features are also well visible in the individual spectra (Sect.~\ref{sec:spectra}). The velocity resolved optical depth is above unity for all spectra and peaks at the [\CII] absorption dip. The averaged optical depth varies along the ring between $\tau_{\mathrm{[\CII]}}\sim3-4$.
        	
        It is known that the rotational transitions of $^{12}$CO often suffer from self-absorption effects, due to the high abundance of the $^{12}$CO molecule \citep{Castets1990}. The absorption effects in RCW~120 are strong enough to be clearly visible in the channel maps around the systemic velocity of the molecular cloud $-7.5\,\mathrm{km s^{-1}}$, see Fig.~\ref{fig:channel_maps}. We observe that most of the emission originating from the southwestern PDR is absorbed. This leads to a visual impression of missing emission along a lane extending diagonally from the inner ring towards the southwestern ring. To determine the strength of the self-absorption we compare the $^{12}$CO with the less abundant $^{13}$CO line, assuming all emission comes from a single layer. \\
        Similar to the [\CII] line, the $^{12}$CO (3$\to$2) line also shows in Fig.~\ref{fig:optical_depth} a double peak structure for all average spectra with a blue- and red-shifted tail. The most prominent red-shifted emission is found for the southwestern PDR region, most likely due to the internal outflow source. A comparison to the $^{13}$CO~(3$\to$2) line clearly reveals that the apparent double peak structure is due to optical depth effects, since the isotope shows a single peak at the location of the emission dip. However, the line-dip is somewhat blue-shifted with respect to the $^{13}$CO (3$\to$2) peak. This is especially visible in the southwestern PDR. A possible explanation is that the observed blue shift of the optical depth maximum with respect to the peak emission is caused by a momentum transfer of the expanding [\CII] bubble into the surrounding molecular cloud. The mechanical impact of the expanding bubble could lead to an accumulation of blue-shifted material along the line of sight and thus a higher optical depths for some velocity channels. We note that the $^{12}$CO optical depth is overall very high, up to a few 10s at the bulk emission velocity, but this is a consequence of the assumption of the single layer model. These unrealistically high values show that it is required to use a two-layer model where different exctiation temperatures are considered. Only with this approach, presented in the next section, it is possible to model self-absorption effects, such as dips, due to a colder foreground material, which is intrinsically not possible with a single-layer model. 
    
    \subsection{Two-layer multicomponent model}\label{sec:two-layer}
    
    \subsubsection{An introduction to the model}
        In the following, we use a generalized version of the model introduced by \cite{Guevara2020} which solves the radiative transfer equation for two gas layers with multiple velocity components that is also adequate for optically thick lines. By assuming a background (bg) layer with a higher excitation temperature than the foreground (fg), $T_{\mathrm{ex},i_\mathrm{bg}} > T_{\mathrm{ex},i_\mathrm{fg}}$, the background is responsible for the bulk emission of the cloud, while the foreground layer, providing mainly absorption, is located in front of the bulk emission. The observed brightness temperature then calculates as 
    
    	\begin{equation}  \label{eq:2layer} 
    		\begin{split}
    			T_{\mathrm{mb}}(\vel) = & \left[ \mathcal{J}_{\nu}(T_{\mathrm{ex},\mathrm{bg}}) \, \left( 1-e^{-\sum_{i_\mathrm{bg}}\tau_{i_\mathrm{bg}}(\vel)}\right) \right] e^{-\sum_{i_\mathrm{fg}} \tau_{i_\mathrm{fg}}(\vel)} + \\
    			&  \mathcal{J}_{\nu}(T_{\mathrm{ex},\mathrm{fg}}) \, \left( 1-e^{-\sum_{i_\mathrm{fg}}\tau_{i_\mathrm{fg}}(\vel)}\right)~,  \\
    		\end{split}	
    	\end{equation} 
	
        \noindent
        with the sum over $i_{\mathrm{bg, fg}}$ components located either in the background or foreground, respectively. The foreground components can be associated with the emitting cloud but they also can be unrelated foreground material. Associated foreground is naturally formed through a decreasing temperature gradient along the line of sight when a background PDR layer around an \HII\ region is connected with a foreground molecular cloud. 
        Because the absorption dip of the more abundant $^{12}$C isotope is always close to the systemic velocity of RCW~120, see Fig. \ref{fig:optical_depth}, we can safely assume that the origin of the cold absorbing components is not an unrelated cloud along the line of sight. In addition, the geometry of RCW~120 with a hot, central \HII\ region, followed by a warm PDR layer and then a cool molecular cloud already suggests a gas temperature gradient. In the following we refer to the absorption by the foreground components as self-absorption. We perform the model fit on both isotopic lines simultaneously, thus the combined line shape is given by:
    	\begin{equation}
    		\Phi_i(\vel) = \phi_i(\vel) + \sum_{\mathrm{F\to F'}}s_{\mathrm{F\to F'}}\phi_i(\vel-\Delta\vel_{\mathrm{F\to F'}})~,
    	\end{equation}  
        \noindent
        where $\Delta\vel_{\mathrm{F\to F'}}$ is the line velocity offset from the main isotope and $s_{\mathrm{F\to F'}}$ the relative intensity of each line (1 for $^{13}$CO and see above for [$^{13}$\CII]). 
        The line profile of each component $\phi_i(\vel)$ is a normalized Gaussian profile: 
    	
    	\begin{equation}
    		\phi_i(\vel)=\frac{2\sqrt{\ln 2}}{w_i\sqrt{\pi}}e^{-4\ln 2~\left(\frac{\vel-\vel_{0,i}}{w_i}\right)^2}~,
    	\end{equation}
	
    	\noindent
        with $\vel_{0,i}$ being the central (LSR) velocity and $w_i$ the line's width (FWHM) of the component $i$. The factor $4\ln2$ stems from the usage of the FWHM $w_i$ instead of the standard deviation $\sigma_i$. The equivalent brightness temperature of a black body emission at a temperature $T_{\mathrm{ex}}$ is 
	
    	\begin{equation}
    		\mathcal{J}_{\nu}(T_{\mathrm{ex},i}) = \frac{T_0}{e^{T_0/T_{\mathrm{ex},i}}-1}~,
    	\end{equation}
    	
        \noindent
        with the equivalent temperature of the transition $T_0=h\nu/k_B$ and $\nu$ the transition frequency. Each component of the model is characterized by four quantities: excitation temperature, optical depth, position (LSR velocity), and width (FWHM). The position of the component in velocity space and line width are confined by the observed line. The excitation temperature and optical depth are not independent from each other. The excitation temperature can only be read from the line intensity in case of high optical depths. We therefore first constrain a suitable 
        excitation temperature from the observed data, which then leaves only the optical depth as a free parameter.   
    	
        Assuming that the warm background is partly shining through the cold layer lying in front of it, we can determine the background excitation temperature around the emission peak main beam temperature $T_{\mathrm{p, mb}}$ by:
        \begin{equation}
    		T_{\mathrm{ex}} = T_0\ln\left(\frac{T_0}{T_{\mathrm{p,mb}}}\left(1-e^{-\tau_{\mathrm{p}}}\right)+1\right)^{-1}~,
    		\label{eq:excitation_temperatur}
    	\end{equation}
        \noindent
        using the optical depth $\tau_{\mathrm{p}}$  determined through equation~\ref{eq:optical_depth}. 
    	
        \subsubsection{Two-layer fit of the [\CII] ring}\label{sec:result_cii} 
	
            \begin{table*}[htb!]
            \centering
                \caption{Physical properties of the emitting and absorbing layers along the ring of RCW 120 from [\CII] using a foreground excitation temperature of 15~K and 30~K, respectively.}
                \label{tab:physical_conditions_cii}
                \begin{tabular}{lc|ccccc|ccccc}
                    \hline
                    \hline
                    \rowcolor[gray]{0.8}\multicolumn{2}{c}{$T_{\mathrm{ex, fg}}=15\,\mathrm{K}$} & \multicolumn{5}{|c}{Background} &  \multicolumn{5}{|c}{Foreground} \\
                    \hline
                    Region & (1) & (2) & (3) & (4) & (5) & (6) & (7) & (8) & (9) & (10) & (11) \\
                    & $A$ & $\tau_{\mathrm{[\CII], bg}}$ &$N_{\mathrm{[\CII], bg}}$ & $N_{\mathrm{H, bg}}$ & $M_{\mathrm{H, bg}}$ & $L_{\mathrm{[\CII], bg}}$ & $\tau_{\mathrm{[\CII], fg}}$ & $N_{\mathrm{[\CII], fg}}$ & $N_{\mathrm{H, fg}}$ & $M_{\mathrm{H, fg}}$ & $L_{\mathrm{[\CII], fg}}$\\
                    & [pc$^2$] & & [$10^{18}\,\mathrm{cm}^{-2}$] & [$10^{21}\,\mathrm{cm}^{-2}$] & [$M_{\sun}$] & [$L_{\sun}$] & & [$10^{18}\,\mathrm{cm}^{-2}$] & [$10^{21}\,\mathrm{cm}^{-2}$] & [$M_{\sun}$] & [$L_{\sun}$]\\
                    \hline
                    Ring & 12.9 & 2.8 & 3.1 & 19.6 & 2018.0 & 357.3 & 0.7 & 0.5 & 3.3 & 344.0 & 1.77\\
                    SW PDR & 0.6 & 2.9 & 3.9 & 24.1 & 118.0 & 31.9 & 0.8 & 0.4 & 2.8 & 14.0 & 0.07\\
                    SE PDR & 0.5 & 3.8 & 4.9 & 30.7 & 114.0 & 21.1 & 0.6 & 0.4 & 2.3 & 8.0 & 0.04\\
                    NE PDR & 1.5 & 3.4 & 2.8 & 17.7 & 212.0 & 38.1 & 0.6 & 0.3 & 1.8 & 21.0 & 0.11\\
                    \hline
                    \rowcolor[gray]{0.8}\multicolumn{2}{c}{$T_{\mathrm{ex, fg}}=30\,\mathrm{K}$} & \multicolumn{5}{|c}{Background} &  \multicolumn{5}{|c}{Foreground} \\
                    \hline
                    Ring & 12.9 & 2.6 & 2.8 & 17.6 & 1818.0 & 333.2 & 1.2 & 1.1 & 6.6 & 685.0 & 58.5\\
                    SW PDR & 0.6 & 2.9 & 3.8 & 23.9 & 118.0 & 31.8 & 1.4 & 0.9 & 5.7 & 28.0 & 2.34\\
                    SE PDR & 0.5 & 3.6 & 4.7 & 29.3 & 109.0 & 20.9 & 0.9 & 0.5 & 3.3 & 12.0 & 1.16\\
                    NE PDR & 1.5 & 3.3 & 2.7 & 17.0 & 203.0 & 37.5 & 1.1 & 0.6 & 3.9 & 47.0 & 4.19\\
                    \hline
                \end{tabular}
                \tablefoot{
                 (1) The area  of the regions, defined by Dendrograms \citep{Rosolowsky2008,Robitaille2019}. \\
                 (2,7) [\CII] peak optical depth.\\
                 (3,8) [\CII] column density (eq.~8). \\
                 (4,9) Hydrogen column density N$_{\rm H}$ = ${\rm N}_{\rm {\HI}} + 2 {\rm N}_{{\rm H}_2}$, using C/H = 1.6 $\cdot$ 10$^{-4}$ \citep{Sofia2004}. \\
                 (5,10) Mass from hydrogen column density (eq.~A.1). \\
                 (6,11) Luminosity for the background and foreground layers (eq.~A.3).
                }
            \end{table*}

            We can determine the [\CII] optical depth for each component of the emitting background as a function of the excitation temperature $T_{\mathrm{ex}}$ and column density $N_{[\mathrm{\CII}]}$  \citep{Guevara2020}  
        
        	\begin{equation}
        		\tau_{\mathrm{[\CII]}, i}(\vel)=\Phi_i(\vel)\frac{g_u}{g_l}\frac{c^3}{8\pi \nu^3}A_{ul} N_{[\mathrm{\CII}], i}\frac{1-e^{-\frac{T_0}{T_{\mathrm{ex},i}}}}{1+\frac{g_u}{g_l}e^{-\frac{T_0}{T_{\mathrm{ex},i}}}}~,
        	\label{eq:NCII}
        	\end{equation}
        
            \noindent with $\nu = 1900.5369\,\mathrm{GHz}$ the rest frequency of the [\CII] fine structure line, $A_{ul} = 2.3\cdot10^{-6}\,\mathrm{s^{-1}}$ the Einstein coefficient for spontaneous emission \citep{Wiese2007}, $T_0=h\nu/\kb=91.25\,\mathrm{K}$ the equivalent temperature of the upper level, and the statistical weights of the [\CII] transition energy levels $g_u=4$ and $g_l=2$. The maximum excitation temperature along the ring in RCW~120, i.e., the background layer, is determined with equation \ref{eq:excitation_temperatur} to be $T_{\mathrm{ex}} = 50 - 70\,\mathrm{K}$. 
        	
        	For the foreground layer, we constrain the lower and upper limits of the excitation temperature of [\CII].
            The lower limit is derived from the [\CII] energy balance. We consider here the case without excess radiative heating, this means the energy input for the ionized carbon gas is provided only by the standard interstellar radiation field and cosmic ray (CR) ionization. In a diffuse optically thin gas the [\CII] line emissivity can be computed by \citep{Ossenkopf2013}:
    
        	\begin{equation}
        		\begin{aligned}
        			\int \epsilon\,\mathrm{d}\vel &= \frac{hA_{ul}c^3}{8\pi \kb \nu^2}\times N_{\mathrm{[\CII]}} \times \frac{g_ue^{\frac{-T_0}{\Tex}}}{g_l+g_ue^{\frac{-T_0}{\Tex}}}\\
        													&= 2.3\cdot10^{-28} \frac{\mathrm{J}}{\mathrm{sr}} \times N_{\mathrm{[\CII]}} \times E({\Tex})~,
        			\label{eq:cii_emisivity}
        		\end{aligned}
        	\end{equation}  
	        \noindent
            where we abbreviate the excitation-temperature dependent ratio as $E({\Tex})$ and $\epsilon$ is the heating efficiency. The total [\CII] cooling is 
        	\begin{equation}
        		\Lambda_{\mathrm{[\CII]}} = \int_{4\pi}\int \epsilon\,\mathrm{d}\vel\,\mathrm{d}\Omega = 2.9\cdot10^{-27}\,\mathrm{W}\times N_{\mathrm{[\CII]}} \times E({\Tex})
        		\label{eq:cii_cooling}
        	\end{equation} 
	
            \noindent and the heating of the cold diffuse cloud is given by the ambient UV field and the CR ionization. The energy density $\varrho_{\mathrm{UV}}$ for an UV field\footnote{The Habing field Go relates to the Draine field $\chi$ by $\chi  = 1.71 G{\rm o}$ where Go is the mean interstellar radiation field from \citet{Habing1968,Draine1978}.} of $\chi = 1$ 
            is $6.8\cdot10^{-21}\,\mathrm{J/cm^3}$. It is absorbed mainly by interstellar dust providing an UV optical depth of $\tau_{\mathrm{UV}}\sim 3 A_{\rm v}/1.08$. In a column with $A_{\rm v} = 1$ the energy heating of dust and gas is thus:
        	\begin{equation}
        		F_{\mathrm{UV}} =\left(1-e^{-\tau_{\mathrm{UV}(A_{\rm v} = 1)}}\right)\varrho_{\mathrm{UV, \chi=1}} c  = 1.9\cdot10^{-6}\mathrm{\frac{W}{m^2}}~.
        	\end{equation}
   
            With a typical gas heating efficiency $\varepsilon = 0.01$ \citep{Okada2013}, and the gas column density of $N_{\mathrm{H},  A_{\rm v} = 1}=1.9\cdot10^{21}\,\mathrm{cm^{-2}}$ for an $A_{\rm v}=1$ we can translate this into a heating rate per hydrogen; and with the carbon to hydrogen ratio $\mathrm{C/H}=1.6\cdot10^{-4}$ \citep{Sofia2004} we obtain a [\CII] heating rate of:
        	\begin{equation}
        		\Gamma_{\mathrm{UV}} = \varepsilon \frac{F_{\mathrm{UV}}}{N_{\mathrm{H}, A_{\rm v} = 1} \times \mathrm{C/H}} \times N_{\mathrm{[\CII]}} = 6.3\cdot10^{-30}\,\mathrm{W} \times N_{\mathrm{[\CII]}}~.
        	\end{equation}
	
            The CR heating rate is given by the CR heating per ionization  $ Q ~ \sim 10\,\mathrm{eV}$ \citep{Glassgold2012} and the CR ionization rate $\eta_{\mathrm{H}}\approx2\cdot10^{-16}\,\mathrm{s^{-1}}$ \citep{Indriolo2015}. Dividing by the C/H-ratio gives the CR [\CII] heating rate:
        	
        	\begin{equation}
        		\Gamma_{\mathrm{CR}} = \frac{Q \times \eta_{\mathrm{H}}}{\mathrm{C/H}} \times N_{\mathrm{[\CII]}} = 2\cdot10^{-30}\,\mathrm{W} \times N_{\mathrm{[\CII]}}~,
        	\end{equation} 
        
            \noindent resulting in a total heating of the diffuse [\CII]:
        	\begin{equation}
        		\Gamma_{\mathrm{[\CII]}} = \Gamma_{\mathrm{UV}} + \Gamma_{\mathrm{CR}} = 8.3 \cdot 10^{-30}\mathrm{W} \times N_{\mathrm{[\CII]}}~.
        		\label{eq:cii_heating}
        	\end{equation}
	
            Energy balance between [\CII] cooling (equation \ref{eq:cii_cooling}) and [\CII] heating (equation \ref{eq:cii_heating}) then provides the limit for the [\CII] excitation temperature in case of low radiation fields:
        	\begin{equation}
        		\begin{aligned}
        			E({\Tex}) = 0.003 = \frac{g_ue^{\frac{-T_0}{\Tex}}}{g_l+g_ue^{\frac{-T_0}{\Tex}}}\\
        			\Rightarrow \Tex = \frac{T_0}{\ln\left( \frac{2}{E} -2\right)} = 14\,\mathrm{K}~.
        			\label{eq:Tex_min}
        		\end{aligned}
        	\end{equation}
        [\CII] as the main cooling line in diffuse material cannot get colder. This is the foundation of the phase separation in the neutral interstellar medium as derived already by \citet{Field1969}, just with updated numbers. The CR ionization rate from \citet{Indriolo2015} is probably a lower limit; a factor 10 higher rate for example leads to an excitation temperature of 17~K. We thus take for simplicity in the following 15~K as the minimum excitation temperature for our foreground layer.
        
        In order to find an upper limit for the C$^+$ excitation temperature of the foreground layer, we use 
        the information from an independent estimate of the hydrogen column density for the foreground which is given by optical absorption measurements towards the ionized gas bubble of the \HII region in \citet{Zavagno2007}. They derived a visual extinction of ${\rm A}_{\rm v}\sim4.36$, corresponding to a hydrogen column density of $8.15\cdot$10$^{21}$ cm$^{-2}$ that in turn translates into a C$^+$ column density of $1.3\cdot10^{18}$ cm$^{-2}$.
        From Fig.~\ref{fig:tex_colden_cii}, we derive then that such a value for the C$^+$ column density is only achieved for a C$^+$ excitation temperature of $\sim$30~K for the foreground. 
        Significantly higher foreground C$^+$ excitation temperatures T$_{\rm ex, [\CII]}$ are thus excluded. In the case of emission lines with strong absorption dips, as observed by \cite{Guevara2020} it is possible to derive an upper limit for the excitation temperature from the depth of the absorption dip. Eventually, at temperatures much below the upper level energy of 91.25~K, the derived gas parameters depend only weakly on the exact foreground excitation temperature \citep{Guevara2020}. 

\begin{figure*}[!h]
    	    \centering
    		\includegraphics[width=.45\textwidth]{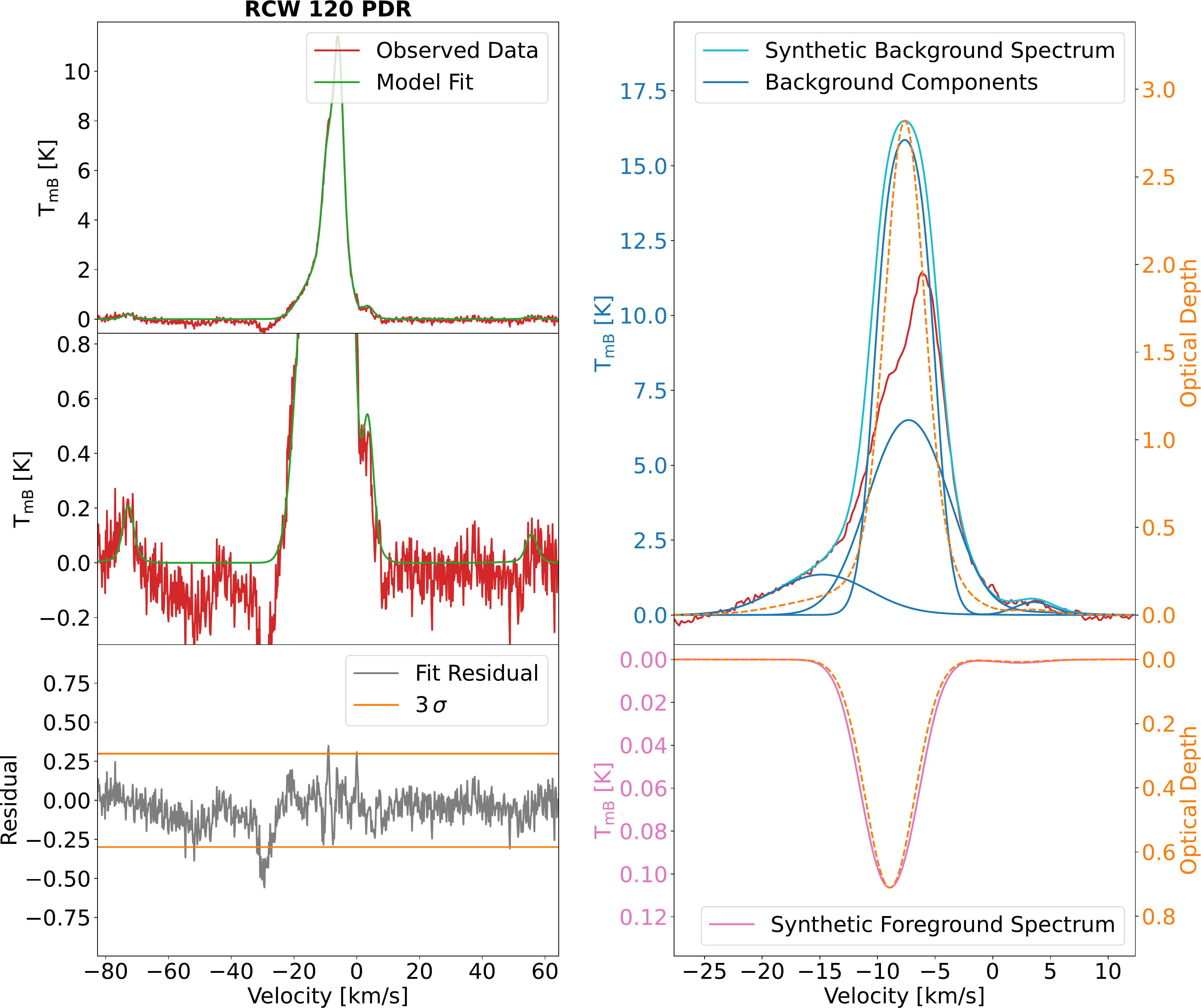}
    		\includegraphics[width=.45\textwidth]{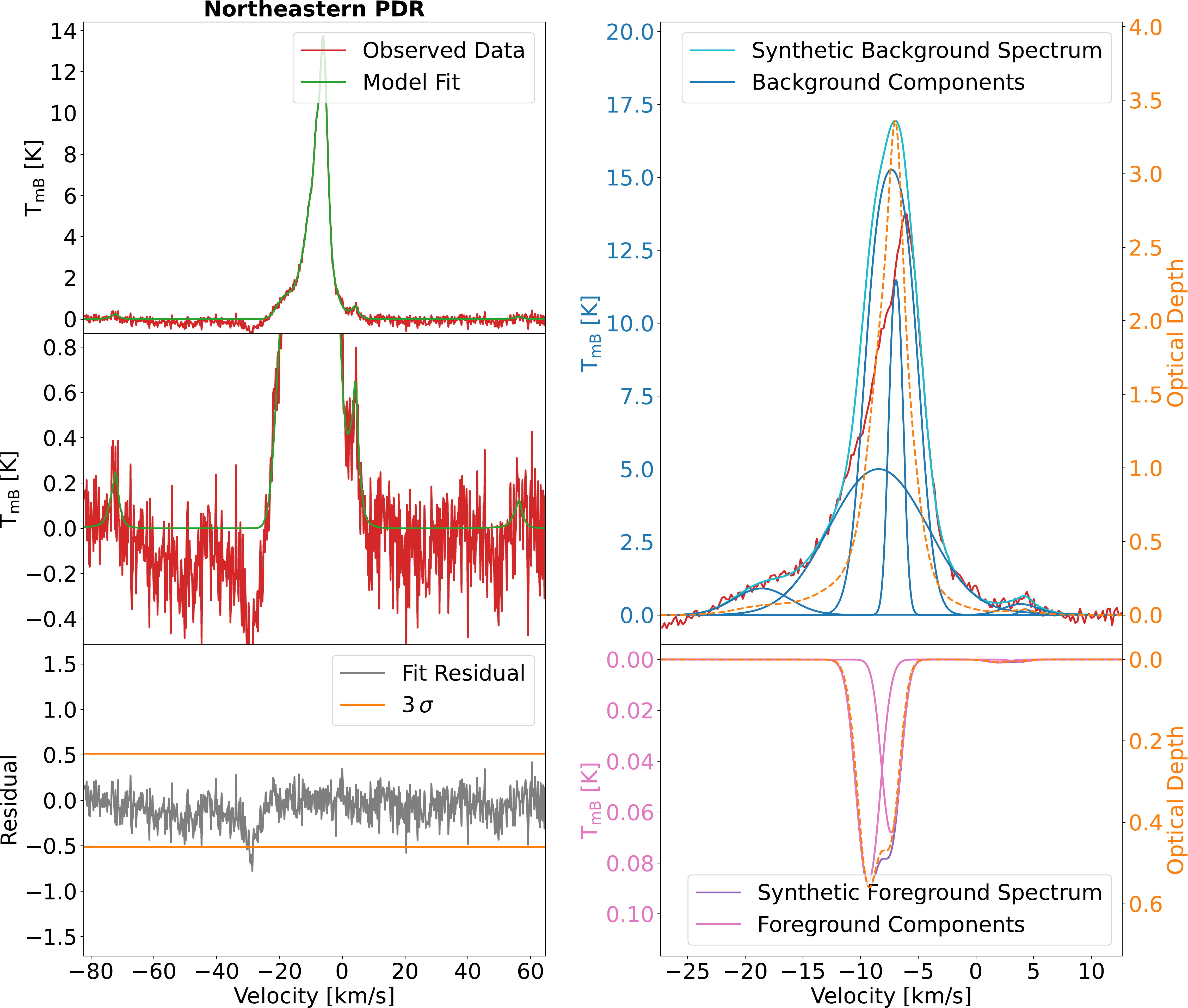}
    		\includegraphics[width=.45\textwidth]{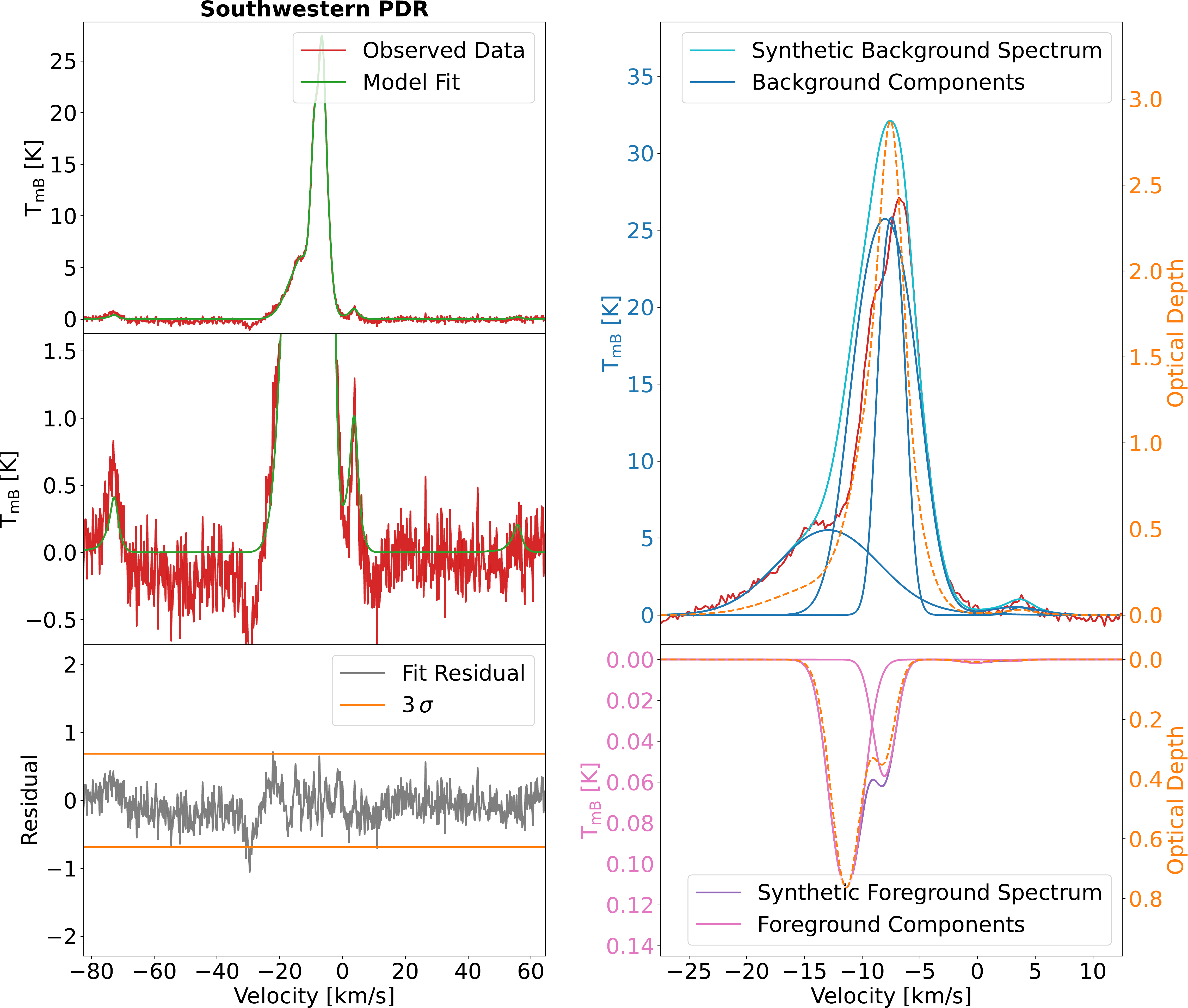}
    		\includegraphics[width=.45\textwidth]{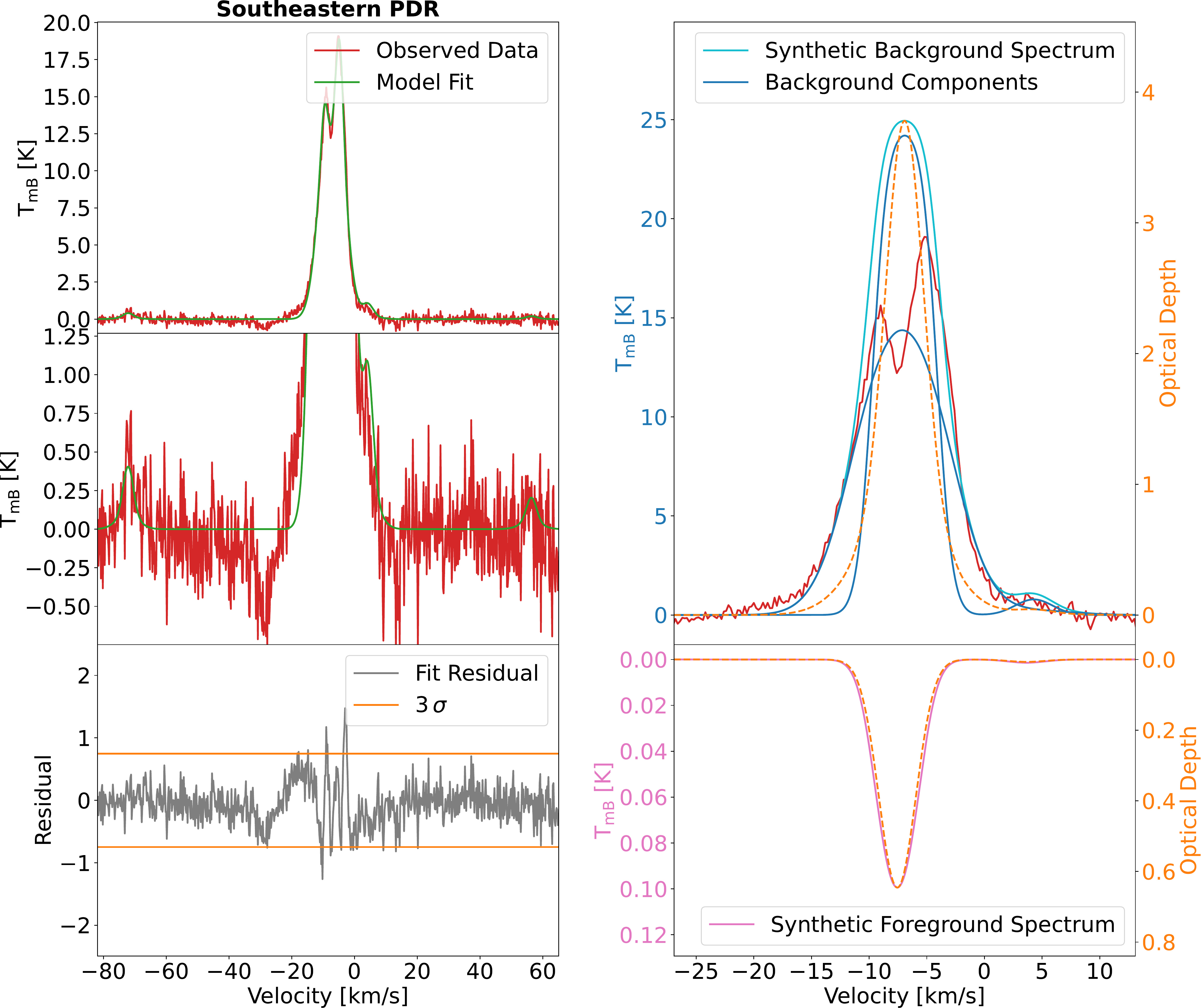}
    		\caption{Two-layer multicomponent model results for the four regions extracted with the Dendogram technique. The left top subfigure is a template for all regions: The observed averaged [\CII] spectrum is indicated in the left upper panel in red and the green curve on top is the resulting model fit. In the left middle panel, the temperature axis range is smaller so that the weak  [$^{13}$\CII] satellites around the main isotope line are clearly visible. The left lower panel shows residuals of the fit in gray and the two horizontal orange lines indicate the $3\sigma$ level. In the right top panel the red spectrum once more displays the observed [\CII] emission and the dark blue lines show the single Gaussian components. The light blue line is the superposition of all synthetic background components. The dashed orange line indicates the velocity resolved optical depth (values on the right axis). The right lower panel shows the pink single foreground Gaussian components and the violet line indicates the superposition of all the synthetic foreground components. The dashed orange line illustrates again the velocity resolved optical depth.} \label{fig:two_layer_cii}
    	\end{figure*}

        The results of the two-layer multicomponent model are displayed in Fig.~\ref{fig:two_layer_cii}.  The red spectra in each subfigure represent the observed averaged [\CII] emission in the RCW~120 PDR ring (top left), the northeastern PDR (top right), the southwestern PDR (bottom left), and the southeastern PDR (bottom right), respectively. The green line shows the model fit result, which always provides a very good match to the data, as demonstrated by the fit residuals (gray data points in the left lower panel in each subfigure). 
        The upper right panel of each subfigure shows the background components in dark blue that sum up to the overall synthetic background. The difference in intensity between this synthetic background spectrum and the observed spectrum (in red) shows how much emission is absorbed and thus missing in the integrated intensity map. The absorbing cold foreground components are shown in the right lower panel by the pink curve. It becomes obvious that there is significant absorption, since the synthetic spectra are a factor ~2 higher than the observed ones.  
    
        Table~\ref{tab:physical_conditions_cii} summarizes the values for [\CII] column density $N_{\mathrm{[\CII]}}$ and hydrogen column density N$_{\rm H}$ = ${\rm N}_{\rm {\HI}} + 2 {\rm N}_{{\rm H}_2}$, mass, and luminosity for the background- and foreground layers. The C$^+$ excitation temperature for the background varies between $\sim$50 and 70~K depending on the region (see Tables in Appendix~\ref{appendix-c1}). For the foreground, we use the lower (15~K) and upper (30~K) limits of the C$^+$ excitation temperature. 

        From Table~\ref{tab:physical_conditions_cii}, it becomes obvious that the background [\CII] column density $N_{\mathrm{[\CII],bg}}$ is typically $3 - 5 \times 10^{18}\,\mathrm{cm^{-2}}$, which corresponds to a hydrogen column density of $2 - 3 \times 10^{22}\,\mathrm{cm^{-2}}$. We anticipate that the C$^+$ emission background emission arises from PDR surfaces of dense, molecular clumps in the UV-illuminated ring of RCW~120 with a total mass of $\sim$2000 M$_\odot$. In this case, H$_2$ molecules are the primary collision partner of C$^+$. We focus here on the emission of the ring but we note that there is little variation in the values for C$^+$ and hydrogen column density for the different regions in RCW~120 (individually listed in the table). 
        In \citet{Luisi2021}, we extracted $\sim$100 clumps from the $^{12}$CO (3$\to$2) data and derived typical clump densities of a few 10$^4$\,cm$^{-3}$, radii between 0.3 - 1\,pc, and a total mass of 2500\,M$_\odot$. From the C$^+$ column densities, assuming a density of a few 10$^4$\,cm$^{-3}$ \citep{Luisi2021}, we roughly estimate a thickness of the C$^+$ surface layer as a sum over these clumps of $\sim$0.1$-$0.2\,pc. 
        This is consistent with C$^+$ in a thin surface layer around the clumps
        but potentially several of these clumps along the line of sight in the beam.
        This calculation, however, ignores beam and volume filling factors, but shows that the PDR surface origin of the C$^+$ emission in the ring is a reasonable approach. 
        It is remarkable that the luminosity in the [\CII] line per cloud mass is quite uniform with a typical ratio of $L_{\mathrm{[\CII], bg}}/M_{\mathrm{H, bg}} \approx 0.18 L_{\sun}/M_{\sun}$ and a maximum of 0.27 for the SW PDR. This means that in the bright background PDRs we have a typical mass to single-line luminosity conversion factor of $0.2 L_{\sun}/M_{\sun}$.

        The foreground C$^+$ column density $N_{\mathrm{[\CII],fg}}$ is typically a factor 3 lower, the corresponding hydrogen column densities are of the order of $2 - 7 \times 10^{21}\,\mathrm{cm^{-2}}$ (depending on T$_{{\rm [\CII],fg}}$ = 15~K or 30~K). However, there is significant mass concentrated in this foreground component, summing up to 344\,M$_\odot$ (or 685 M$_\odot$) in the PDR ring area. The nature of the foreground gas is a priori unknown, but we assume that it constitutes an extended, low-density, cool atomic envelope. To settle this issue, we study the \HI\ emission and absorption properties in Sect.~\ref{hisa} and come back to this point in the discussion in Sect.~\ref{sec:discussion}.

    \subsection{Clustering of CO spectra} \label{sec:gmm}
        \subsubsection{The Gaussian mixture model} \label{sec:gmm-sub}

            In contrast to the [\CII] data, the signal-to-noise ratio is sufficiently high in both CO isotopes at most positions of the map to run the two-layer multicomponent model for every spectrum of the entire CO data cube. The model requires an initial guess (number of components, velocity, line width, excitation temperature and optical depth) as an input to converge towards a physically meaningful solution. While the selection of reasonable initial guesses is a simple exercise for a single spectrum, doing so for thousands of spectra requires an unreasonable amount of time.  
            We thus need a different approach here leveraging the fact that the shape of each spectrum in a data cube is not random, but is confined by the local physical conditions. The idea is to apply identical initial guesses for similar spectra.	To do so, we first need to group all spectra into clusters in which each cluster represents a typical spectrum assembled by similar spectra. In a second step, the average cluster spectrum is fed into the two-layer multicomponent model and fitted. The values from those fits are then used as initial guesses for all spectra of the corresponding cluster. 
            
             We employ a Gaussian Mixture Model (GMM), which groups similar objects by linear combinations of multidimensional Gaussian distributions. 
             This is an unsupervised task and the GMM employed in this study is one of the possible options chosen among many others \citep{Brunton2019}. A GMM is a probabilistic approach to clustering and has been proven to be a robust methodology in a variety of real world problems, for example by  \cite{Jones2019}. Treating each cluster as Gaussians allows for decomposing the whole dataset as a linear combination of Gaussians; it is then a convenient and simple choice which allows for an inexpensive and robust algorithm. Different initialization may result in somewhat different clusters. However, the GMM is only used here as a preprocessing step for the two-layer multicomponent model and we verified that small differences in the identified clusters do not alter our conclusions.
     
             Let $\bm{S}$ be a set of $n$ spectra. Each spectrum $\bm{s}\in\bm{S}$ is defined as a vector of temperature values depending on the velocity v. Given a number of channels $D$ in velocity space (hereafter feature space) a spectrum $\bm{s}$ is therefore a $D$-dimensional point.
    	
            A GMM aims at describing the probability distribution of the data set $P(\bm{s})$, as a weighted, linear combination of multiple Gaussian distributions $\mathcal{N}(\bm{s}$):
            \begin{equation}
                P(\bm{s}) = \sum_{c} \bm{\phi}_{c} ~ \mathcal{N}(\bm{s}; \bm{\mu}_{c}, \Sigma_{c})~,
            	\label{eq:gmm1}
            \end{equation}
            in which the multidimensional Gaussian distribution is defined as 
            \begin{equation}
            	\mathcal{N}(\bm{s}; \bm{\mu}_{c}, \Sigma_{c}) = \frac{\exp{(-\frac{1}{2}(\bm{s} - \bm{\mu}_{c})^{T}}\Sigma_{c}^{-1}(\bm{s} - \bm{\mu}_{c}))}{\sqrt{(2\pi)^{D} |\Sigma_{c}|}}\,,
            	\label{eq:gmm2}
            \end{equation}
            with mean $\bm{\mu}_{c}$, covariance matrix $\Sigma_{c}$ and weight $\bm{\phi}_{c}$. The probability of cluster $c$ is $P(c) = \phi_{c}$ and the normalization is given as 
            \begin{equation}
                \sum_{c} \bm{\phi}_{c} = 1~,
            	\label{eq:gmm3}
            \end{equation}
            $\bm{\mu}_{c}$ is a $D$-dimensional vector and $\Sigma_{c}$ is a matrix of size $D \times D$. 
    	
            The GMM starts with a random guess for the $k$ clusters to discover and it solves an Expectation-Maximization (EM) algorithm \citep{Dempster1977} to converge to an optimal solution. The EM algorithm iteratively updates the mean $\bm{\mu}_{c}$ and covariance matrix $\Sigma_{c}$ of each cluster $c$ in order to maximize the following measure of likelihood:
            	\begin{equation}
            		\log P(\bm{S}) = \sum_{i}^{n} \log (\sum_{c}^{k} \bm{\phi}_{c} ~ \mathcal{N}(\bm{s}; \bm{\mu}_{c}, \Sigma_{c}))~,  
            		\label{eq:likelihood}
            	\end{equation}
            The EM algorithm always guarantees convergence to a \textit{local} optimum. While we can simply provide a list of spectra to the GMM algorithm  without further doing, we remove the dimensions dominated by noise. A dimension of the feature space is removed when the rms of this dimension is lower than $3\,\langle\sigma\rangle_{\mathrm{cube}}$, with $\langle\sigma\rangle_{\mathrm{cube}}$ the average rms of the spectral cube. Furthermore, we scale the intensity of each spectra by
        	\begin{equation}
        	    T' = \frac{T_{\mathrm{mb}} - T_{\mathrm{mb, min}} }{T_{\mathrm{mb, max}} - T_{\mathrm{mb, min}}}~,
        	\end{equation}
            \noindent so that the temperature range of every spectra is confined between 0 and 1. This forces the GMM to cluster the spectra based on the line shape regardless of the intensity. The amplitude of each spectrum is proportional to the column density (or optical depth) which has a unique solution for a given excitation temperature. However, the shape of the spectra determines the amount of Gaussian components, which is not found by the model itself but is a model input parameter. 
        \subsubsection{Bayesian information criterion} \label{sec:bic}
	
        	\begin{figure}
             \centering
        		\includegraphics[width=.45\textwidth]{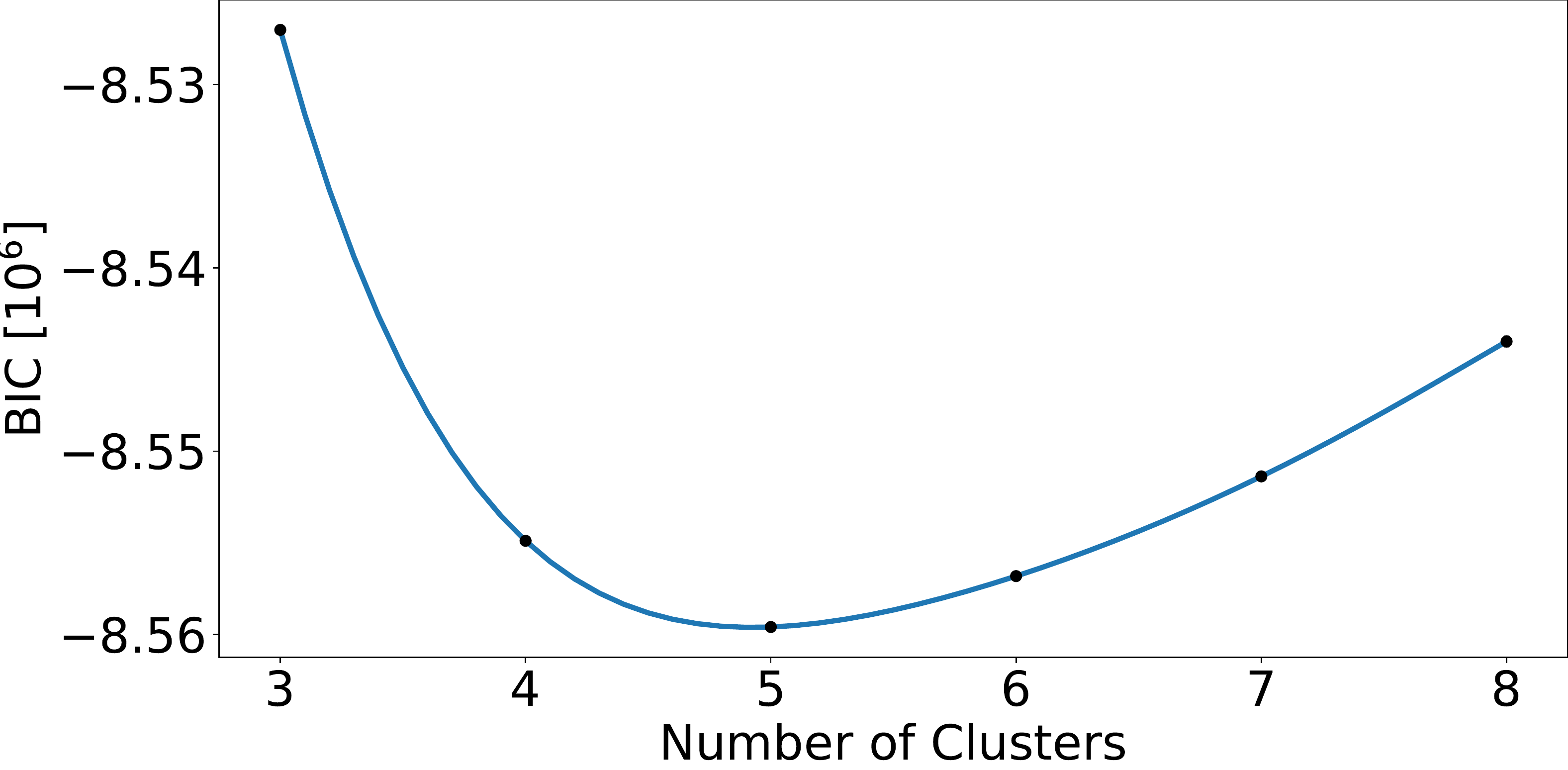}
        		\caption{Bayesian information criterion. The black points represent the mean BIC value for each cluster $k_i$ of the $^{13}$CO spectral cube. The error bars (which are very small and thus almost not visible) show the standard deviation. A parabolic fit is also shown (see blue line) visualizing the trend of the data points.} \label{fig:bic}
        	\end{figure}

            \begin{figure}[!h]
                \centering
        		\includegraphics[width=.45\textwidth]{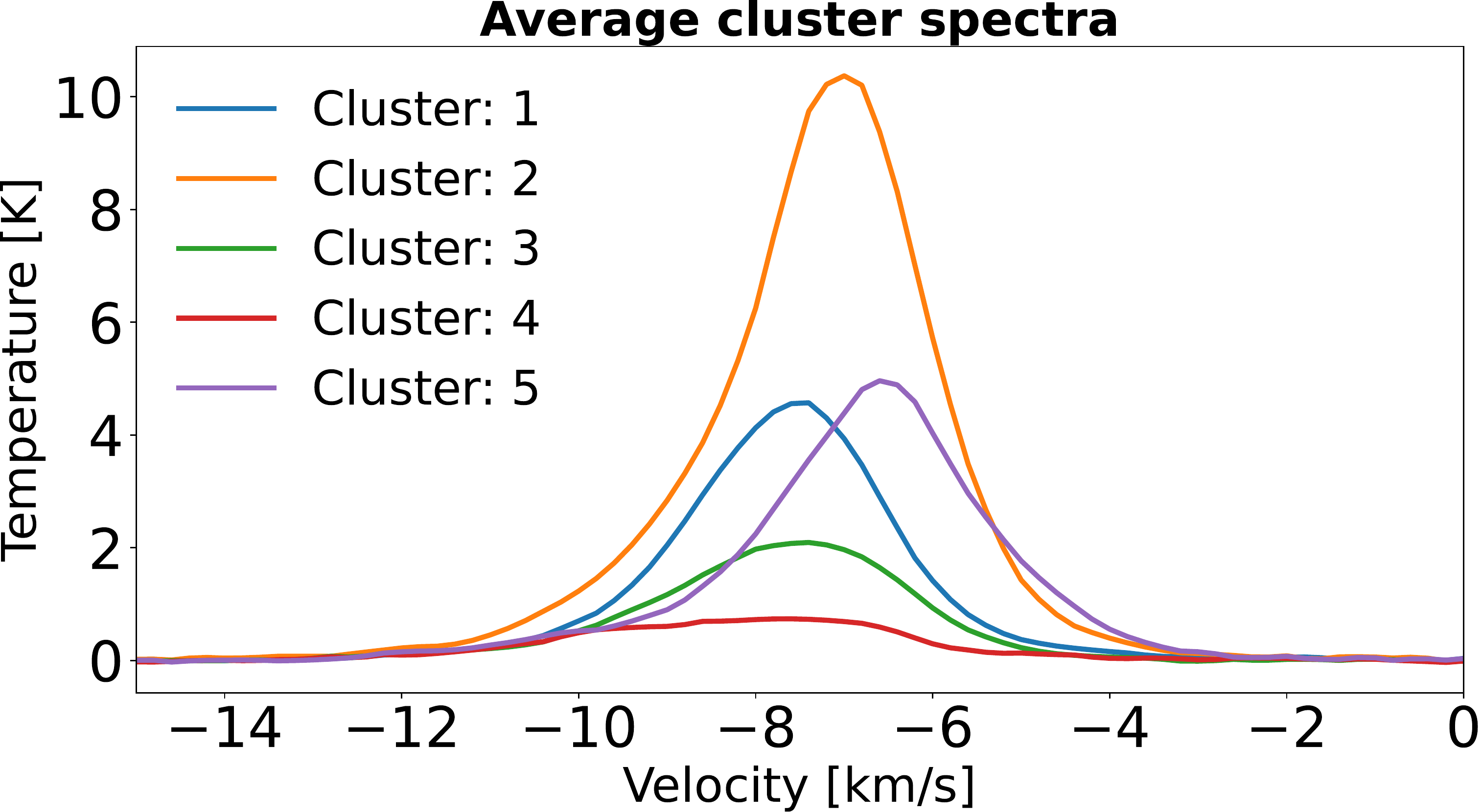}
        	    \includegraphics[width=.45\textwidth]{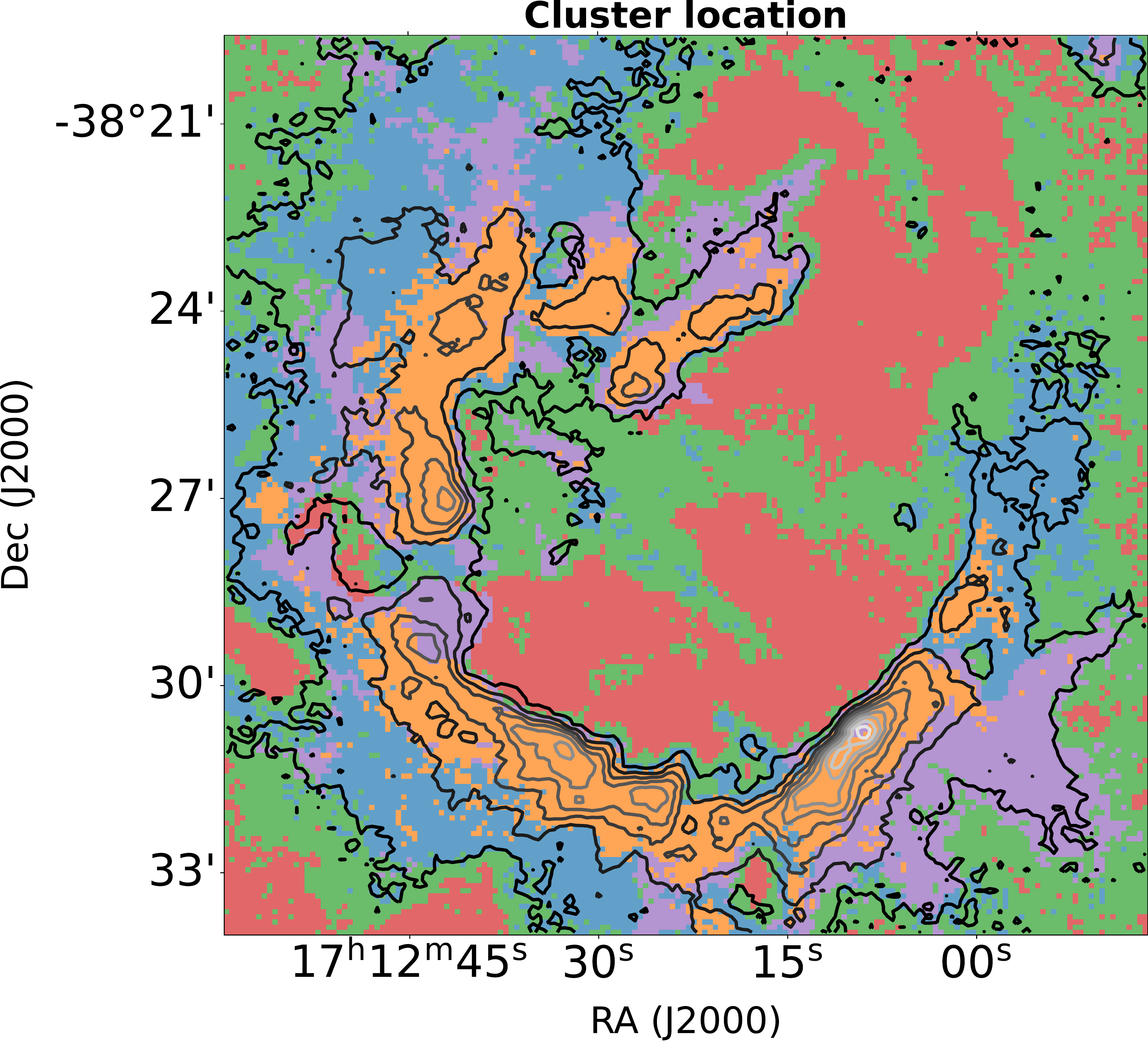}
        	    \caption{GMM clusters. Top panel: The average spectra of the clusters are plotted in different colors. Bottom panel: The corresponding cluster 
        	    locations, adopting the same color scheme, are shown. The integrated $^{13}$CO (3$\to$2) emission is indicated by the contours. } \label{fig:cluster_spectra}
            \end{figure} 

             To select an appropriate number of clusters $k$ we employ the Bayesian information criterion (BIC), a useful statistical test for model selection \citep{Schwarz1978}. Given the total number of spectra $n$ and clusters $k$, the BIC is defined as:
	
        	\begin{equation}
        		\begin{aligned}
        			\mathrm{BIC}(k) &= N(k) \log(n) - 2\mathcal{L}(k)~;\\
        			N(k) &= k - 1 + k~D + \frac{k~D(D-1)}{2}~.
        			\label{eq:BIC}
        		\end{aligned}
        	\end{equation}
	
        	\noindent
        	We note that $\mathcal{L}$ is a measure of likelihood (see Eq. \ref{eq:likelihood}) and $D$ is the dimensionality of the feature space.
        	
            Over a range of candidate numbers $\bm{K}= ( k_{1},k_{2},... )$, we choose the $k = k_{i}$ which minimizes the BIC. This allows to select the simplest model (i.e., fewest parameter) and avoids model-overfitting. Since the GMM starts in an initial random state, which affects the final outcome to some extend, we calculate the BIC for each $k_i$ multiple times (100 times), following the procedure in \cite{Jones2019}. Finally, the mean BIC value and the standard deviation is derived for each $k_i$. 
            	
            We apply the GMM on the molecular carbon isotope $^{13}$CO~(3$\to$2) since it is not strongly affected by foreground absorption effects which "artificially" change the spectral shape, i.e., dips in the spectrum. In addition, the $^{13}$CO emission does not show high-velocity wings. The determined BIC for the $^{13}$CO~(3$\to$2) data cube is shown in Fig.~\ref{fig:bic}. The black dots represent the mean BIC value and the error bars the standard deviation (which is small and thus almost not visible in the plot). We reach the minimum value at $k=5$, implying that the 3D data cube can be clustered into 5 different regions. The average spectra of each cluster, along with their locations, are shown in Fig.~\ref{fig:cluster_spectra}. The strong ring emission is shown by the orange cluster 2 and the corresponding averaged spectrum of the ring is indicated by the orange spectrum in the upper panel of Fig.~\ref{fig:cluster_spectra}. Cluster 1 and 5 surround cluster 2 as shown in the lower panel of Fig.~\ref{fig:cluster_spectra}. The superposition of their average spectra almost coincides with the average spectrum of cluster 2.  
            All three clusters form the emission by the ring in RCW~120 as indicated by the contours. The velocity shift between cluster 1 and 5 indicate that even the dense ring has a complex velocity structure. Cluster 3 (green) is mainly visible in the two lanes at the center of the bubble and around the ring. The two lanes are also visible in H$\alpha$ in absorption and thus are located at the nearside of the region. Finally, cluster 4 (red) mainly shows the weak CO emission from the inner part of the bubble.
	
    \subsection{Physical properties of the CO (3$\to$2) emission} \label{sec:result_co} 
        \subsubsection{Average CO spectra}	

            \begin{table*}[htb!]
                \centering
                \caption{Physical properties of the emitting and absorbing layers along the ring of RCW~120 from CO~(3$\to$2). }
                \label{tab:physical_conditions_co}
                \begin{tabular}{lc|ccccc|ccccc}
                    \hline
                    \hline
                    \multicolumn{2}{c}{} & \multicolumn{5}{|c}{Background} &  \multicolumn{5}{|c}{Foreground} \\
                    \hline
                    Region & (1) & (2) & (3) & (4) & (5) & (6) & (7) & (8) & (9) & (10) & (11)\\
                           & $A$ & $\tau_{\rm CO, bg}$ & $N_{\mathrm{CO, bg}}$ & $N_{\mathrm{H_2, bg}}$ & $M_{\mathrm{H_2, bg}}$ & $L_{\mathrm{CO, bg}}$ & $\tau_{\rm CO, fg}$ & $N_{\mathrm{CO, fg}}$ & $N_{\mathrm{H_2, fg}}$ & $M_{\mathrm{H_2, fg}}$ & $L_{\mathrm{CO, fg}}$\\
                     & {\small [pc$^2$]} & & {\small [$10^{18}\,\mathrm{cm}^{-2}$]} & {\small [$10^{21}\,\mathrm{cm}^{-2}$]} & {\small [$M_{\sun}$]} & {\small [$L_{\sun}$]} & & {\small [$10^{18}\,\mathrm{cm}^{-2}$]} & {\small [$10^{21}\,\mathrm{cm}^{-2}$]} & {\small [$M_{\sun}$]} & {\small [$L_{\sun}$]} \\
                    \hline
                    Cluster 1 & 9.9 & 9.0 & 0.4 & 4.3 & 685.0 & 2.5 & 1.7 & 0.2 & 1.8 & 284.0 & 0.1\\
                    Cluster 2 & 6.2 & 23.4 & 0.9 & 10.5 & 1035.0 & 2.0 & 1.7 & 0.2 & 1.8 & 180.0 & 0.1\\
                    Cluster 3 & 15.7 & 4.0 & 0.2 & 2.2 & 561.0 & 3.5 & 1.9 & 0.2 & 1.9 & 491.0 & 0.2\\
                    Cluster 4 & 11.2 & 1.4 & 0.1 & 0.9 & 169.0 & 1.7 & 2.4 & 0.1 & 1.7 & 310.0 & 0.1\\
                    Cluster 5 & 6.3 & 9.5 & 0.4 & 5.0 & 510.0 & 1.9 & 1.8 & 0.2 & 2.0 & 205.0 & 0.1\\
                    \hline
                \end{tabular}
            \tablefoot{
            (1) Cluster size of the regions defined by the GMM.\\
            (2,7) CO~(3$\to$2) peak optical depth. \\
            (3,8) CO column density, determined using Eq.~\ref{eq:tau_co}. \\
            (4,9) Molecular hydrogen column density, using CO/H$_2$ = 8.5 $\cdot$ 10$^{-5}$ \citep{Tielens2010}.\\
            (5,10) Molecular hydrogen mass of the layers.\\
            (6,11) Luminosity for the background and foreground layers.
            }
            \end{table*}

            Similar to the [\CII] spectra we fit the radiative transfer equation (\ref{eq:2layer}) for multiple components distributed between two layers to the observed CO (3$\to$2) spectra. The optical depth as a function of column density and excitation temperature is given by \citep{Mangum2015}: 
        	\begin{equation}
        		\tau_{\mathrm{CO}, i}(\vel) = \phi_i(\vel)N_{\mathrm{CO}, i}\frac{ 8\pi^{3}\mu^2J_u }{3h} \left( \frac{\kb T_{\mathrm{ex},i} }{hB}+\frac{1}{3} \right)^{-1}e^{\frac{-T_u }{T_{\mathrm{ex},i} }}\left( e^{\frac{T_0}{T_{\mathrm{ex},i} }}-1 \right)~,
        		\label{eq:tau_co}
        	\end{equation}
	
        	\noindent
            with the dipole moment $\mu$, the upper J-level $J_u$, the rotational constant $B$  and  the temperature of the upper level $T_u$, which can be approximated as: 
        	\begin{equation}
        		T_u = \frac{hBJ_u(J_u+1)}{\kb}~.
        	\end{equation}
        	
            The average (over the PDR ring) excitation temperature of the warm emitting background is determined with equation (\ref{eq:excitation_temperatur}) to be $\sim40\,\mathrm{K}$. This value is the same value as in \cite{Luisi2021}. For the cold absorbing foreground, we assume a temperature of  $6\,\mathrm{K}$ as a typical value for cold molecular clouds. We note that for larger foreground temperatures the model fit goodness decreases drastically unable to reproduce the observed CO (3$\to$2) lines. The assumptions of constant excitation temperature for all velocity components in the  background and foreground, respectively, is justified: for the background, the velocity components that contribute the most to the emission are the ones around the bulk emission of the cloud (around $-$7.5 km s$^{-1}$) and thus all arise from the same volume of gas. The foreground velocity components are slightly more scattered in velocity, but here, the assumption that we have uniform conditions in the cold gas phase is straightforward and justified.

            To determine the emitting background layer, we fit the less abundant isotopologue $^{13}$CO (3$\to$2) with the model first. The corresponding $^{12}$CO (3$\to$2) emitting layer is then determined by scaling up the $^{13}$CO column density  with  the carbon abundance ratio. However, there are components which are only visible in the main isotope but lost in the noise of the weaker  $^{13}$CO (3$\to$2). These background components are added in the second step. Finally, the model fit is used on both molecular CO lines simultaneously, including the foreground layer, while the previously determined background layer is fixed.  This simultaneous fit ensures that the additional background components are only visible in the main isotope and that the determined absorbing foreground is only affecting the $^{12}$CO (3$\to$2) line. 
        	
\begin{figure*}
    		\begin{subfigure}[c]{0.28\textwidth}
    			\includegraphics[width=1.\textwidth]{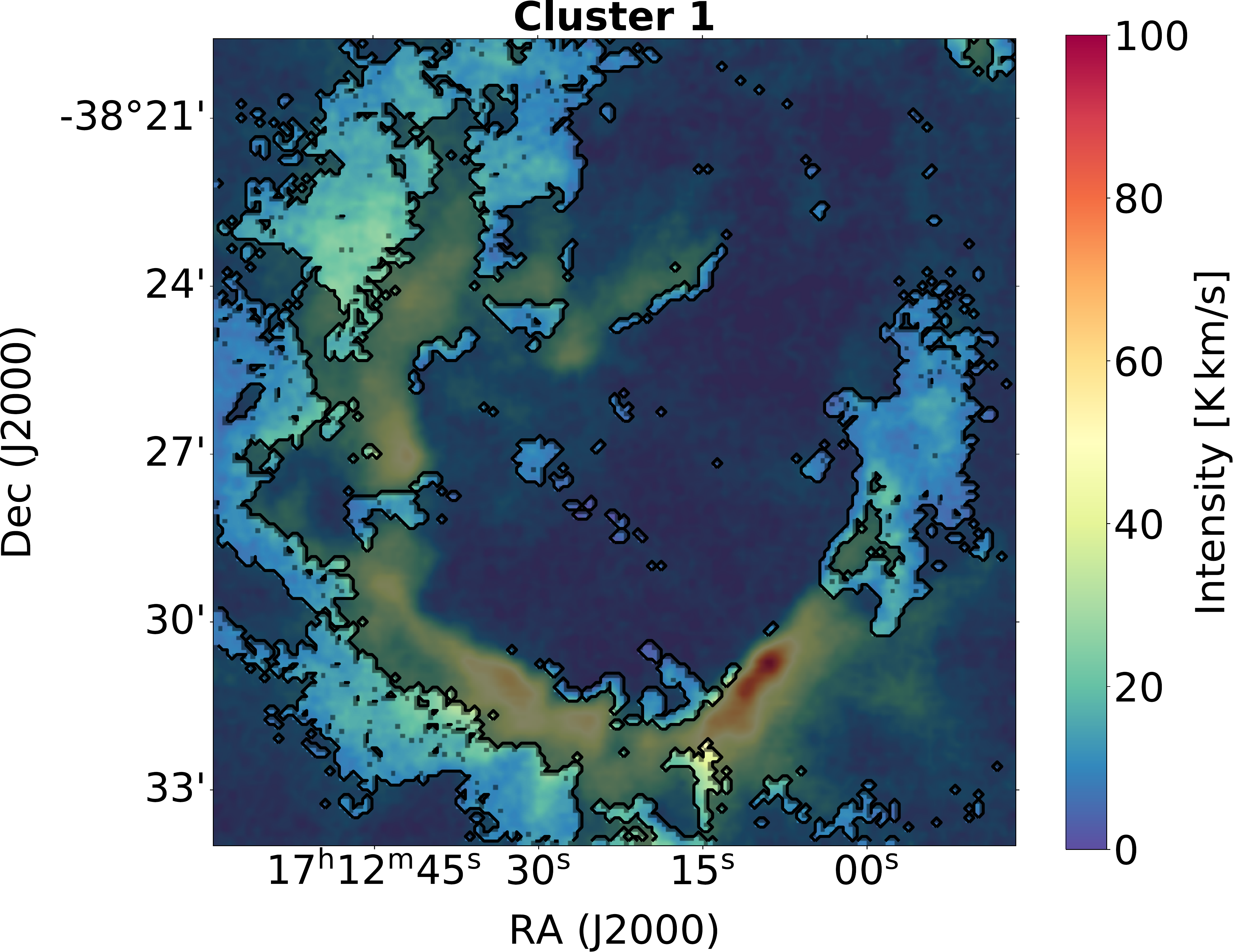}	
    		\end{subfigure}
    		\begin{subfigure}[c]{0.28\textwidth}
    			\includegraphics[width=1.\textwidth]{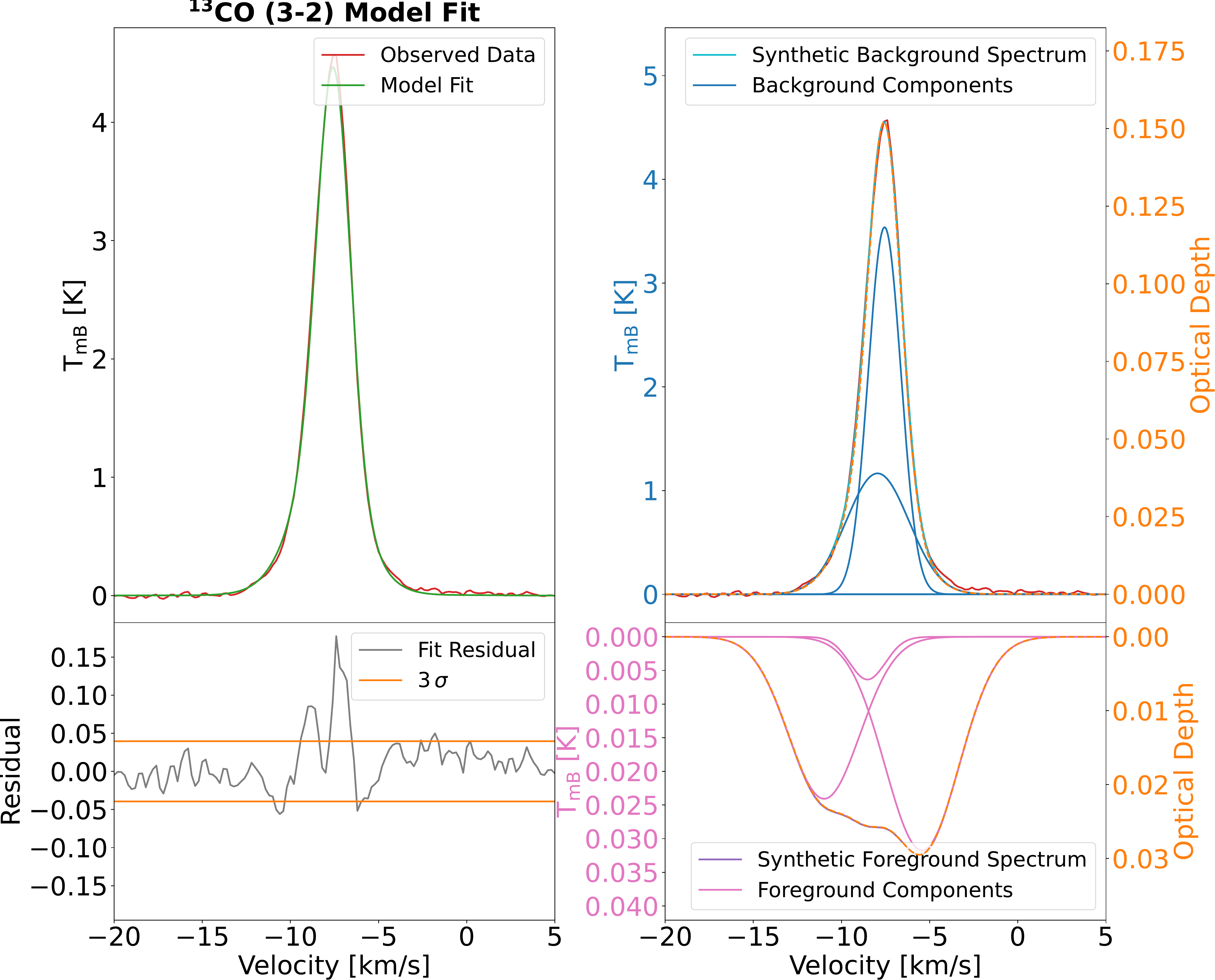}
    		\end{subfigure}
    	 	\begin{subfigure}[c]{0.28\textwidth}
    	 		\includegraphics[width=1.\textwidth]{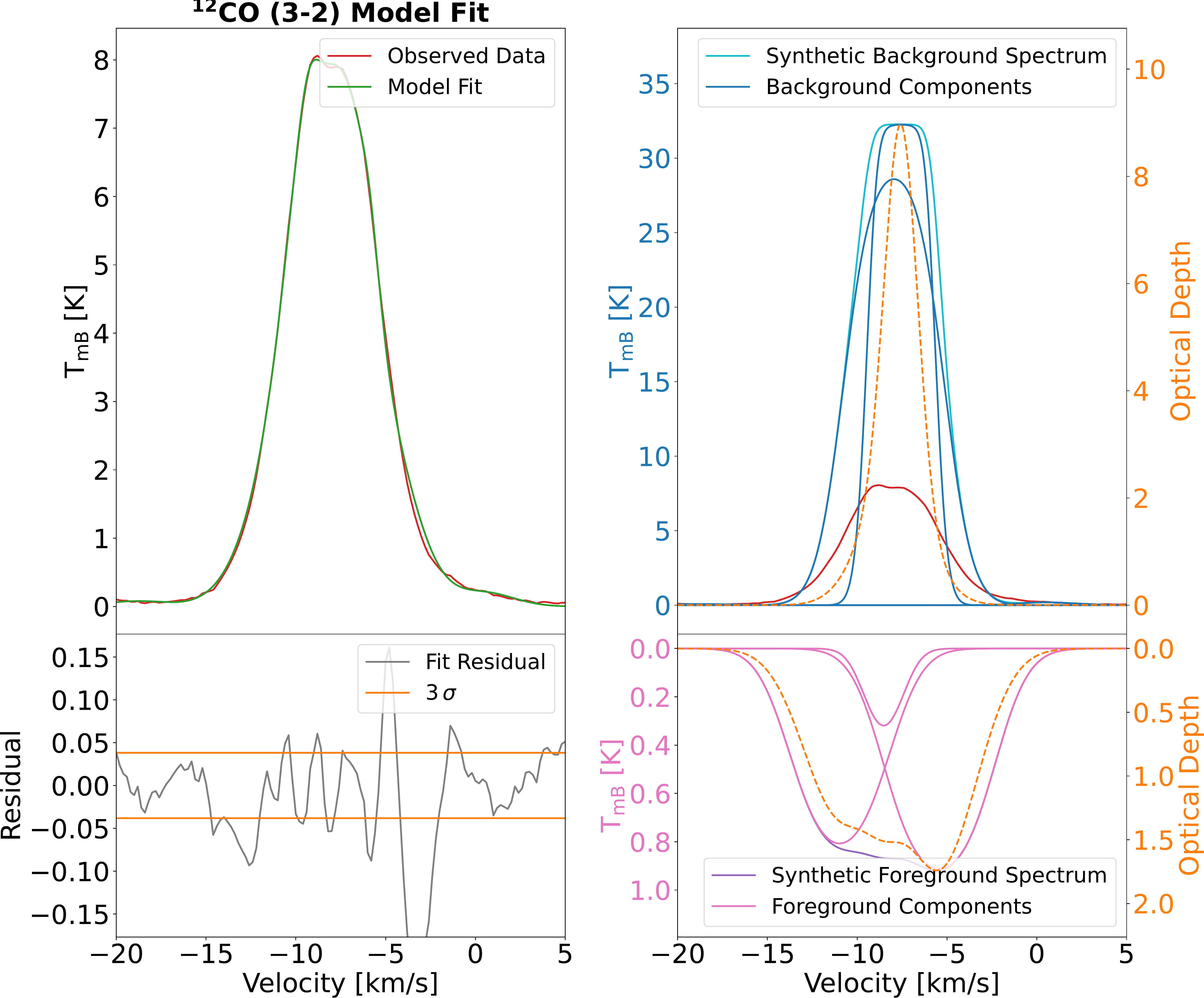}
    	 	\end{subfigure}
     	
     		\begin{subfigure}[c]{0.28\textwidth}
     			\includegraphics[width=1.\textwidth]{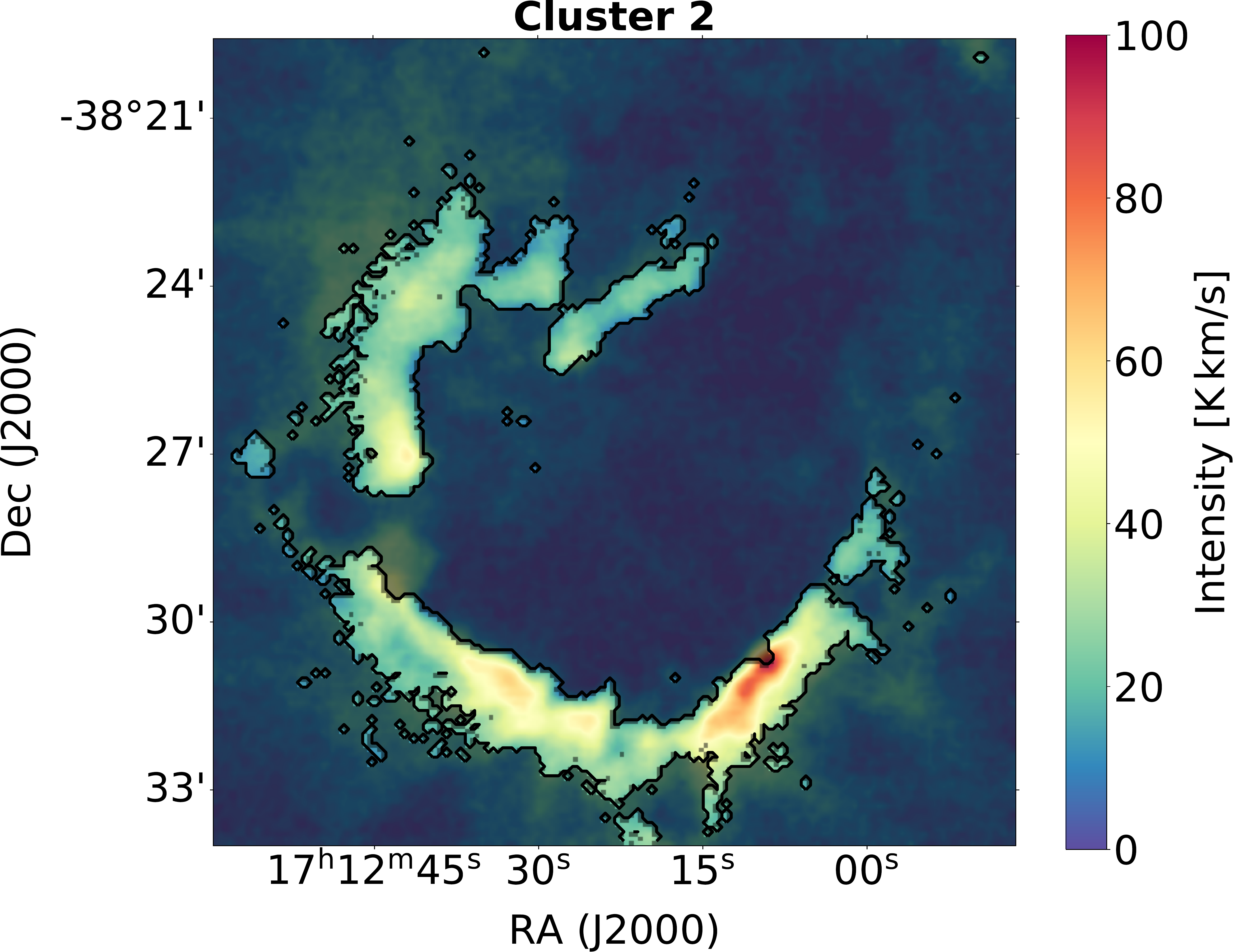}	
     		\end{subfigure}
     		\begin{subfigure}[c]{0.28\textwidth}
     			\includegraphics[width=1.\textwidth]{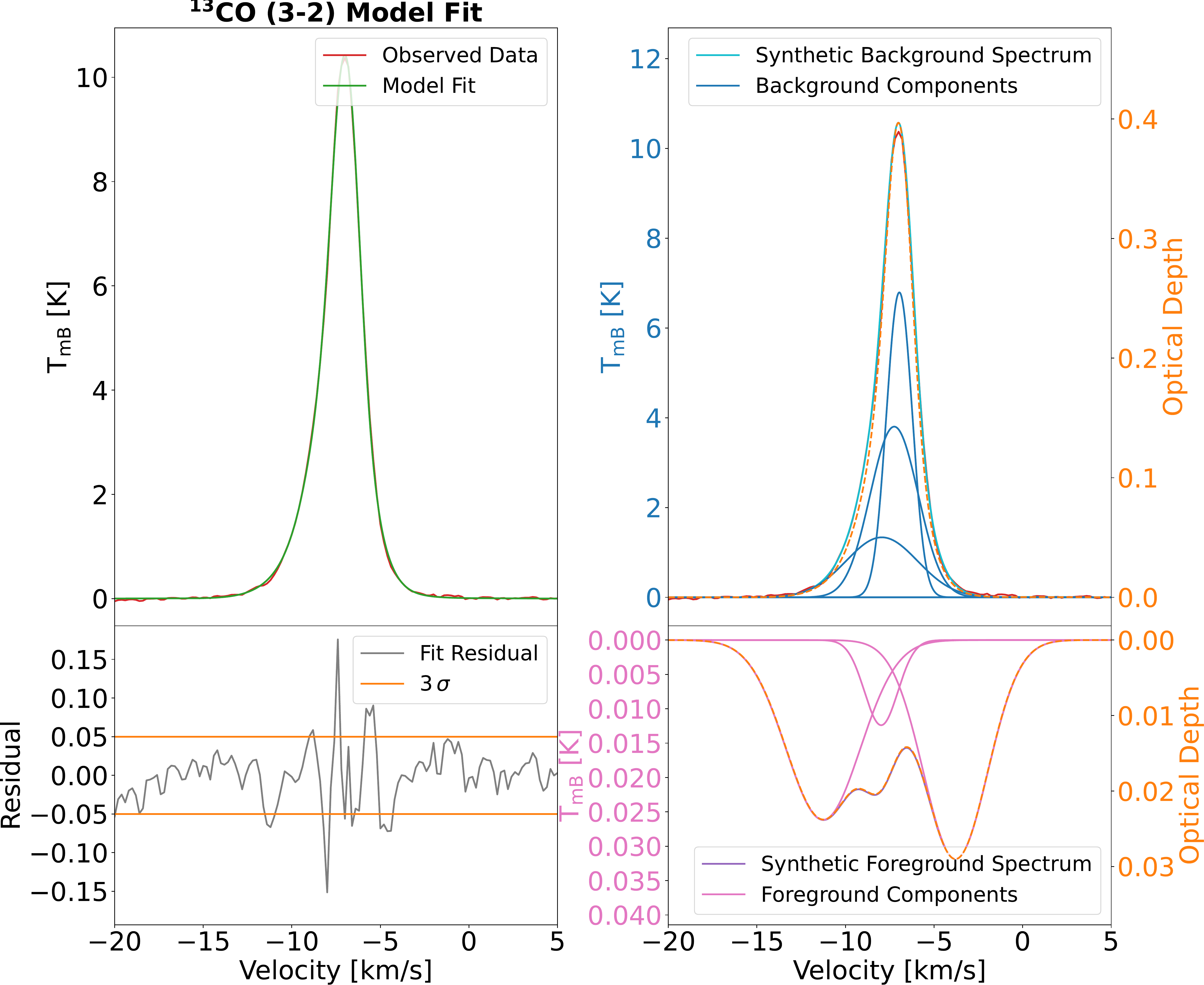}
     		\end{subfigure}
     		\begin{subfigure}[c]{0.28\textwidth}
     			\includegraphics[width=1.\textwidth]{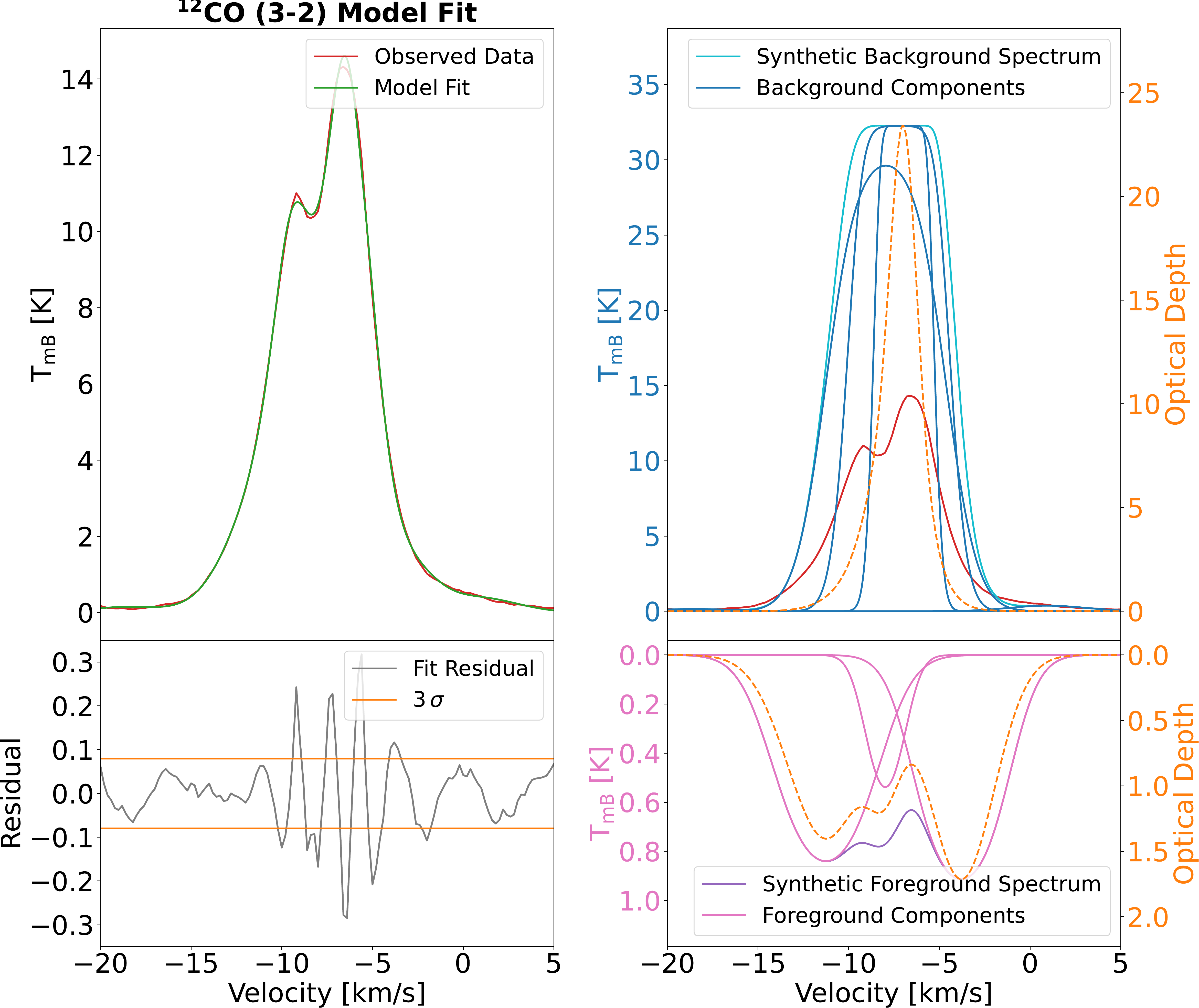}
     		\end{subfigure}
     	
     		\begin{subfigure}[c]{0.28\textwidth}
     			\includegraphics[width=1.\textwidth]{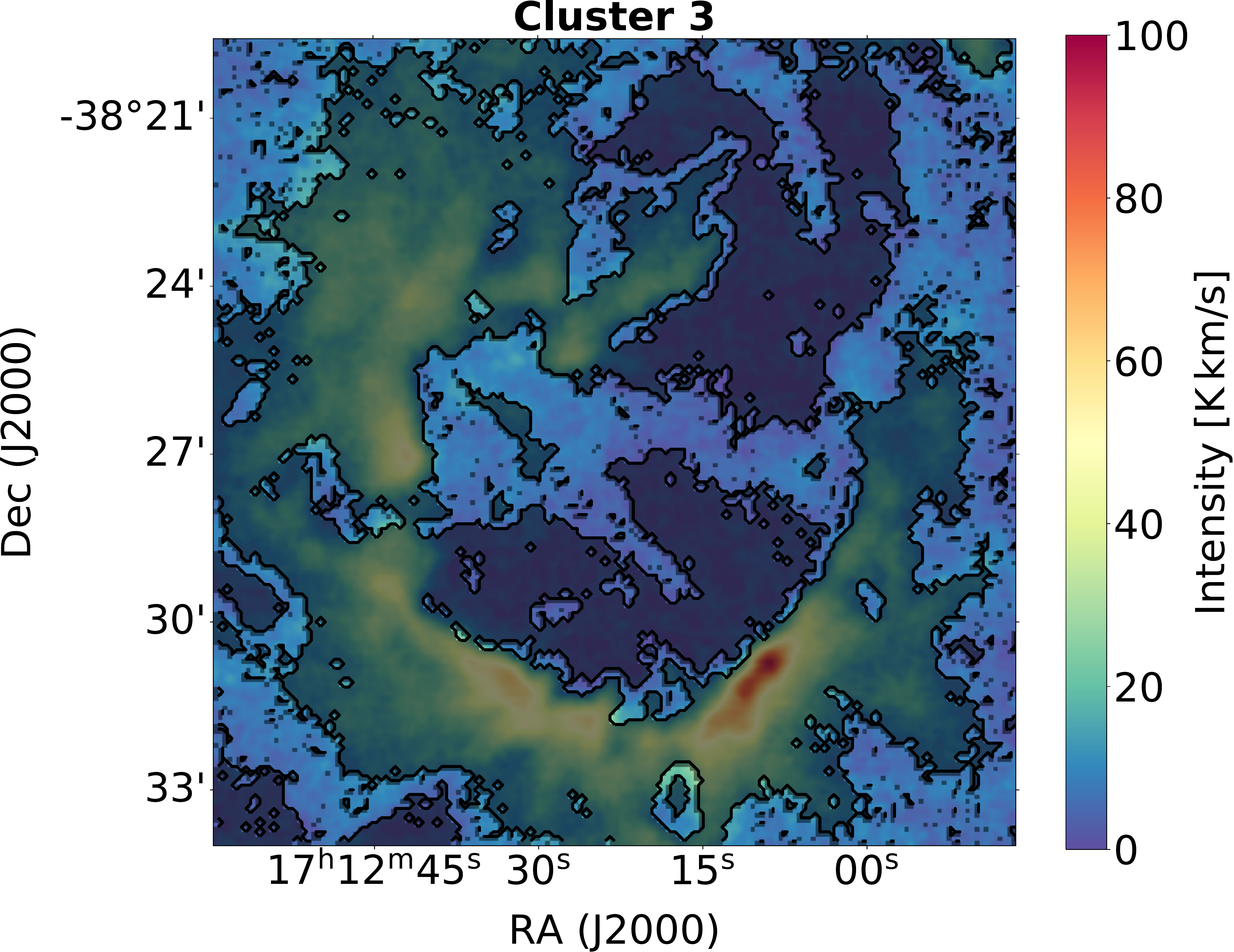}	
     		\end{subfigure}
     		\begin{subfigure}[c]{0.28\textwidth}
     			\includegraphics[width=1.\textwidth]{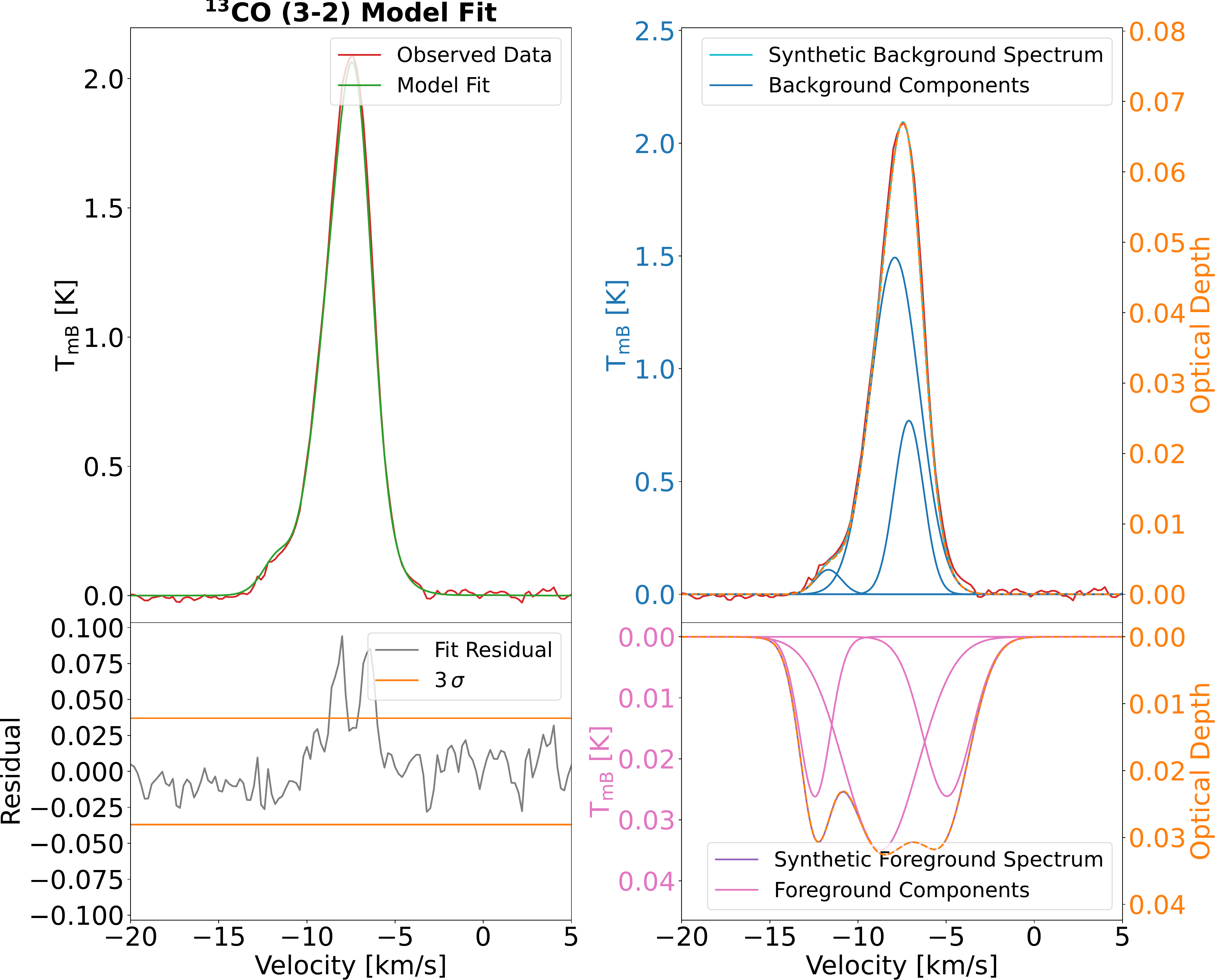}
     		\end{subfigure}
     		\begin{subfigure}[c]{0.28\textwidth}
     			\includegraphics[width=1.\textwidth]{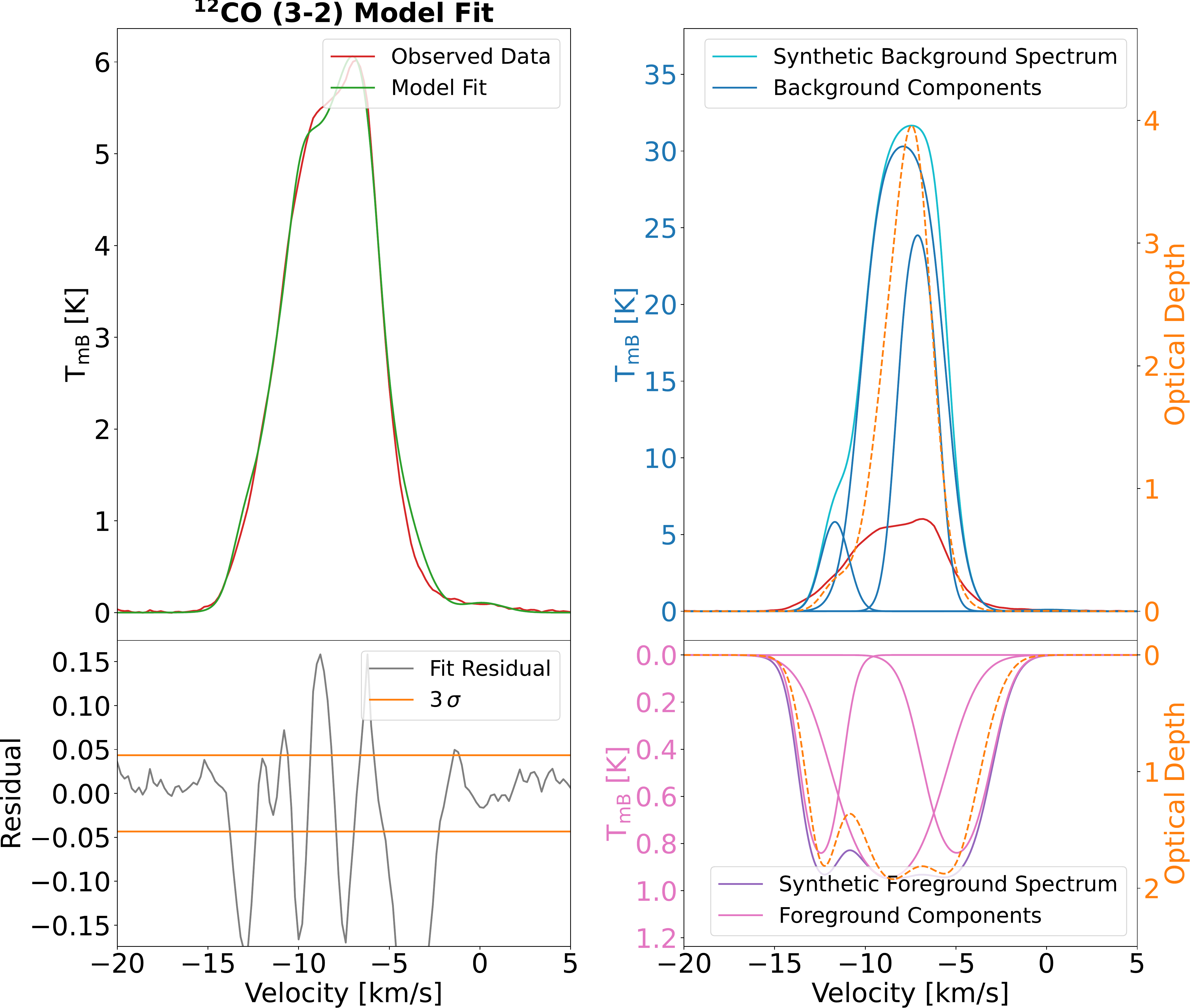}
     		\end{subfigure}
     	
     		\begin{subfigure}[c]{0.28\textwidth}
     			\includegraphics[width=1.\textwidth]{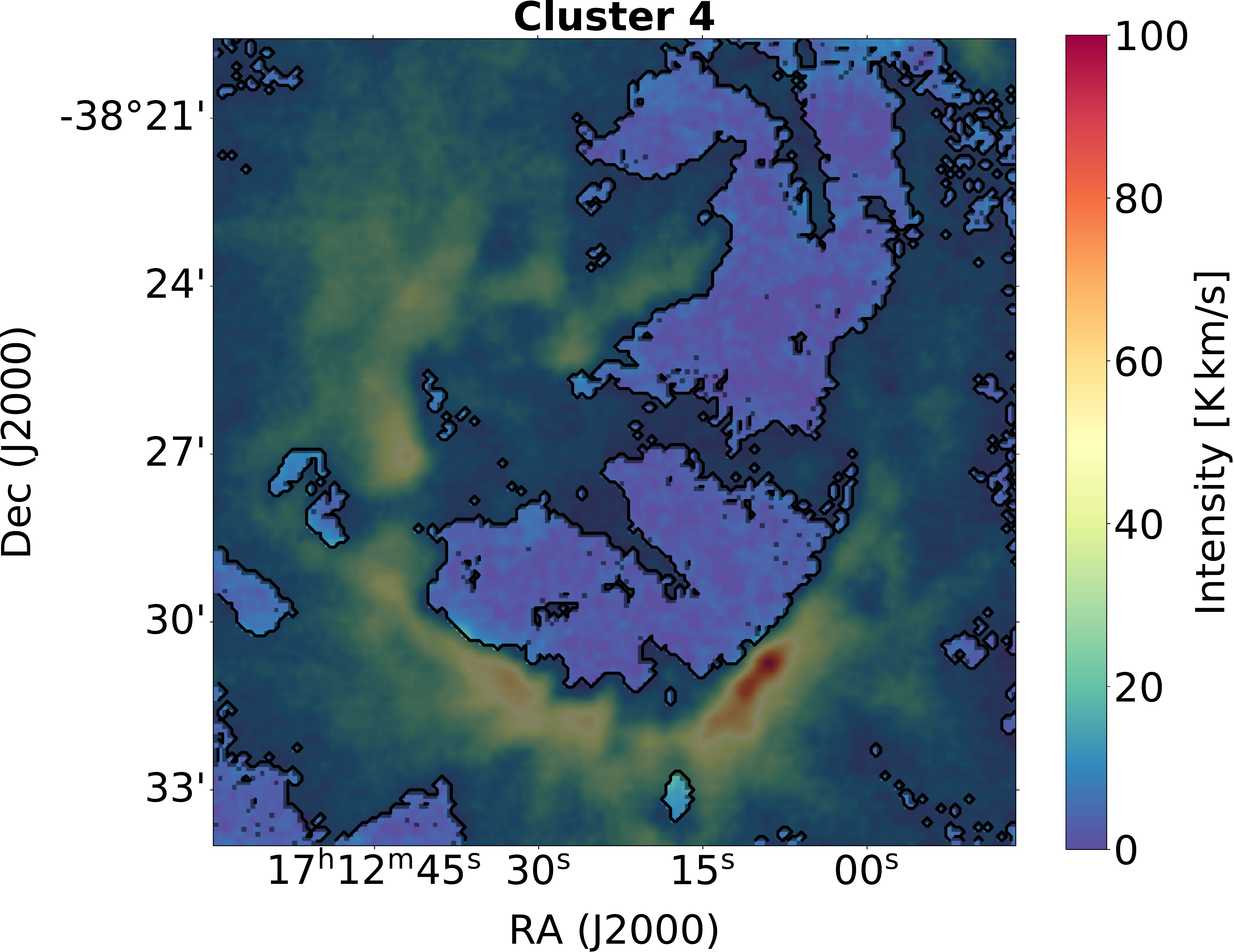}	
     		\end{subfigure}
     		\begin{subfigure}[c]{0.28\textwidth}
     			\includegraphics[width=1.\textwidth]{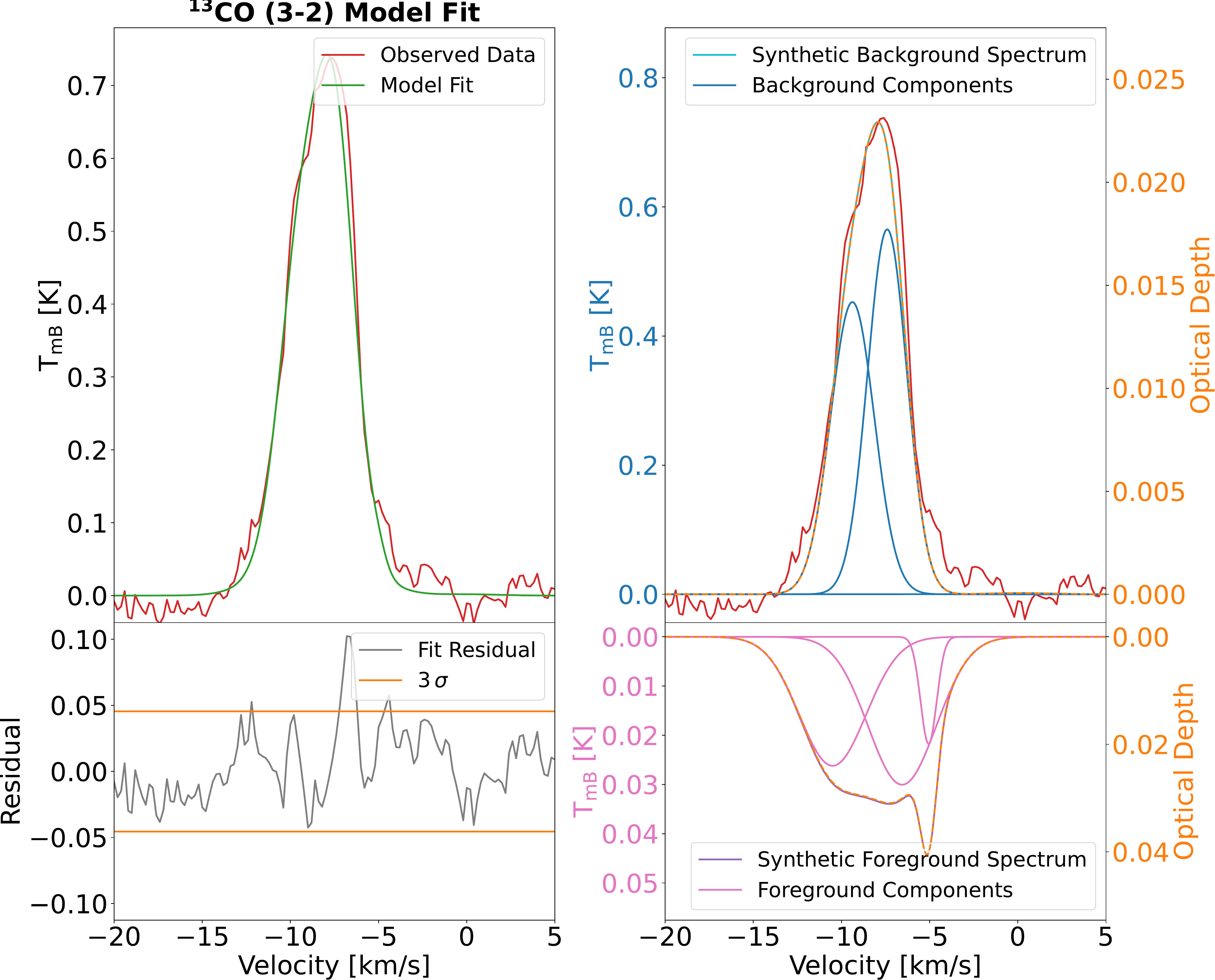}
     		\end{subfigure}
     		\begin{subfigure}[c]{0.28\textwidth}
     			\includegraphics[width=1.\textwidth]{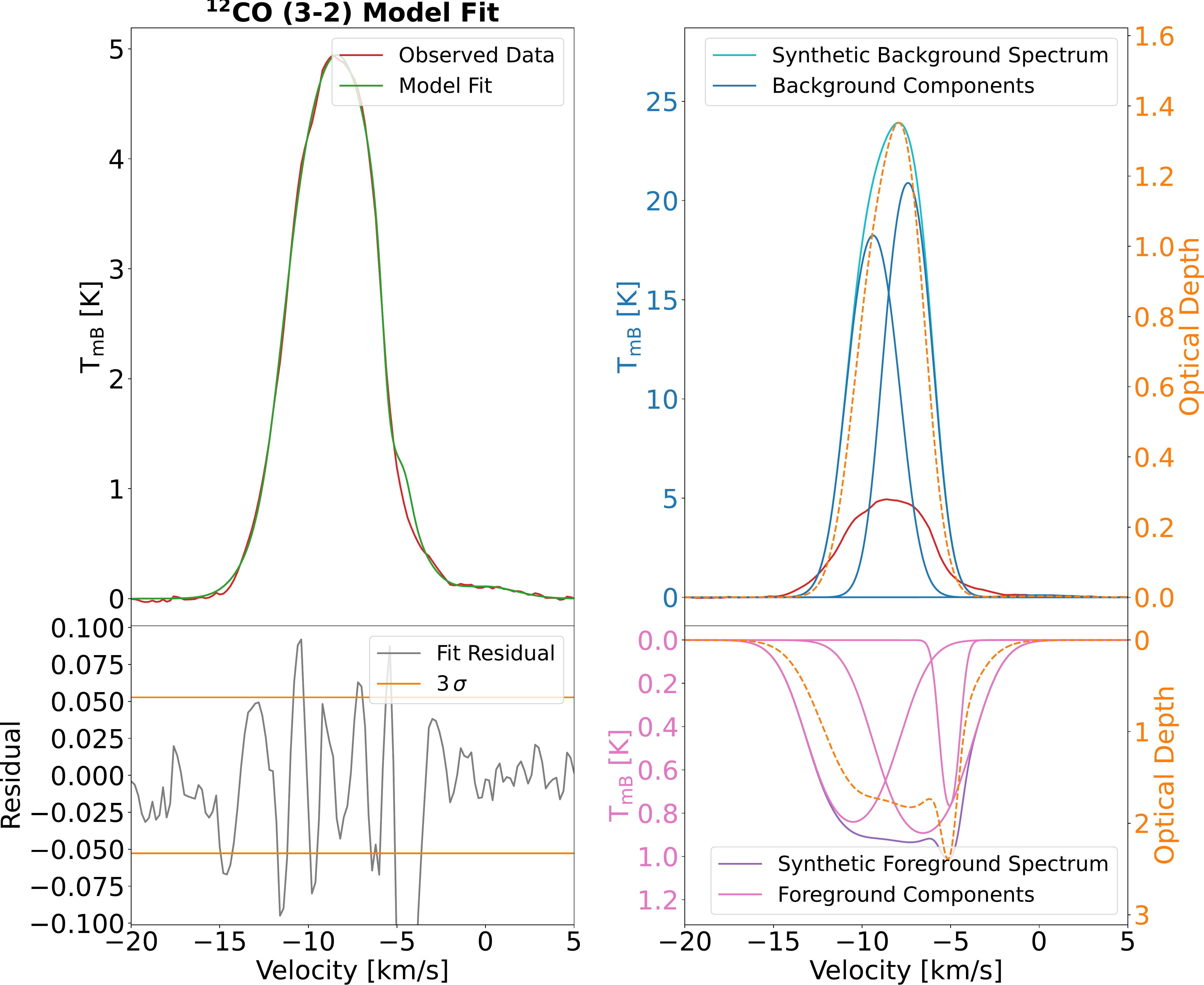}
     		\end{subfigure}
     	
     		\begin{subfigure}[c]{0.28\textwidth}
     			\includegraphics[width=1.\textwidth]{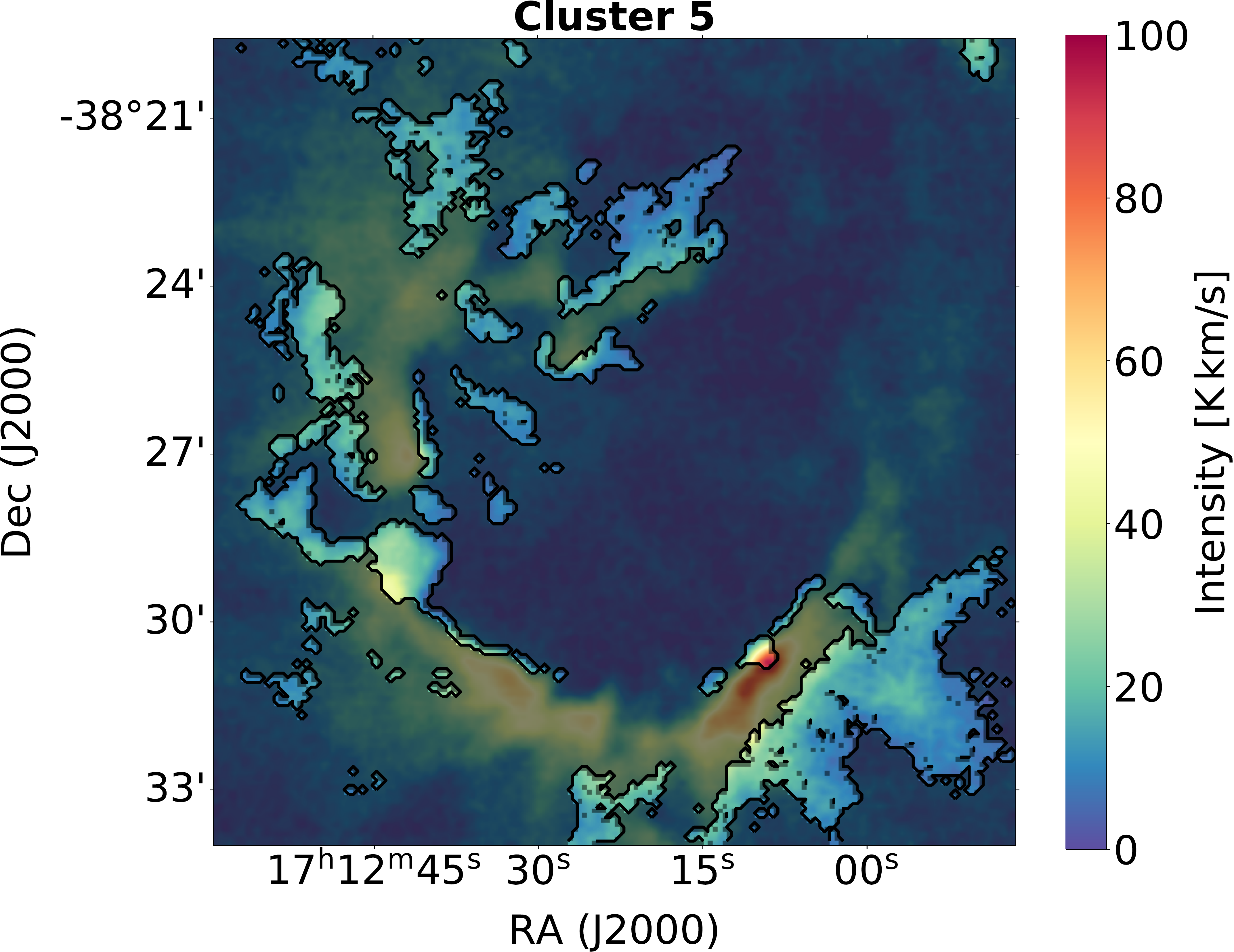}	
     		\end{subfigure}
     		\begin{subfigure}[c]{0.28\textwidth}
     			\includegraphics[width=1.\textwidth]{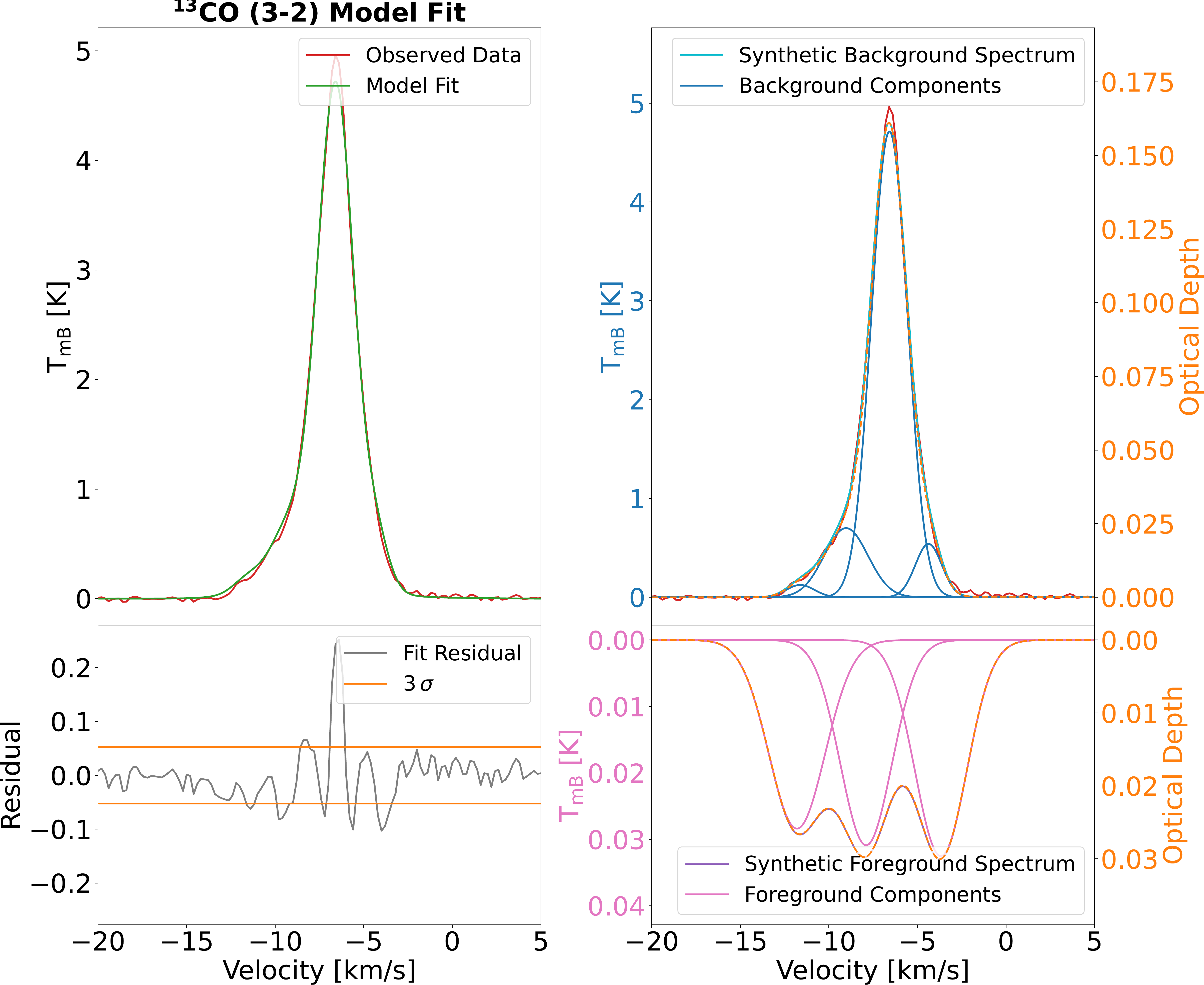}
     		\end{subfigure}
     		\begin{subfigure}[c]{0.28\textwidth}
     			\includegraphics[width=1.\textwidth]{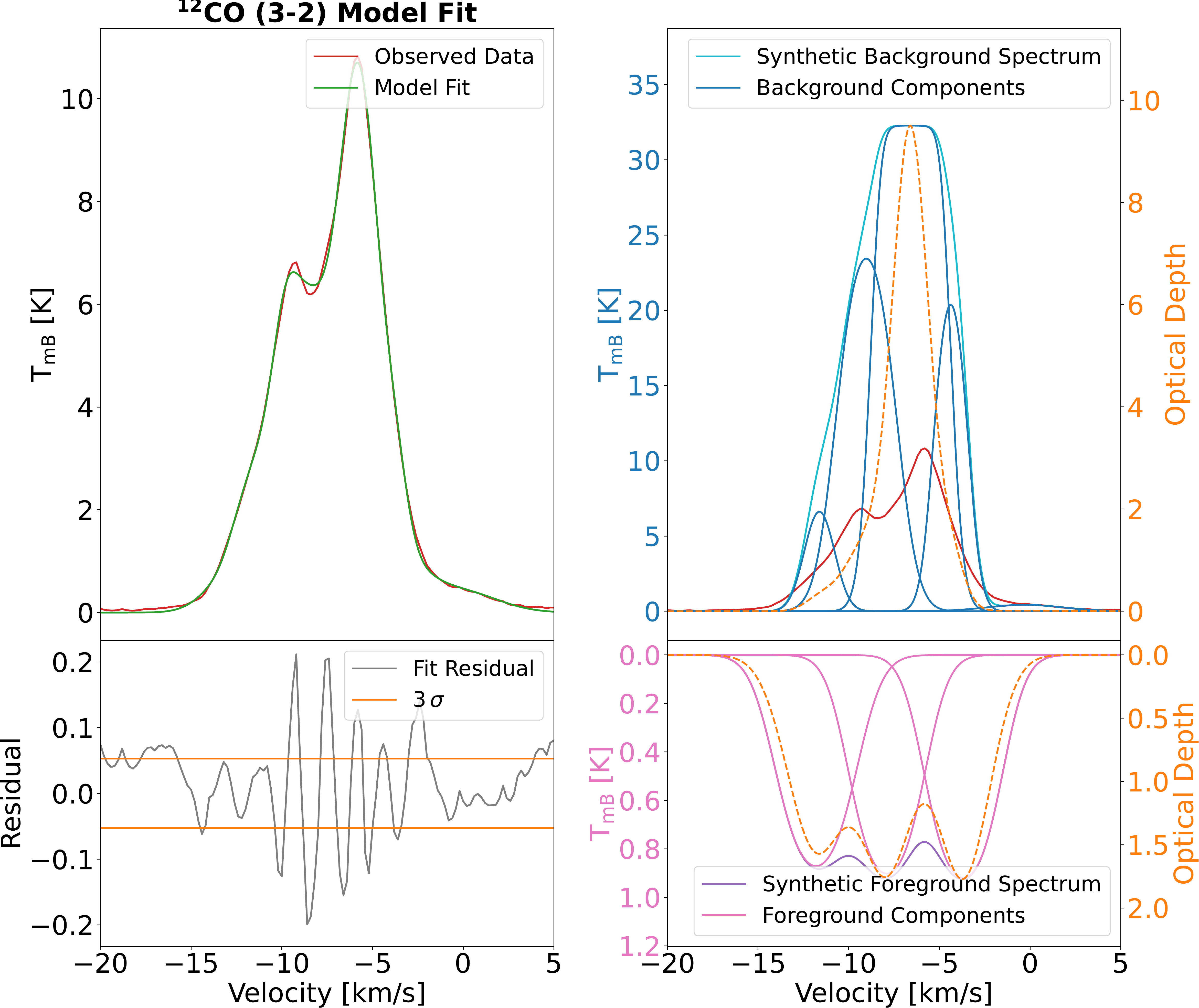}
     		\end{subfigure}
     	
     	 	\caption{Two-layer multicomponent model fit to the average spectra of each cluster determined by the GMM. Left column: The black contours indicate the spatial location of the clusters. Area that is located outside the clusters is dark shaded. Following the middle and right columns: $^{13}$CO and $^{12}$CO (3$\to$2) model fits. All other parameters are as shown in Fig.~\ref{fig:two_layer_cii}. The red spectra in the upper left panels indicate the observed CO lines and the green curve is the resulting model fit. In the panel below, the gray data points show the fit residual and the two horizontal orange lines indicate the $3\sigma$ level. In the upper right panels, the dark blue lines show the single Gaussian components and the light blue line the superposition of all synthetic background components. In the lower right panels, the pink line displays the single foreground Gaussian components and the magenta line shows the superposition of all  synthetic foreground components. The dashed orange line indicates the velocity resolved optical depth. } \label{fig:cluster_two_layer}
        \end{figure*}

            The two-layer multicomponent fit for the averaged spectra is shown in Fig. \ref{fig:cluster_two_layer} and the physical properties for each cluster in Table~\ref{tab:physical_conditions_co}. The left column in Fig. \ref{fig:cluster_two_layer} displays the spatial cluster location, the middle and right columns show the $^{13}$CO (3$\to$2) and $^{12}$CO (3$\to$2) model fits, respectively. Similar to the findings for the ionized carbon, the emitting background layer is overshooting the observed main isotope line significantly. This requires a cold absorbing foreground which absorbs a significant amount of the $^{12}$CO line. The optical depths for $^{12}$CO for the background emission are still high (Table~\ref{tab:physical_conditions_co}), but more of the order of a few and not of a few 10s like in the single-layer model. The background column density of the molecular gas is of the order of $1 - 10 \times 10^{21}$ cm$^{-2}$.  
            The molecular ring surrounding the \HII\ region is described by cluster 1,2 and 5, which gives a total mass of $\sim2200\,$M$_{\sun}$ for the emitting background and $\sim670\,$M$_{\sun}$ for the cold absorbing foreground. This is a lower limit, since we do not see the material behind the warm emitting layer. 

        While the numbers of components varies for the emitting background (see Tables in Appendix \ref{appendix-c2}), we find that the foreground is best described by three components: a component approximately located at the systemic velocity of RCW~120 around $-$7.5 km s$^{-1}$ and a blue-shifted and red-shifted component. The absorbing component at the systemic velocity can be attributed to a temperature gradient between the observer and the warm emitting ring, formed due to the decreasing stellar radiation with distance from the ionizing star. However, the two additional velocity-shifted components suggest a more complex geometry and dynamics than the simple assumption of a spherical molecular cloud enveloping the ionizing star. We already assumed from the opacity calculations for the single layer model (Sec.~\ref{sec:tau}) that dynamics play an important role in understanding the CO emission features and we propose the following scenario: \\
        \citet{Luisi2021} reported a fast ($\sim$15 km s$^{-1}$) expanding [\CII] bubble that impacts the surrounding molecular cloud. 
        The expanding bubble drives a shock wave into the surrounding gas, sweeping it up into a shell. The inside of this shell is ionized and separated from the warm atomic gas in the PDR by the ionization front. For a large enough column, the PDR layers further away from the UV source will become molecular and this molecular gas would move at the velocity of the shell. There is no communication between the gas in front of the shock front and behind it. The momentum transfer occurs in the shock front associated with the outer boundary of the shell.
        However, the red-shifted component can not be simply attributed to the red-shifted expanding shell, since a possible absorbing layer would be located at the rear side of the region where absorption features can not be detected.
        Thus, a red-shifted component with respect to the systemic velocity indicates an inflow of cold molecular material towards RCW~120.  

        \subsubsection{Entire spectral CO cube} \label{CO-cube}

             In the following we use the model fits of the average spectra as an input for the entire spectral cube to resolve the spatial distribution of the two layers. Like for the average cluster spectra we first fit the $^{13}$CO (3$\to$2) data cube with the two-layer multicomponent model assuming weak foreground absorption. The resulting model fit is scaled up by the local carbon abundance ratio to determine the  $^{12}$CO (3$\to$2) emitting background layer. However, for the full cube we need to take into account that the noise in each pixel is much higher than for the average cluster spectra. Simply using the number of background components determined from the averaged spectra might lead to 'over-fitting', thus assuming more background components than visible in a single spectrum can result in fitting of noise fluctuations which scaled up by the carbon abundance ratio are visible as artifacts in the resulting synthetic map. We thus perform the model fit in a sequential way: the model fit to each pixel is started with the strongest background component, afterwards the goodness of the model fit is tested. If the reduced chi-squared  is approximately $\chi_{\mathrm{R}}^2 \sim 1$, no further background component is added. Otherwise the sequential model fit is continued. 

        	\begin{figure*}[!ht]
        	\centering
        	    \includegraphics[width=0.3\textwidth]{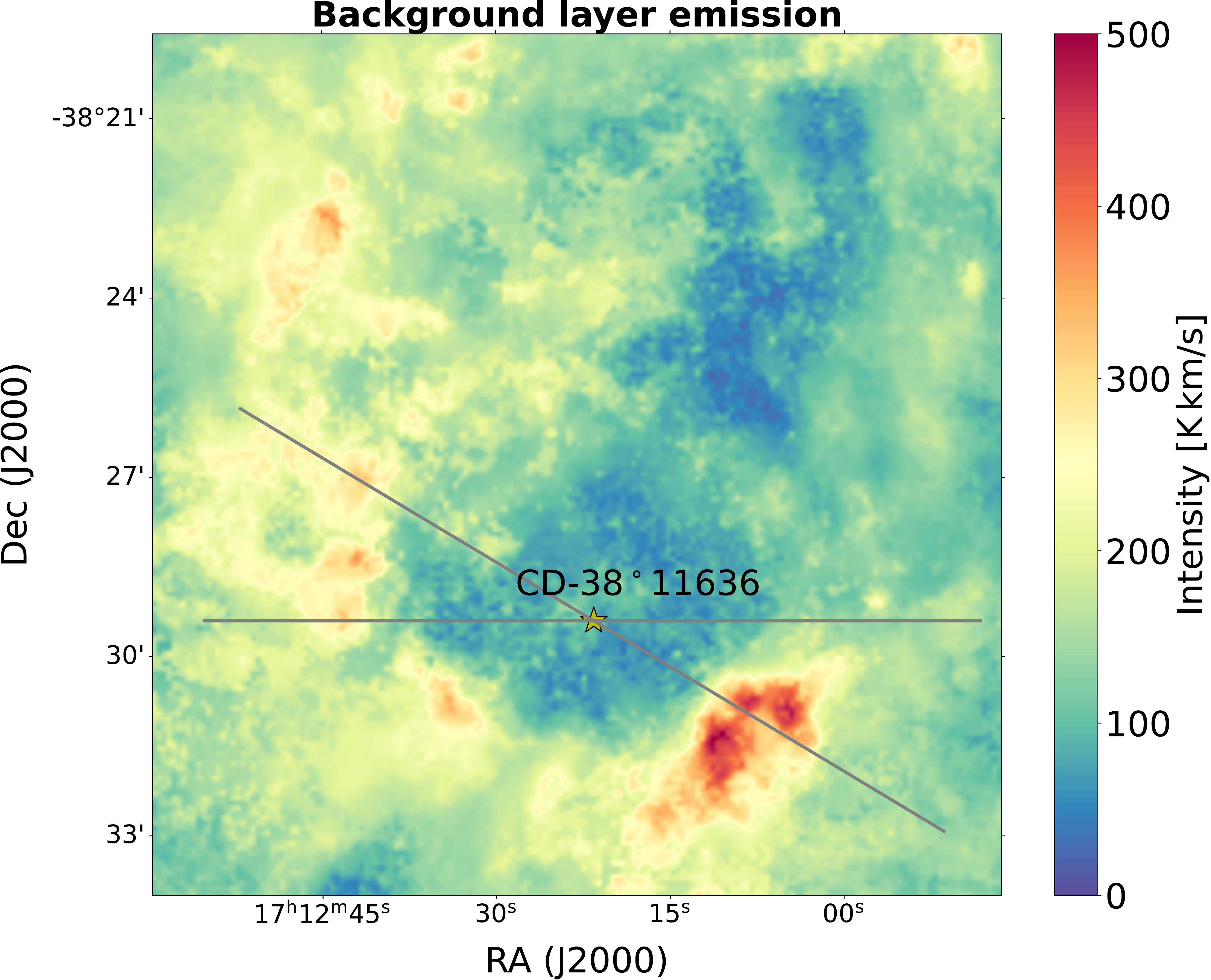}
        		\includegraphics[width=0.3\textwidth]{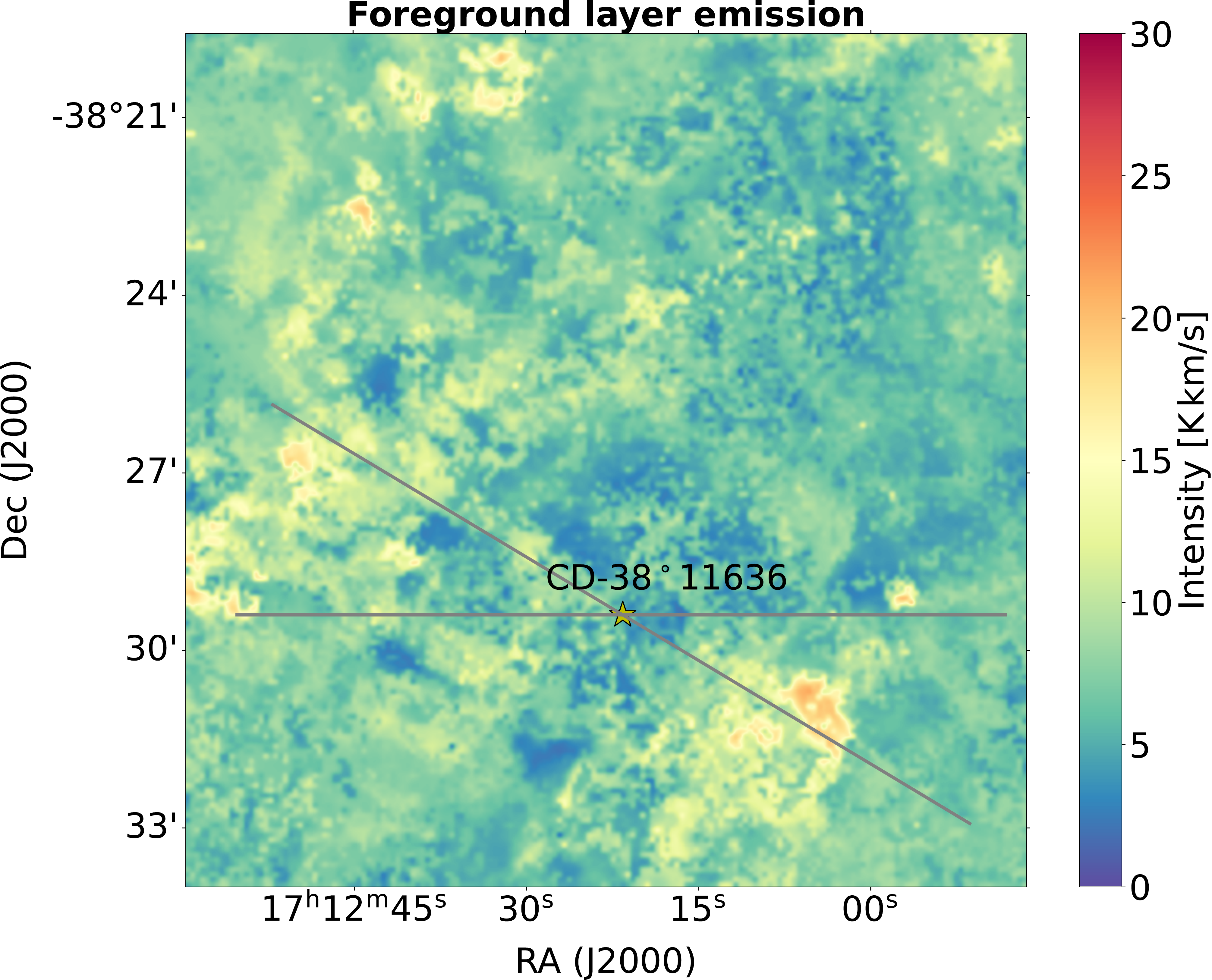}
        		\includegraphics[width=0.25\textwidth]{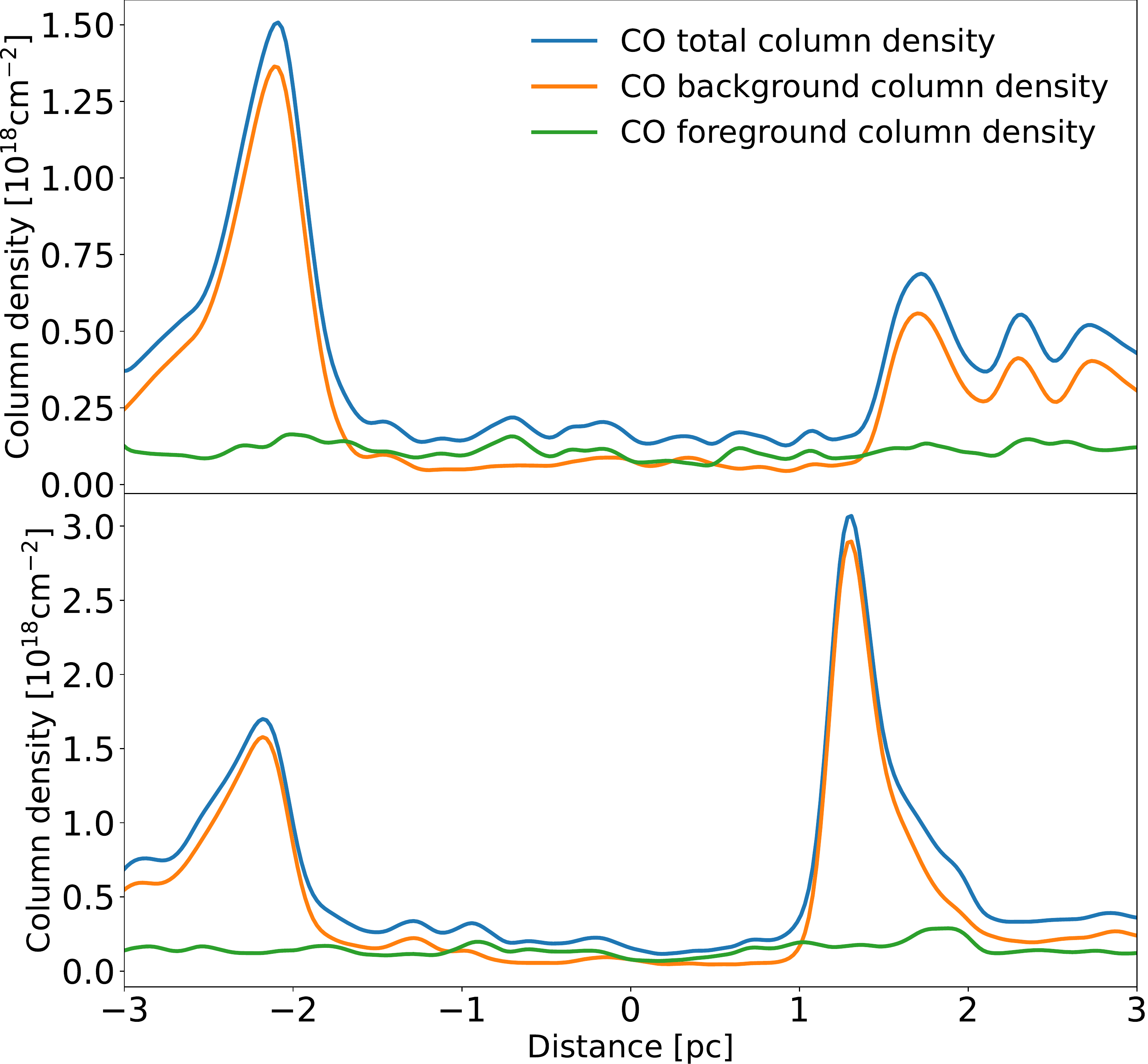}
        		\caption{Line integrated $^{12}$CO (3$\to$2) synthetic maps and cuts. Left panel: Spatial distribution of the warm background emitting layer. Middle panel: Spatial distribution of the cold foreground absorbing layer. Right panel: Position-intensity diagram along the two cuts indicated by the horizontal gray lines crossing the ionizing star. The upper panel shows the horizontal cut and the lower panel shows the diagonal
        		cut. The orange curve is showing the column density of the emitting background layer, the green curve shows the column number density of the absorbing foreground layer and the blue curve indicates the combined column density of both layers.} 
        		\label{fig:synthetic_maps}
        	\end{figure*}
	
        	\begin{figure*}[!ht]
        	    \centering
        		\includegraphics[width=0.9\textwidth]{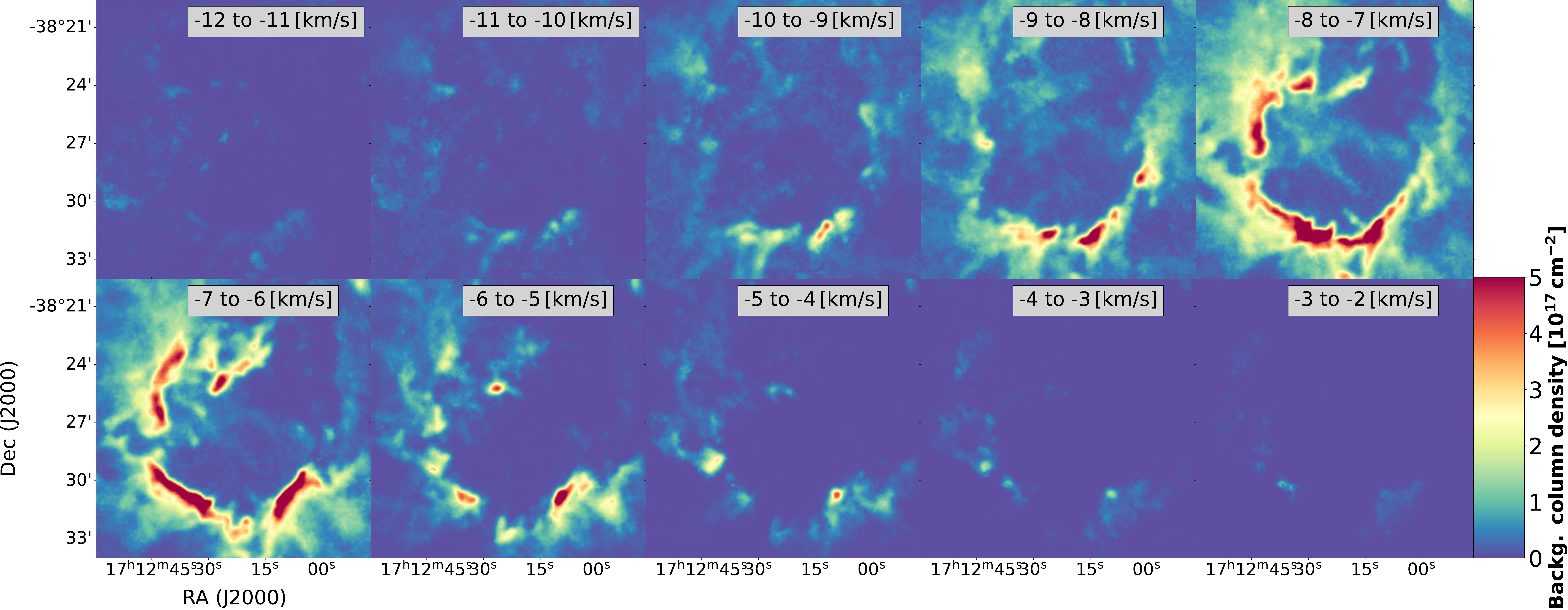}
        		\includegraphics[width=0.9\textwidth]{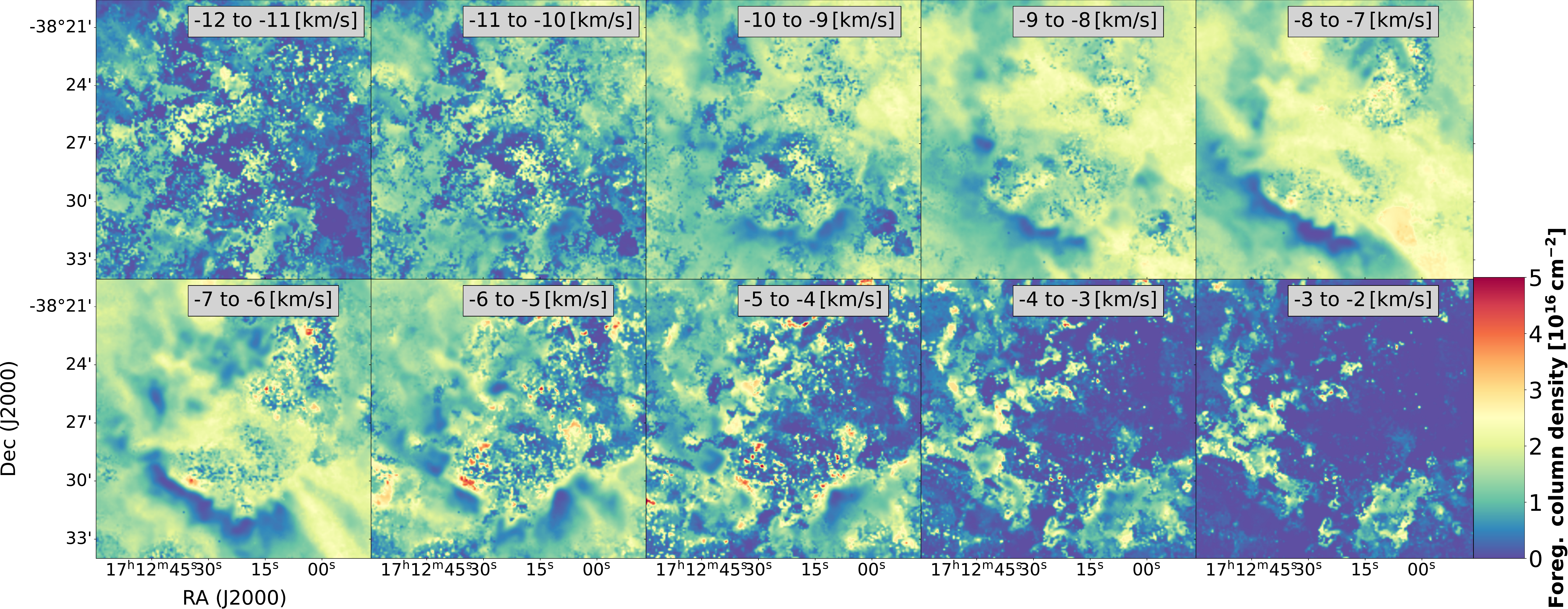}
        	    \caption{Velocity resolved $^{12}$CO column density maps of the warm emitting background layer (top) and the cold absorbing foreground layer (bottom).}
        		\label{fig:channel_synthetic_maps}
        	\end{figure*} 

            The resulting synthetic $^{12}$CO (3$\to$2)  map of the emitting background layer is shown in the left panel of Fig. \ref{fig:synthetic_maps}. Even though the map is derived from the observed $^{13}$CO (3$\to$2) spectral cube, the map reveals a different spatial distribution because
            the  $^{12}$CO line is optically thick. 
            In order to reveal the "real" $^{12}$CO distribution (unaffected by optical depth effects), we determine the $^{12}$CO column density using equation~(\ref{eq:tau_co}) and show channel maps of this synthetic column density in the top panel of Fig. \ref{fig:channel_synthetic_maps}. The velocity resolved $^{12}$CO column density of the warm emitting layer now has a similar spatial distribution as the $^{13}$CO~(3$\to$2) emission (compare to Fig.~\ref{fig:channel_maps}). 

            \begin{figure*}[!ht]
                \centering
                \includegraphics[width=.3\textwidth]{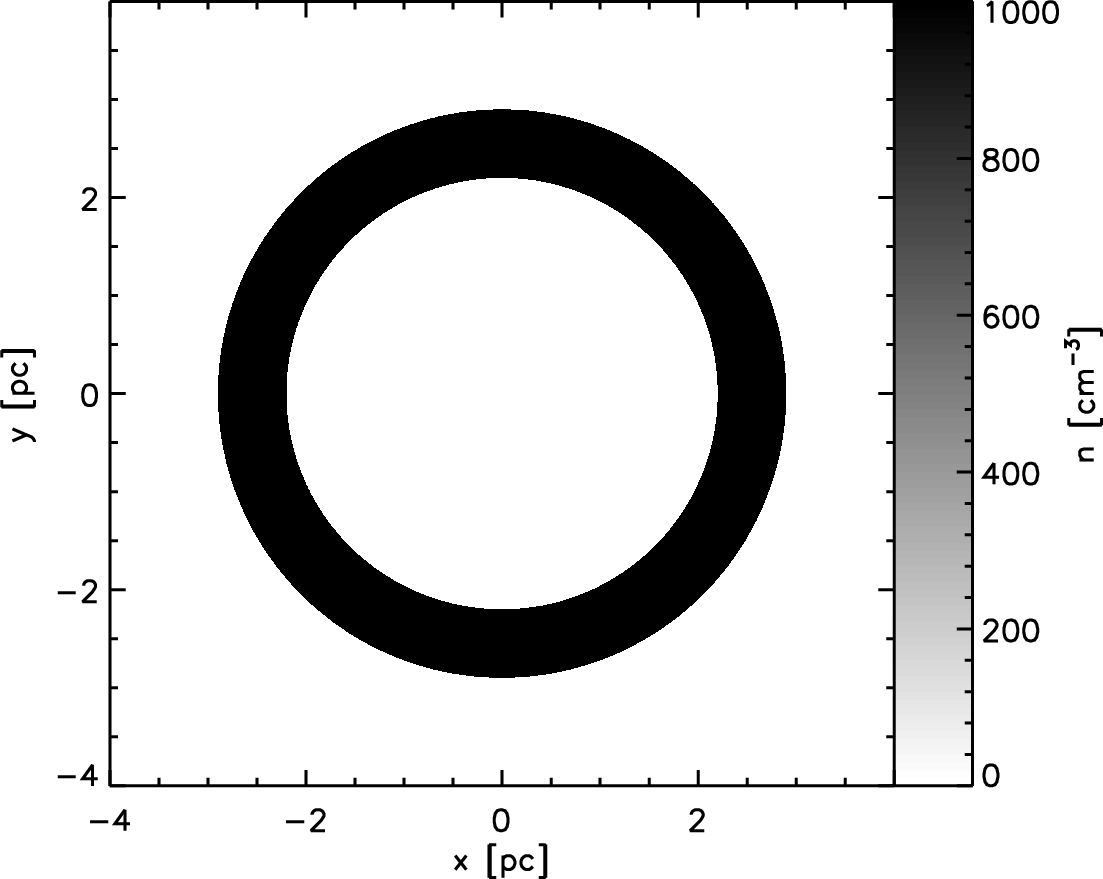}
            	\includegraphics[width=.32\textwidth]{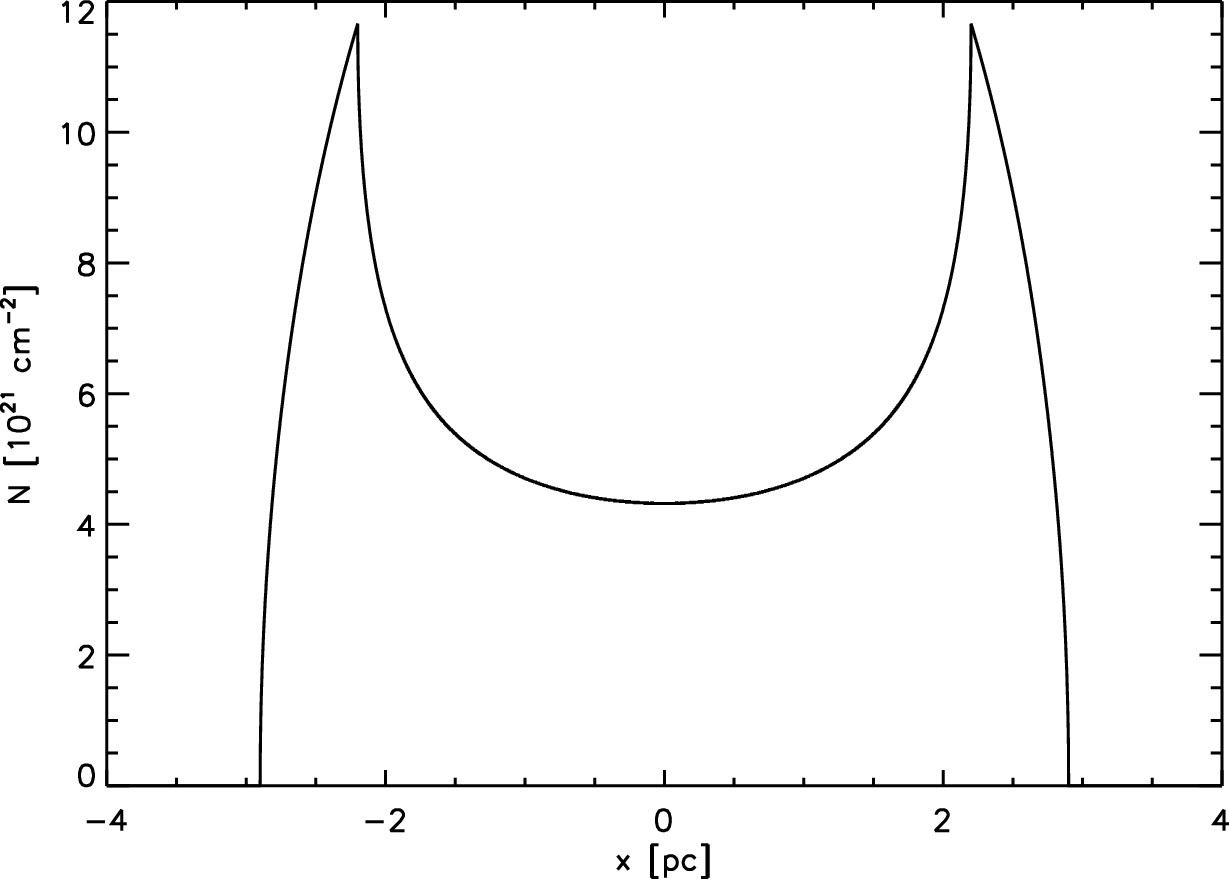}
            	\includegraphics[width=.34\textwidth]{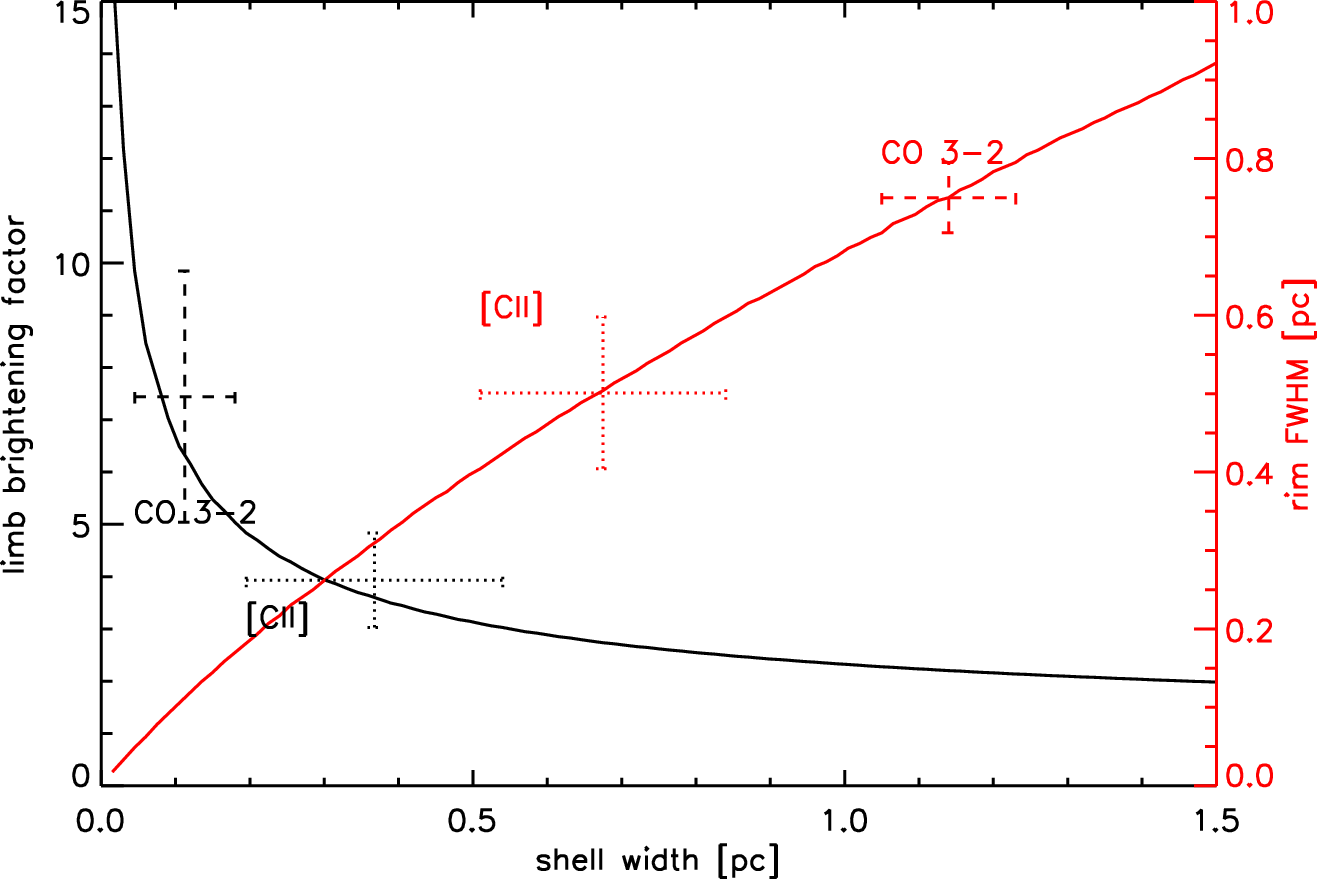}
                \caption{Geometrical effects of a spherical shell. Left panel: Visualization of the radial profile that would provide the projected column number density distribution in the central panel along a cut through the bubble center. Right panel: Dependence of the apparent shell width (red) and the limb brightening factor (black) on the geometric shell width. The observed [\CII] and CO $(3\to 2)$ shell widths and limb brightening factors are overplotted. The error bars indicate the variation of both quantities along the observed ring.} \label{fig:bubble_geometry}
            \end{figure*} 

            \begin{figure}[!ht]
                \centering
            	\includegraphics[width=.4\textwidth]{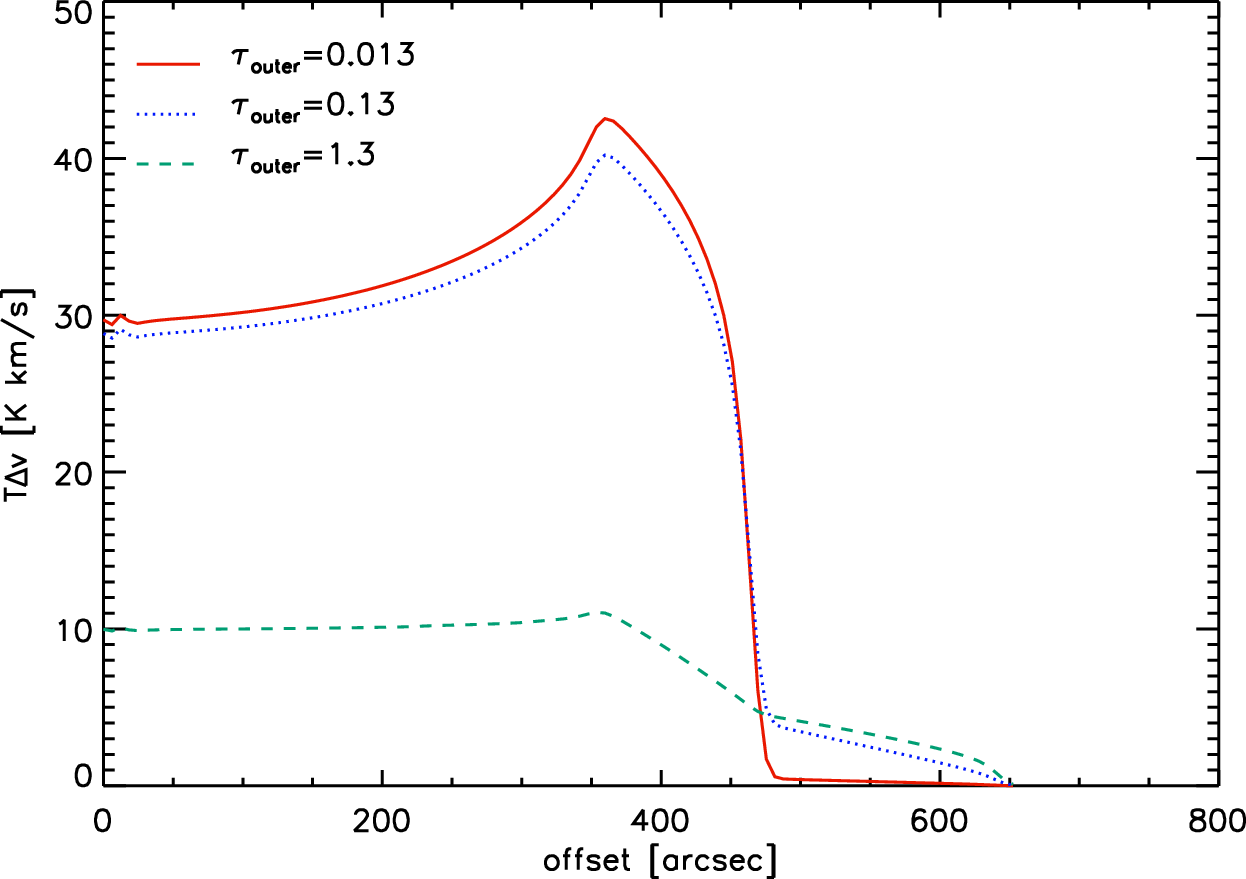}
            	\includegraphics[width=.4\textwidth]{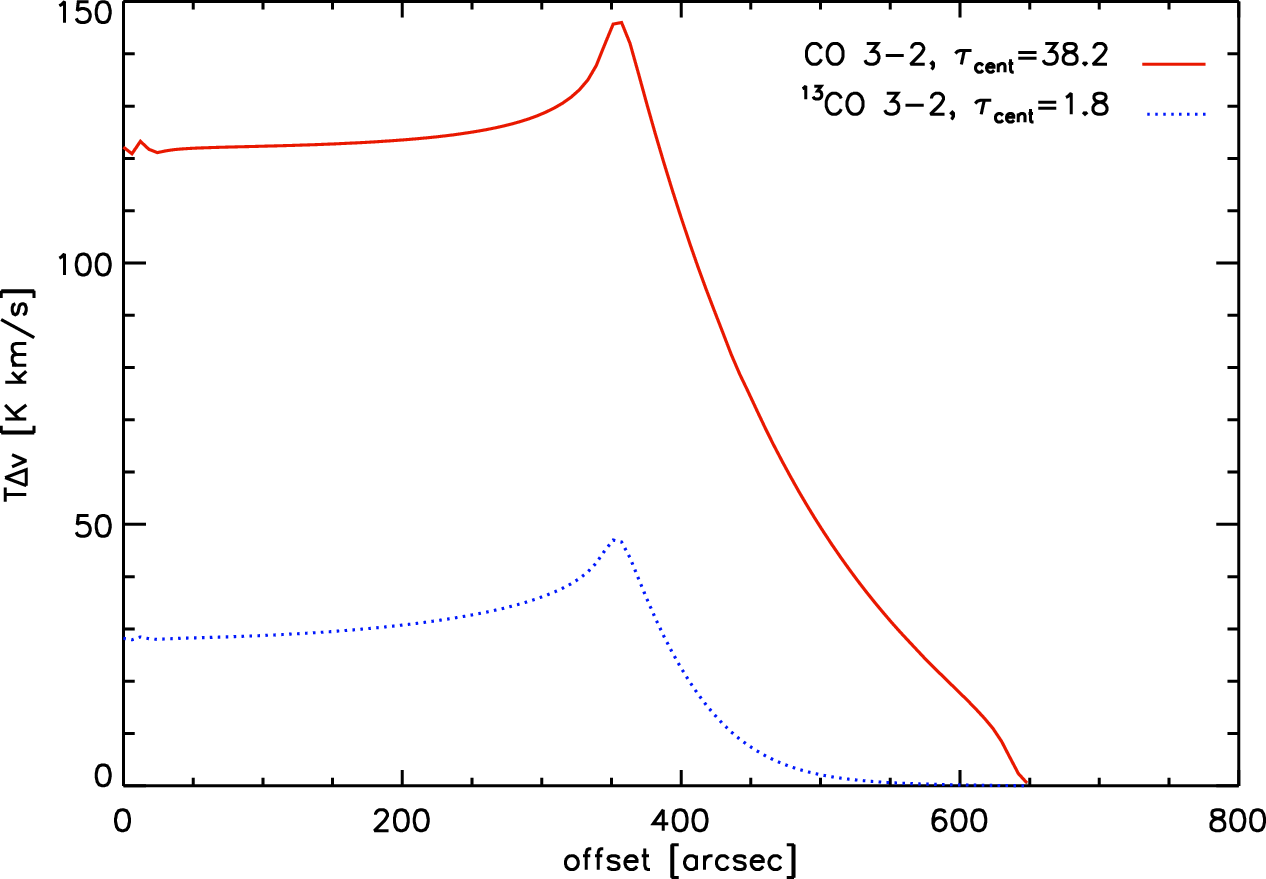}
            	\caption{Modeled intensities from the radiative transfer in a spherically symmetric bubble. Top panel: Modeled [\CII] intensity for simple shell geometries (Fig.~\ref{fig:bubble_geometry}) for various optical depths of surrounding cold material. Bottom panel: Modeled $^{12}$CO and $^{13}$CO~(3$\to$2) intensities for a slightly more realistic configuration with a continuous density and temperature gradient. The modeled intensities are determined by solving the radiative transfer equations with SimLine \citep{Ossenkopf2001}.} \label{fig:bubble_radeq}
            \end{figure} 

            In a second step, we determine the spatial distribution of the cold absorbing foreground layer. We run the two-layer multicomponent model on both CO isotopologues simultaneously and use the previously determined background cube as input. Again, possible additional background components (only visible in $^{12}$CO~(3$\to$2)) and the foreground components, derived from the averaging spectra of the clusters, are added iteratively until the model fit goodness is sufficient.
            	
            The resulting emission map of the cold foreground layer is shown in the middle panel of Fig.~\ref{fig:synthetic_maps}. In contrast to the warm emitting background, the ring-like structure characteristic for RCW~120 is barely visible. Moreover,  we observe an almost homogeneous emission pattern. However, we learned from the analysis of the background map that the true distribution of the molecular gas is better revealed in the column density channel maps, which are shown in Fig. \ref{fig:channel_synthetic_maps}. Between the velocity range of $-12$ to $-9$\,km\,s$^{-1}$,  we observe a rather homogeneous density distribution with a slight increase towards the inner ring. At velocities between $-9$ to $-6$\,km\,s$^{-1}$, thus close to the systemic velocity of RCW~120, we find an increased foreground correlating with the absorption features visible in the $^{12}$CO~(3$\to$2) channel maps in Fig. \ref{fig:channel_maps}. 
            
            However, the observed increase of foreground material toward the center of the \HII\ region does not have a significant impact on the observed molecular deficit. The CO column density for the background and foreground layer along two cuts through the ionizing star are shown in the right panel of Fig.~\ref{fig:synthetic_maps}. The column density of the cold absorbing layer is indicated by the green curve, which 
            shows little variation along the cuts and has a low contribution to the overall column density. Thus, most of the molecular column density is found in the dense PDR ring.  

            For velocities red-shifted from the systemic velocity of the cloud, the ring-like structure emerges in the foreground column density. This may suggest some inflow, i.e., the ring is still accumulating mass which then serves as a reservoir for further star-formation. \citet{Figueira2017} reported an over-density of compact sources in the ring with respect to the whole molecular cloud. A total of at least 35 compact, prestellar  sources were found. We note that the red-shifted feature is only visible in the $^{12}$CO maps that were extracted using both molecular isotopes in order to disentangle self-absorption effects from kinematic features.
	
    \subsection{Geometric modeling of RCW~120}\label{sec:simline} 

        For star-formation embedded in the parental molecular cloud we expect that the produced \HII\ region expands isotropically as long as the surrounding medium is isotropic on the size scale of the \HII\ region. By focusing here on [\CII] and CO (3$\to$2) emission, tracing only the warm gas around the \HII\ region, we characterize the structure and velocity in the direct vicinity of the \HII\ region unaffected by the geometry of the molecular cloud on larger scales. In case of a spherically symmetric expanding \HII\ region inside a molecular cloud one would indeed expect to observe a ring-like intensity structure like in RCW~120 due to limb brightening at the edge of the bubble. However, in the following we show that the parameters of the observed structure are inconsistent with a spherically symmetric configuration.
    	
        A simple estimate is possible based on purely geometric arguments. The intensity profiles in Fig.~\ref{fig:pi_diagram} show a width of the bright rim of about 0.7--0.8~pc (FWHM) for CO (3$\to$2) and $^{13}$CO (3$\to$2). In [\CII] it is somewhat narrower (0.4--0.6~pc). The intensity contrast between the rim peak and the inner part of the bubble is 5--10 in CO (3$\to$2), above 10 in $^{13}$CO (3$\to$2) and between 3 and 5 in [\CII]. If we assume the most simple configuration of a spherical shell, the actual width of the shell will affect both the apparent width and the intensity contrast between shell center and rim. A thinner shell will allow for a stronger limb brightening providing a higher intensity contrast, but it will also lead to a narrower apparent shell width. This is illustrated in Fig. \ref{fig:bubble_geometry}. The left and central panels show the geometry and projected column for an example shell with a width of 0.7~pc. The limb brightening from the LOS integration provides a maximum contrast between rim and center of a factor of 2.7. Such a setup would provide approximately the observed rim width, but a too low contrast compared to the observed values. The right panel visualizes the outcome of a parameter scan when changing the shell width. The black line shows the resulting intensity contrast, the red line the apparent rim width. For comparison we added the observed ranges for CO $(3\to2)$ and [\CII]. The error bars indicate the variations of both quantities along the ring. One sees that for [\CII] a shell width of about 0.6~pc could provide a marginal fit to both the rim width and the intensity contrast while for $^{12}$CO~(3$\to$2) both numbers are mutually exclusive. We find a strong discrepancy between the observed $^{12}$CO~(3$\to$2) limb brightening factor and the observed shell width. This discrepancy is even stronger for the molecular isotope $^{13}$CO~(3$\to$2), where we detect a limb brightening factor of larger than >10 for a shell with of about $\sim 0.7\,\mathrm{pc}$. Of course, the simple homogeneous shell is a nonrealistic over-simplification, but as one can build any radial distribution from a series of such shells, it is obvious that other radial profiles typically only worsen the discrepancy as they will have an even lower contrast ratio compared to the shell in this geometric example.
	
        For optically thin lines, that basically provide a LOS integration, the data are therefore incompatible with a spherical structure. However, one might ask whether LOS absorption may provide the observed contrast ratio enhancement. To answer this question we ran full radiative transfer computations using SimLine \citep{Ossenkopf2001}. We did not try to perform a detailed $\chi^2$-fit for best cloud parameters but we only show the fundamental radiative transfer effects for an example of parameters that are in rough agreement with the overall geometry. Results are shown in Fig. \ref{fig:bubble_radeq}. The upper panel displays an experiment based on the simple geometry discussed above using gas temperatures from Sect.~\ref{sec:two-layer}.
        A homogeneous shell with a thickness of 0.7~pc, a density of $10^3$~cm$^{-3}$, and a kinetic temperature of 80~K is surrounded by a shell with the lower kinetic temperature of 15~K. The C$^+$ abundance in that outer shell is varied to create a variable level of absorption. In the low optical depth case we recover an intensity distribution similar to the geometric picture from Fig.~\ref{fig:bubble_geometry}, however, with a lower limb brightening compared to the idealized geometric integration. Adding surrounding material with an increasing optical depth actually worsens the situation by absorbing relatively more from the bright rim than in the center of the configuration. The marginal match for the [\CII] data in Fig.~\ref{fig:bubble_geometry} therefore vanishes if one uses full radiative transfer computations.
        To show that this conclusion is not limited to the artificial two-shell setup, the lower plot uses a continuous density and temperature structure with power law profiles for temperature and density. The thin shell is mimicked by a very steep radial density decrease with a power of -8. The temperature drops from 80~K with a radial exponent of -2. Here, we visualize the CO (3$\to$2) lines\footnote{The setup ignores that for CO the shell is somewhat larger and the inner temperature somewhat cooler than for [\CII{}] (see Sect.\ref{sec:result_co}) but rather tries to allow for a direct comparison with the two-shell setup in the upper panel of Fig. \ref{fig:bubble_radeq}}. 
        The picture confirms the same behaviour; an increasing optical depth lowers the contrast between rim and center in all cases instead of increasing it. Radiative transfer cannot solve the contradiction found in Fig.~\ref{fig:bubble_geometry}, it rather worsens it. The same holds when including a velocity field for the expansion of the bubble. It tends to increase the absorption in the rim while decreasing it in the center. All of our SimLine experiments with more realistic parameters for RCW~120 in a spherically symmetric configuration lead to line predictions that deviated even more strongly from the observed lines than in the simple geometric picture described above.
    
        We conclude that the large limb brightening factor for a ring of the observed thickness can not be explained by a spherical symmetric geometry, but it requires rather a torus-like structure surrounding the expanding \HII\ region. The warm molecular emission must come from a primordially flat structure forming the ring, excluding an isotropic configuration. Even though we detect an expanding bubble in the velocity resolved [\CII] spectra \citep{Luisi2021}, we find that the observed [\CII] limb brightening factor is not compatible with the observed shell width. The [\CII] emission must also originate, at least partially, from the surface of the molecular torus forming a PDR (see also Sect.~\ref{sec:result_cii}). The observed 2D ring structure is a superposition of the expanding shell and the torus surrounding the \HII\ region. 
    
        \begin{figure}[!ht]
            \centering
        	\includegraphics[width=.45\textwidth]{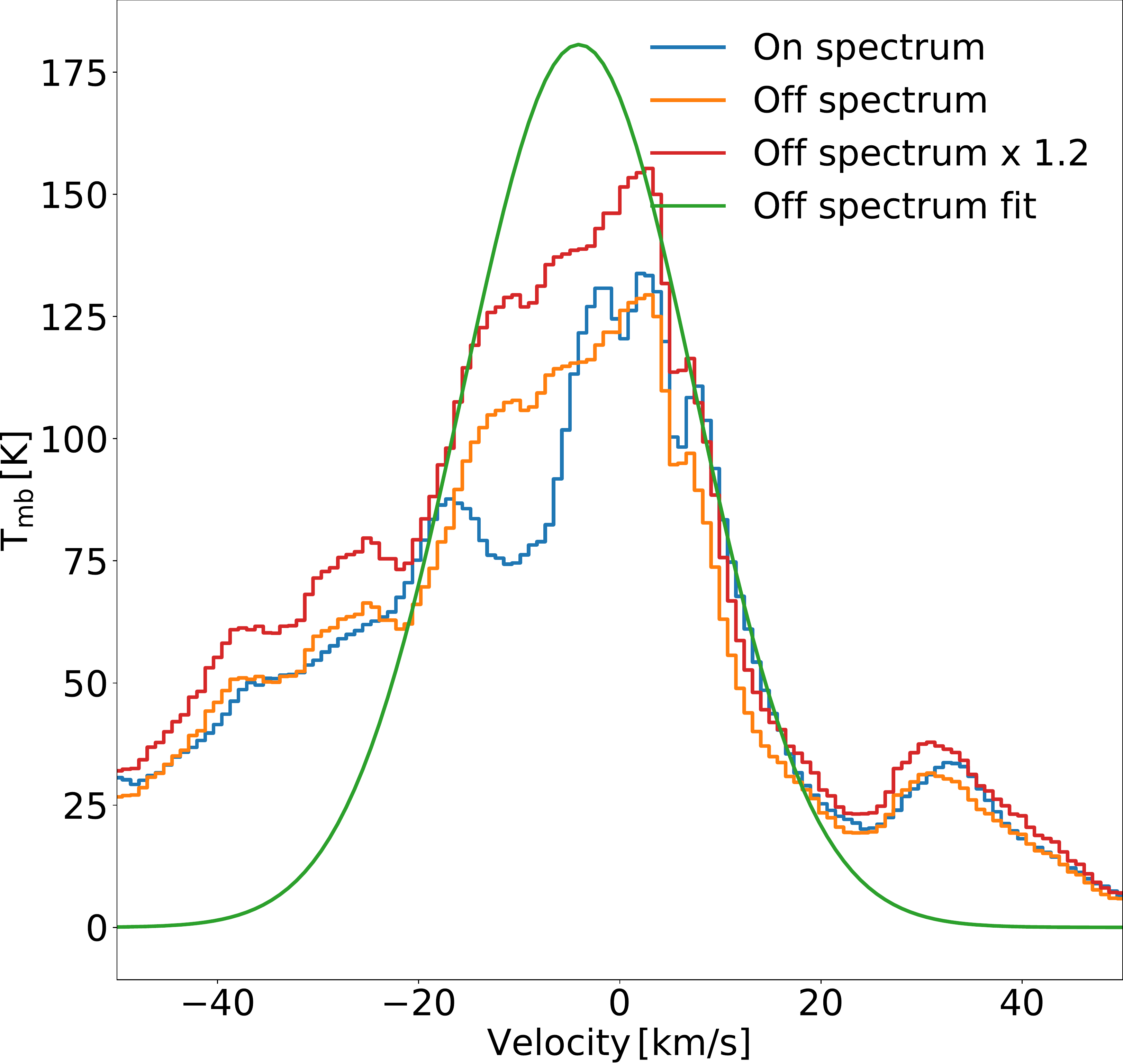}
            \caption{Observed and scaled \HI\ spectra. The blue curve shows the averaged \HI\ emission from the PDR ring area observed in [\CII]. The orange  spectrum is the Off-position (without an absorption dip) that is indicated in Fig.~\ref{fig:intensity_maps_co_j10}. The red spectrum is the same spectrum, scaled up by a factor of 1.2 in order to match the wing emission of the average On-spectrum between $-$20 and 20\,km\,s$^{-1}$. The green curve displays a Gaussian fit to the scaled up Off-position. }	\label{fig:HI_spectra}
        \end{figure}
	

    	\begin{figure*}[!ht]
    	    \centering
            \includegraphics[width=0.4\textwidth]{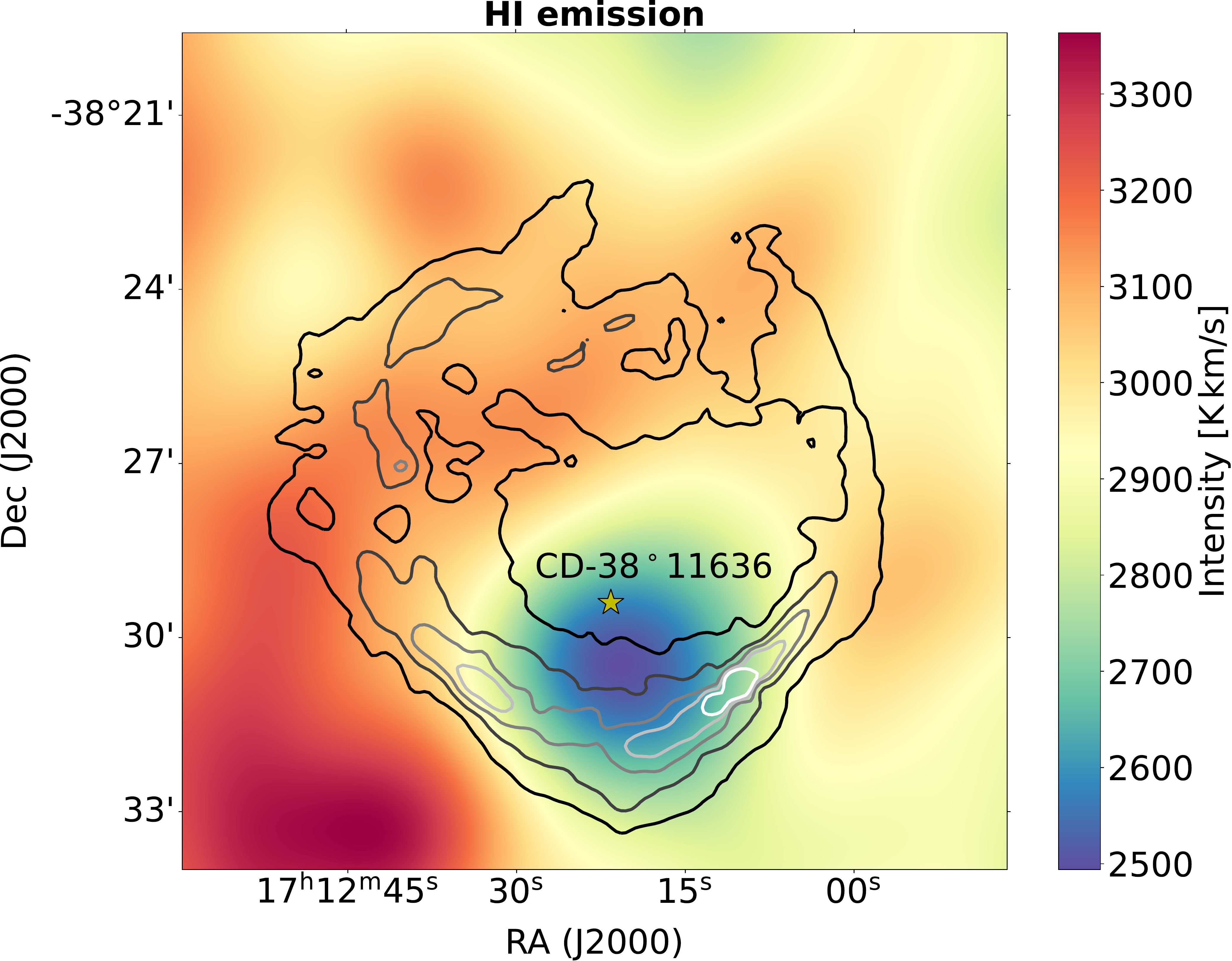}
    		\includegraphics[width=0.4\textwidth]{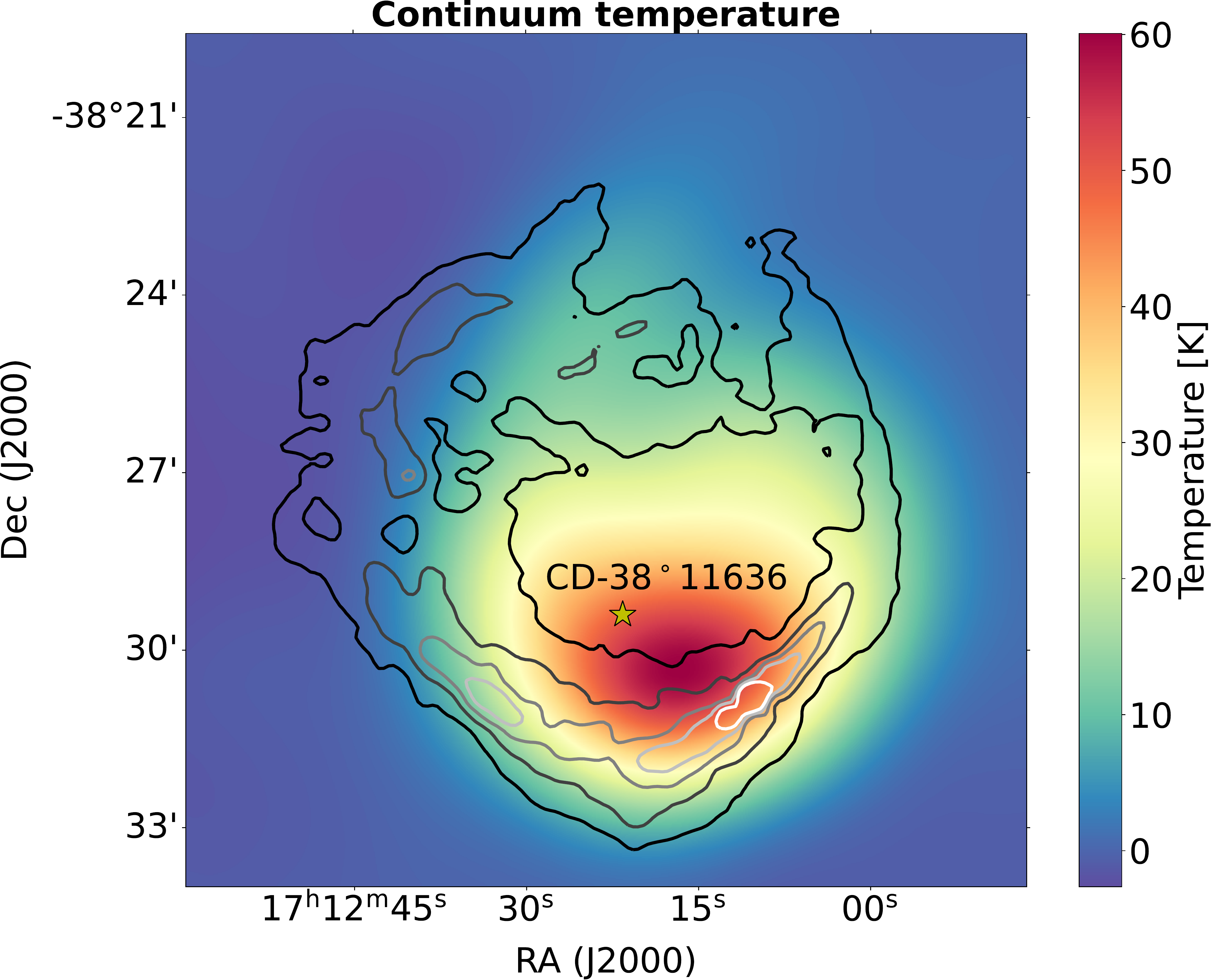}
    		\includegraphics[width=0.4\textwidth]{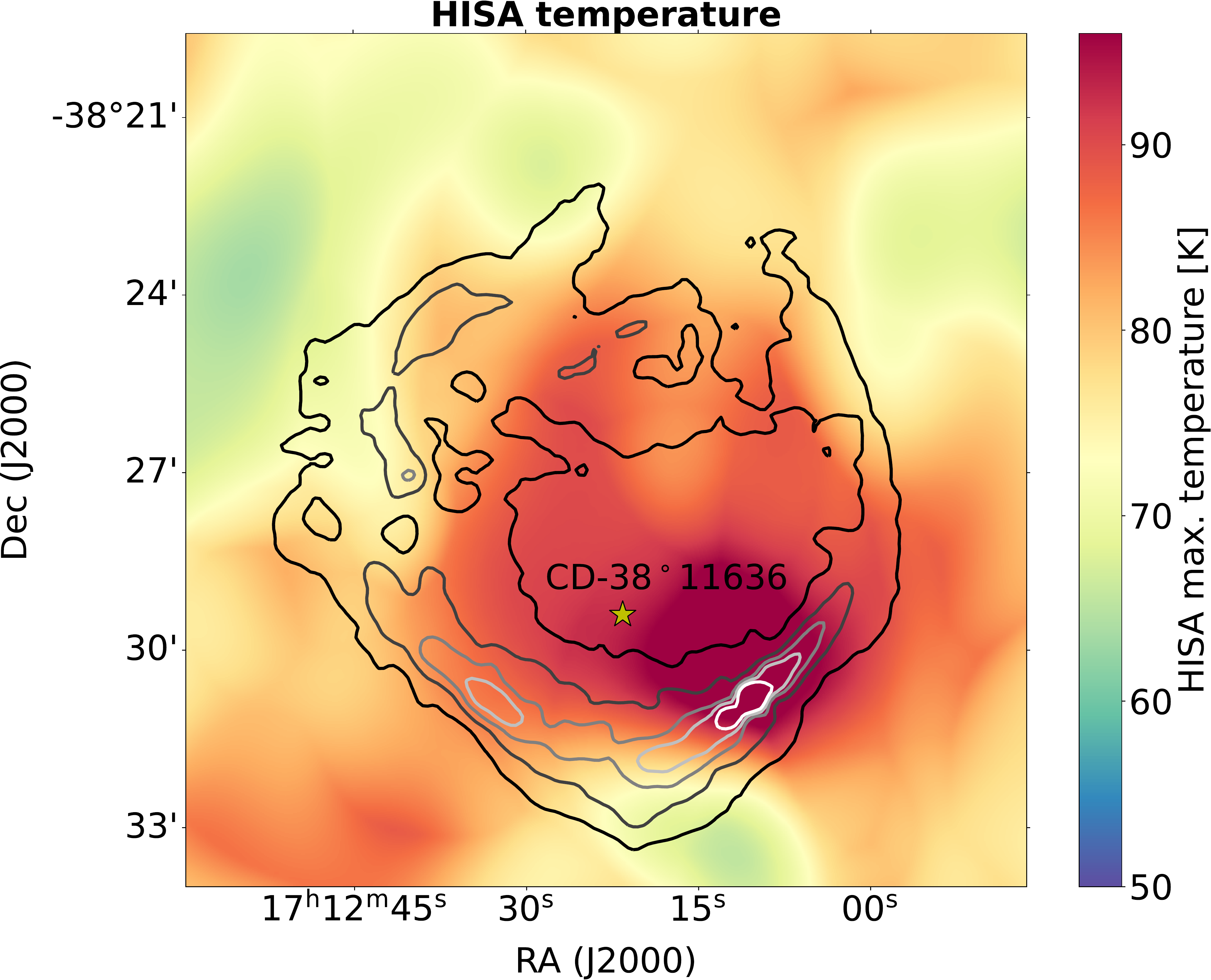}
    		\includegraphics[width=0.4\textwidth]{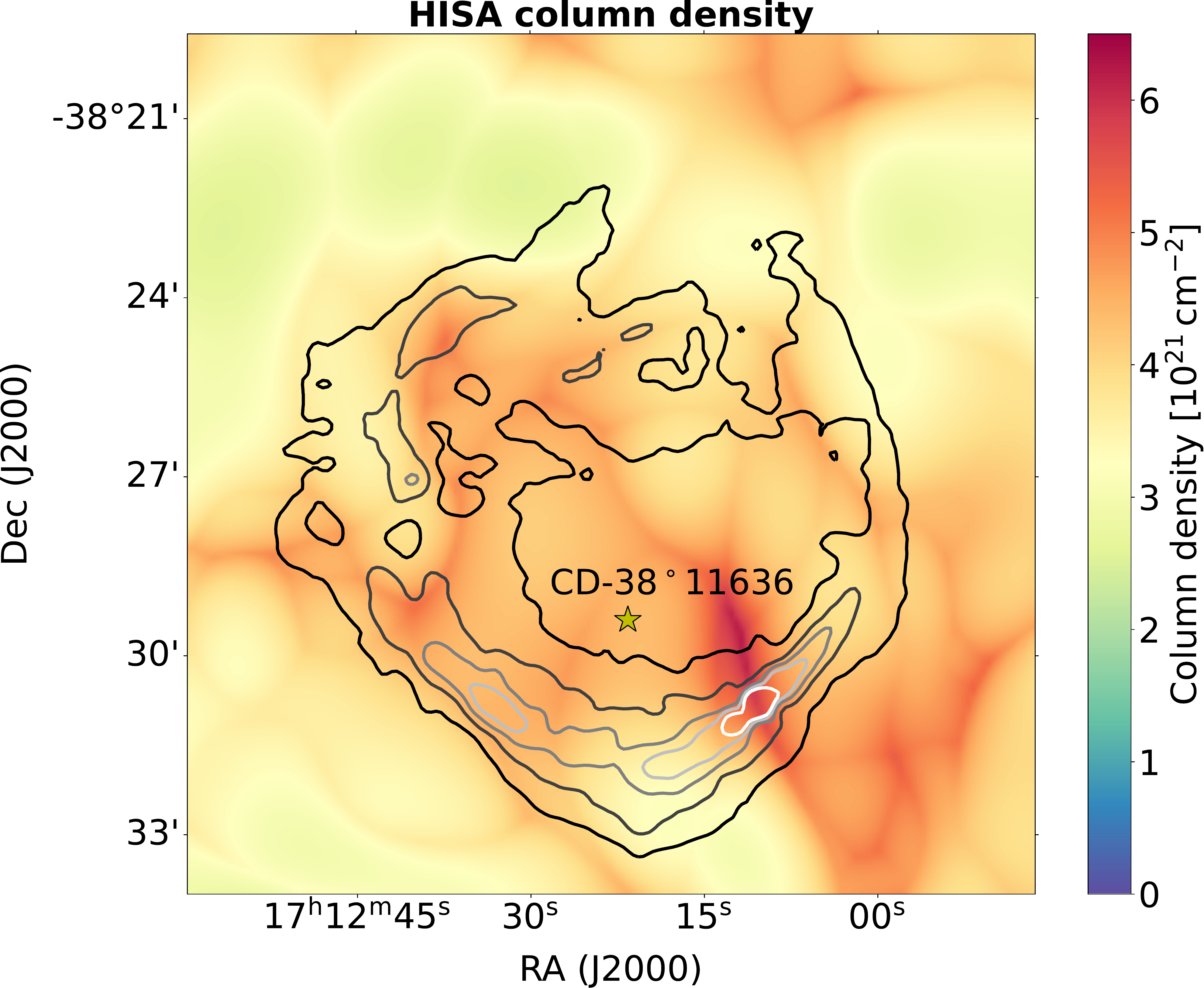}
    		\caption{HISA properties. Top left: Line integrated \HI\ emission between -20 to 10\,km\,s$^{-1}$. Top right: Continuum temperature map from 1.4\,GHz emission. Lower left: Upper limit of HISA temperature. Lower right: HISA column density distribution. The contours give the line integrated [\CII] intensity levels 60, 100, 140, 180, 220\,K\,km\,s$^{-1}$, ranging from black to light gray. } \label{fig:HISA_distribution}
    	\end{figure*} 
	 
    \subsection{Hydrogen self-absorption} \label{hisa}

        The presentation of the \HI\ spectra in Sect.~\ref{sec:HI-spectra} revealed that \HI\ self-absorption (HISA) is found at the same velocities where [\CII] and CO~(3$\to$2) show self-absorption features (broad self-absorption across a velocity range of around -6 to -11 km s$^{-1}$ and a strong dip at $\sim -7.5$ km s$^{-1}$). From the [\CII] self-absorption analysis (Sect.~\ref{sec:result_cii}), we derived that substantial column densities of cold foreground material must be responsible for the absorption. In the following, we assess if cold, atomic gas can be this gas component.  
            
        Figure~\ref{fig:HI_spectra} shows in blue the \HI\ spectrum averaged over the PDR ring where we have reliable values of the [\CII] column density (Table~\ref{tab:physical_conditions_cii}). As in the single spectra, we observe a dip in the spectrum near the systemic velocity of RCW~120. To quantitatively determine the absorption we first need to determine the strength of the warm background emission. To do so we assume that the cold absorbing hydrogen cloud is a local feature compared to the extended background emission, similar as it was done in \cite{Wang2020}. An appropriate Off-position, free of HISA, needs to be close enough to the source to be exposed to the same background radiation, but far enough to be free of absorption features. The orange spectrum in Fig.~\ref{fig:HI_spectra} was selected as a reasonable Off-position (centered at RA(2000)=17$^h$12$^m$21.6$^s$, Dec(2000)=$-$38$^\circ$00$'$35.9$''$, see Fig.~\ref{fig:intensity_maps_co_j10}) and represents an average over an area three times the beam size $\pi(150\cdot3/2)^2\,\mathrm{arcsec^2}$. However, the \HI\ emission outside RCW~120 is not homogeneous but shows some variation in intensity.
        Therefore, we tested different approaches to deal with the resulting uncertainty. We scaled up the Off-spectrum by a factor of 1.2, so that the wings of the Off-spectrum, in the range of $[-20, 20]$\,km\,s$^{-1}$ match the wings of the blue On-spectrum. The resulting spectrum is shown in red in Fig.~\ref{fig:HI_spectra}. The shape of the Off-spectrum also shows some variation and a small dip at -9\,km\,s$^{-1}$ might indicate that the spectrum is not completely free of HISA. Thus we also performed a single Gaussian component fit to the scaled spectrum (green curve in Fig.~\ref{fig:HI_spectra}). In the following we use the parameters for position, width and intensity of this Gaussian for the calculations but will propagate the uncertainties from the insufficient knowledge on the Off-spectrum in the discussion. 
    
        The velocity resolved intensity difference between the On- and Off-spectra $T_{\mathrm{on-off}}(\vel) =  T_{\mathrm{on}}(\vel) - T_{\mathrm{off}}(\vel)$ can be written as a function of the HISA optical depth $\tau_{\mathrm{HISA}}(\vel)$ and  temperature $T_{\mathrm{HISA}}$, describing the excitation temperature of the absorbing \HI{} material, \citep{Wang2020}:
    
        \begin{equation}
            T_{\mathrm{on-off}}(\vel) = (T_{\mathrm{HISA}}-pT_{\mathrm{off}}(\vel)-T_{\mathrm{cont}})\cdot(1-e^{-\tau_{\mathrm{HISA}}(\vel)})~,
        \end{equation}
        
        \noindent
        with the background continuum temperature $T_{\mathrm{cont}}$ and the dimensionless parameter $p$:
        
        \begin{equation}
            p = \frac{T_{\mathrm{bg}}\left(1-e^{-\tau_{\mathrm{fg}}}\right) }{T_{\mathrm{off}}}~,
        \end{equation}
    
        \noindent
        which is defined between [0, 1] and accounts for possible foreground emission. A value of $p=1$ means that there is no foreground emission; for $p =0$, there is no background emission \citep[see also][for a more detailed derivation and discussion of the equations]{Wang2020}. In the following we assume $p=1$, thus all emission originates from the warm background. The main background is  the overall warm Galactic \HI\ emission background \citep{Jackson2002,Li2003}. The continuum temperature is derived from the SGPS continuum map at 1.4\,GHz and displayed in the top right panel of Fig.~\ref{fig:HISA_distribution}. 
        For a given HISA temperature we can determine the velocity resolved optical depth
        
        \begin{equation}
            \tau_{\mathrm{HISA}}(\vel) = -\ln\left(1-\frac{T_{\mathrm{on-off}}(\vel)}{T_{\mathrm{HISA}} -T_{\mathrm{off}}(\vel)-T_{\mathrm{cont}} } \right)
            \label{eq:tau_HISA}
        \end{equation}
    
        \noindent
        and then the HISA column density \citep{Wilson2009}:
    
        \begin{equation}
            \frac{N_{\mathrm{HISA}}}{\mathrm{cm^{-2}}} = 1.8224\cdot10^{18}\frac{T_{\mathrm{HISA}}}{\mathrm{K}}\int\tau_{\mathrm{HISA}}(\vel)\,d\left( \frac{\vel}{\mathrm{km/s}} \right)~.
        \end{equation}

        Apart from the background intensity, the HISA temperature is another uncertainty factor, since we can not determine it independently from the optical depth. 
        An upper limit for the possible HISA temperature can be derived from equation~(\ref{eq:tau_HISA}) for $\tau_{\mathrm{HISA}} \rightarrow \infty$:
    
        \begin{equation}
            T_{\mathrm{HISA,max}} = T_{\mathrm{on, min}} + T_{\mathrm{cont}}~, 
        \end{equation}
        \noindent
        with the absorption dip minimum temperature $T_{\mathrm{on, min}}$. 
        The upper limit HISA temperature distribution is shown in the lower left panel of Fig.~\ref{fig:HISA_distribution} which varies between $60\,\mathrm{K}$ and $100\,\mathrm{K}$.
        Simply using $T_{\mathrm{on, min}}$ is not feasible, since it would result in infinitely large column densities. Thus, we subtract the rms from the On-position $\Delta T_{\mathrm{on}} $:
        \begin{equation}
            T'_{\mathrm{HISA,max}} = T_{\mathrm{HISA,max}} - \Delta T_{\mathrm{on}}~. 
        \end{equation}  
        \begin{figure}[!ht]
            \centering
            \includegraphics[width=.45\textwidth]{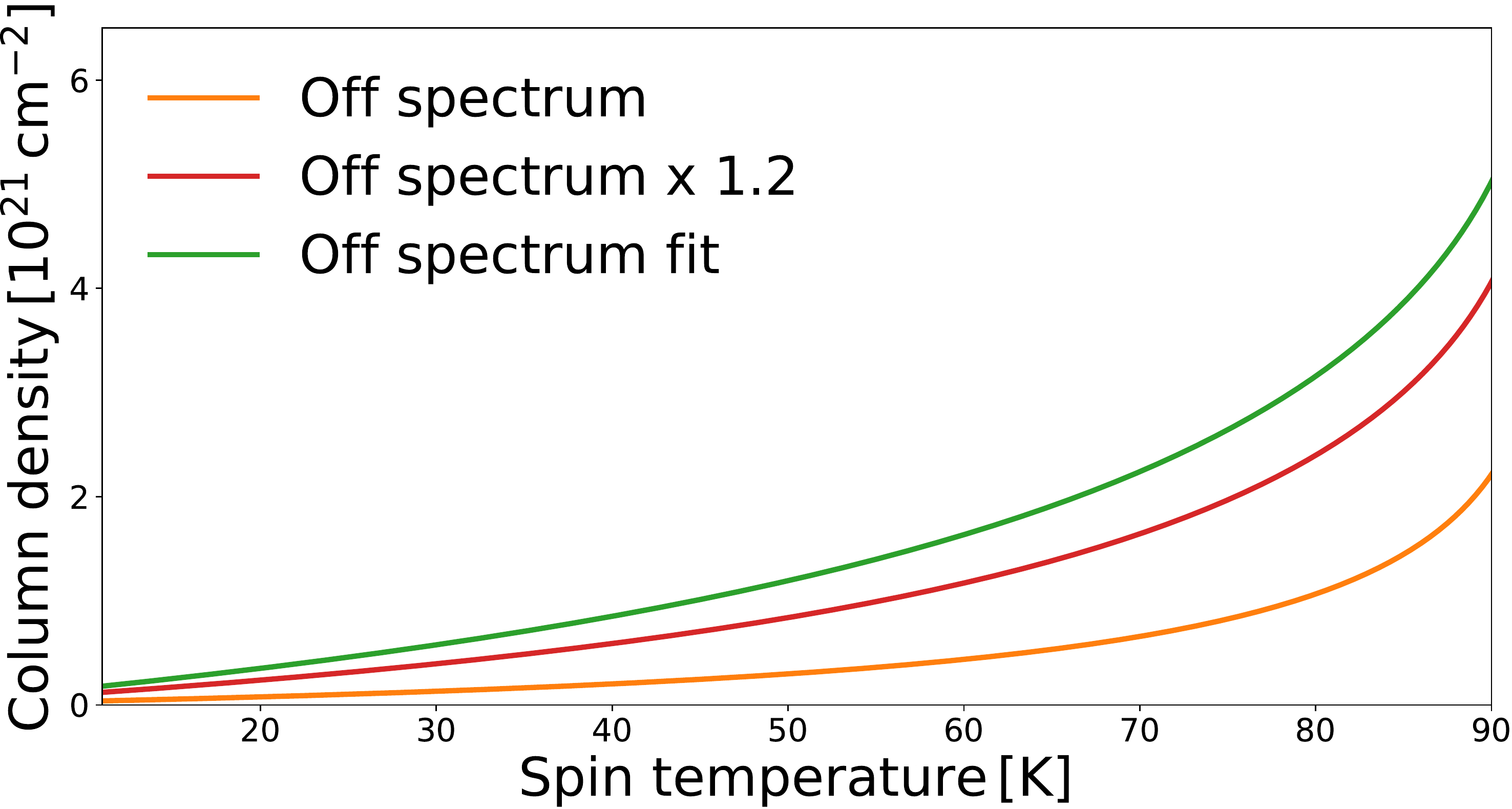}
            \caption{Column densities of absorbing \HI. The orange, red and green curves show the HISA column density as a function of temperature for the Off-position, the scaled up version, and the Gaussian fit, respectively (see Fig.~\ref{fig:HI_spectra}).} \label{fig:aver_hisa_properties}
        \end{figure}

        \begin{figure*}[!ht]
            \centering
            \includegraphics[width=0.9\textwidth]{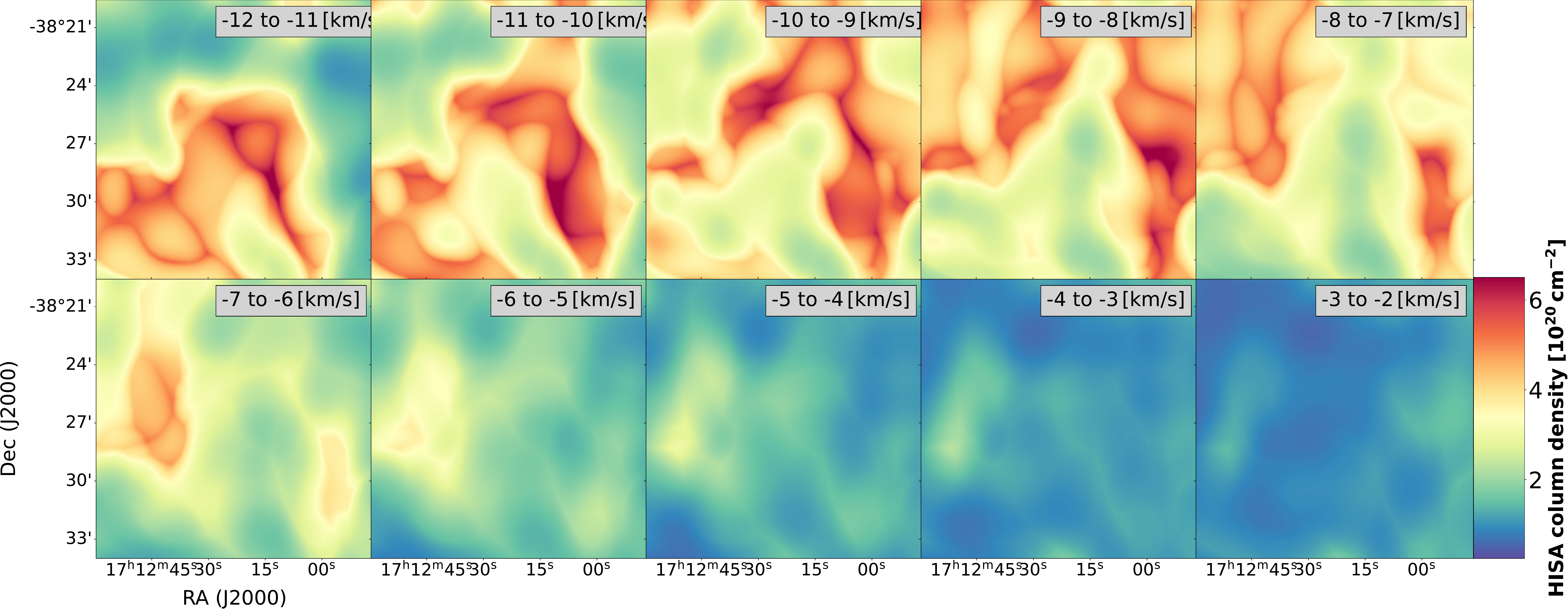}
            \caption{Channel maps of velocity resolved HISA column density.} \label{fig:HISA_channel_maps}
        \end{figure*}
        \begin{figure}[!ht]
        \centering
        	\includegraphics[angle=90,width=\columnwidth]{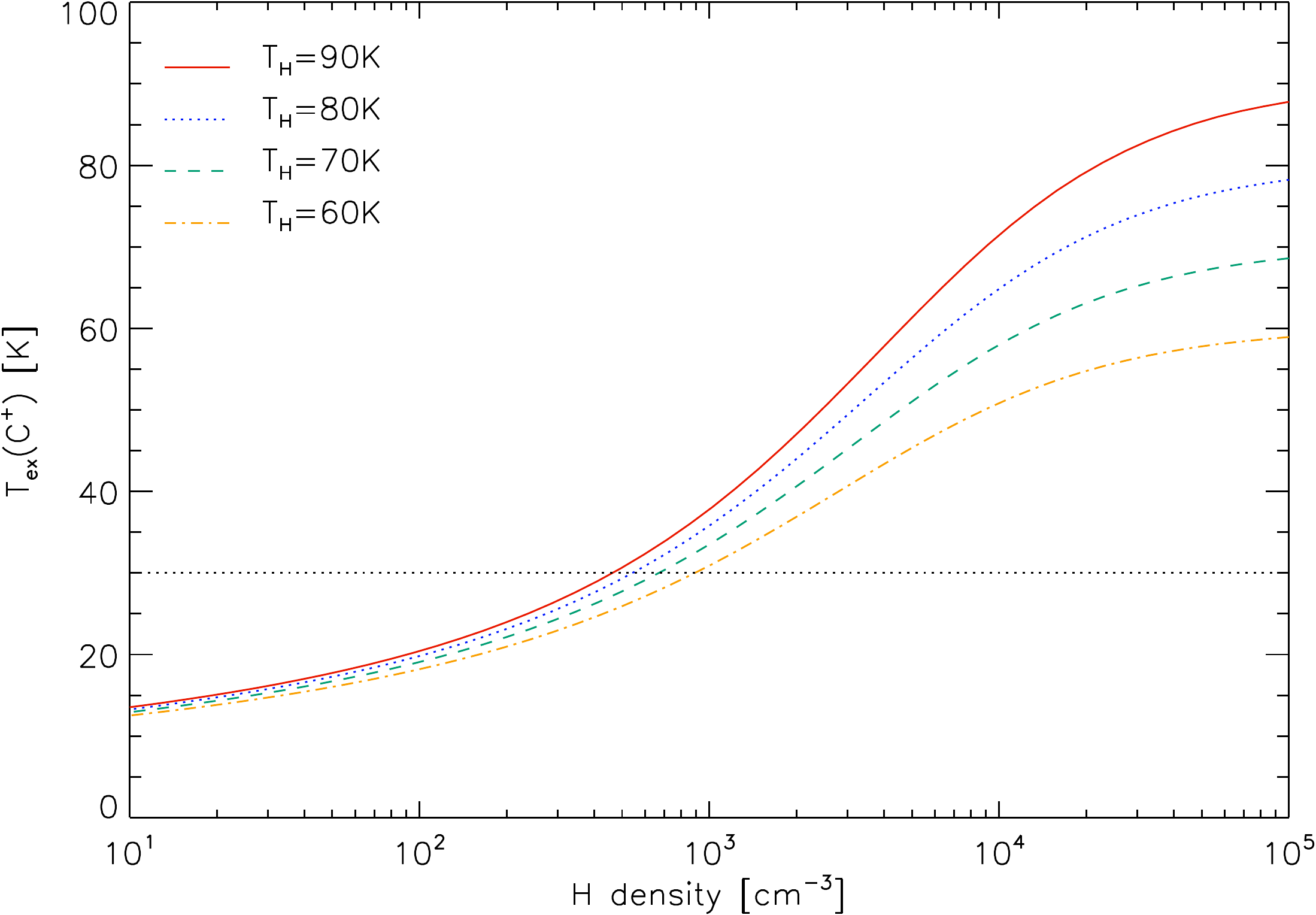}
        	\caption{Hydrogen density in the absorbing foreground layer (x-axis) and C$^+$ excitation temperature (y-axis) as a function of the HISA temperature. } \label{fig:density}
        \end{figure}

        \noindent
        The derived column density becomes very temperature sensitive close to $T_{\mathrm{HISA,max}}$ and thus sensitive to the noise level as illustrated in Fig.~\ref{fig:aver_hisa_properties}.  
        The determined HISA column density distribution is shown in the lower right panel of Fig.~\ref{fig:HISA_distribution}, the values range between $3-6\cdot10^{21}\,\mathrm{cm^{-2}}$. Even though the column density variations are small, they lead to the formation of filamentary structures in the cold absorbing hydrogen layer. However, these structures need to be treated with caution because the map gridding is very high, matching the resolution of the [\CII] and CO maps used in this work, with respect to the low resolution of the \HI\ data ($\sim$2$'$). 

        \begin{figure*}[!ht]
        \centering
        	\includegraphics[width=0.8\textwidth]{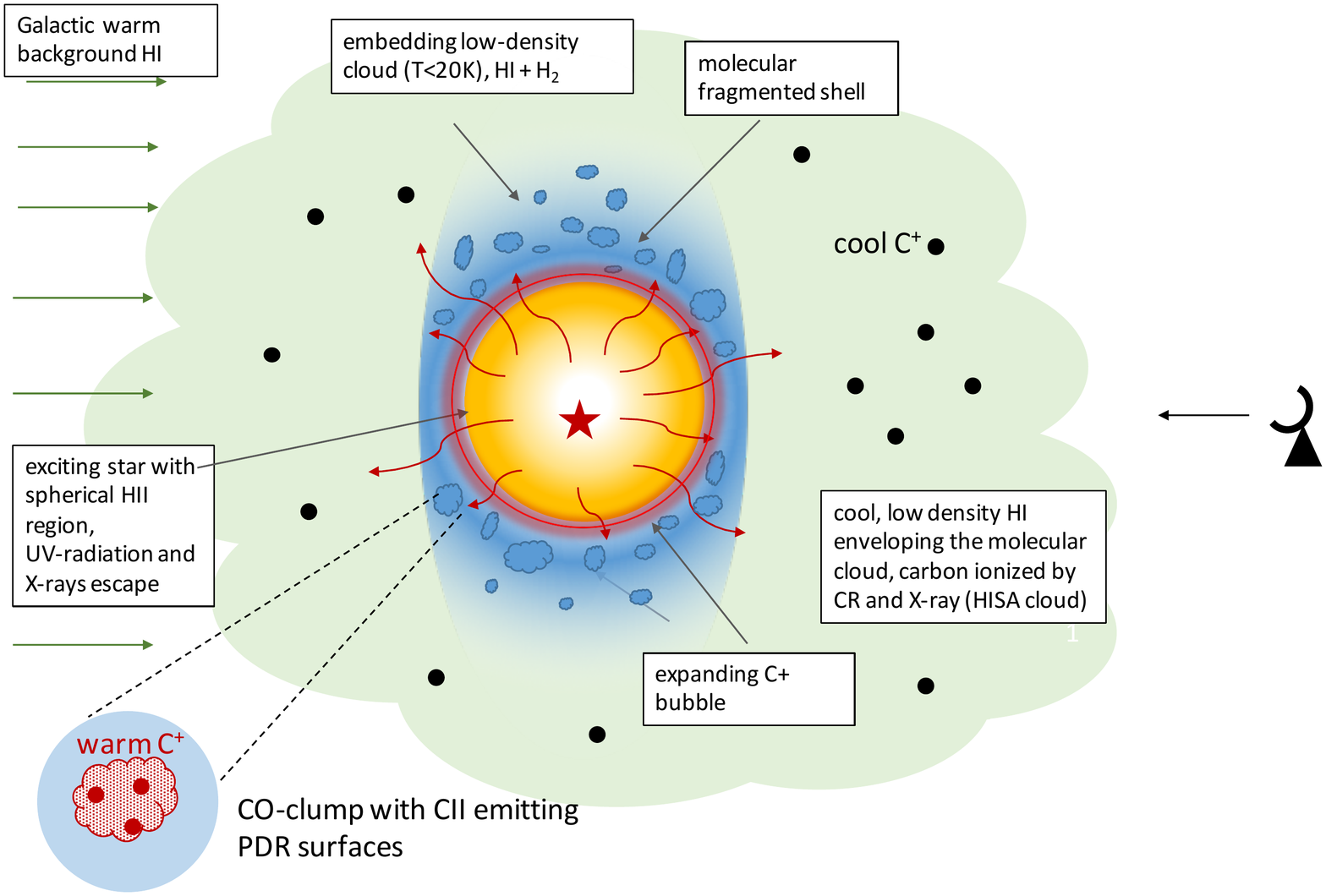}
        	\caption{Cartoon of the RCW~120 region where the observer is located on the right and sees a circular shaped \HII\ region in yellow (which is a 3D 
        	bubble in reality), and an expanding C$^+$ shell (red), surrounded by a molecular ring that is fragmented into clumps (blue). The sketch represents a 'cut', there is also molecular material in front of the \HII\ region, this becomes clearer in Fig.~\ref{fig:cartoon2}. CO emission arises from the clump interiors and [\CII] from the UV-illuminated clump surfaces and the expanding C$^+$ bubble. UV-radiation and X-rays leak into the surrounding molecular cloud and into the enveloping \HI\ cloud (green). This \HI\ layer is our HISA cloud in which carbon is mostly ionized by cosmic rays. The Galactic background (green arrows on the left) is responsible for warm \HI. } \label{fig:cartoon}
        \end{figure*}

        \begin{figure}[!ht]
        \centering
        	\includegraphics[width=0.4\textwidth]{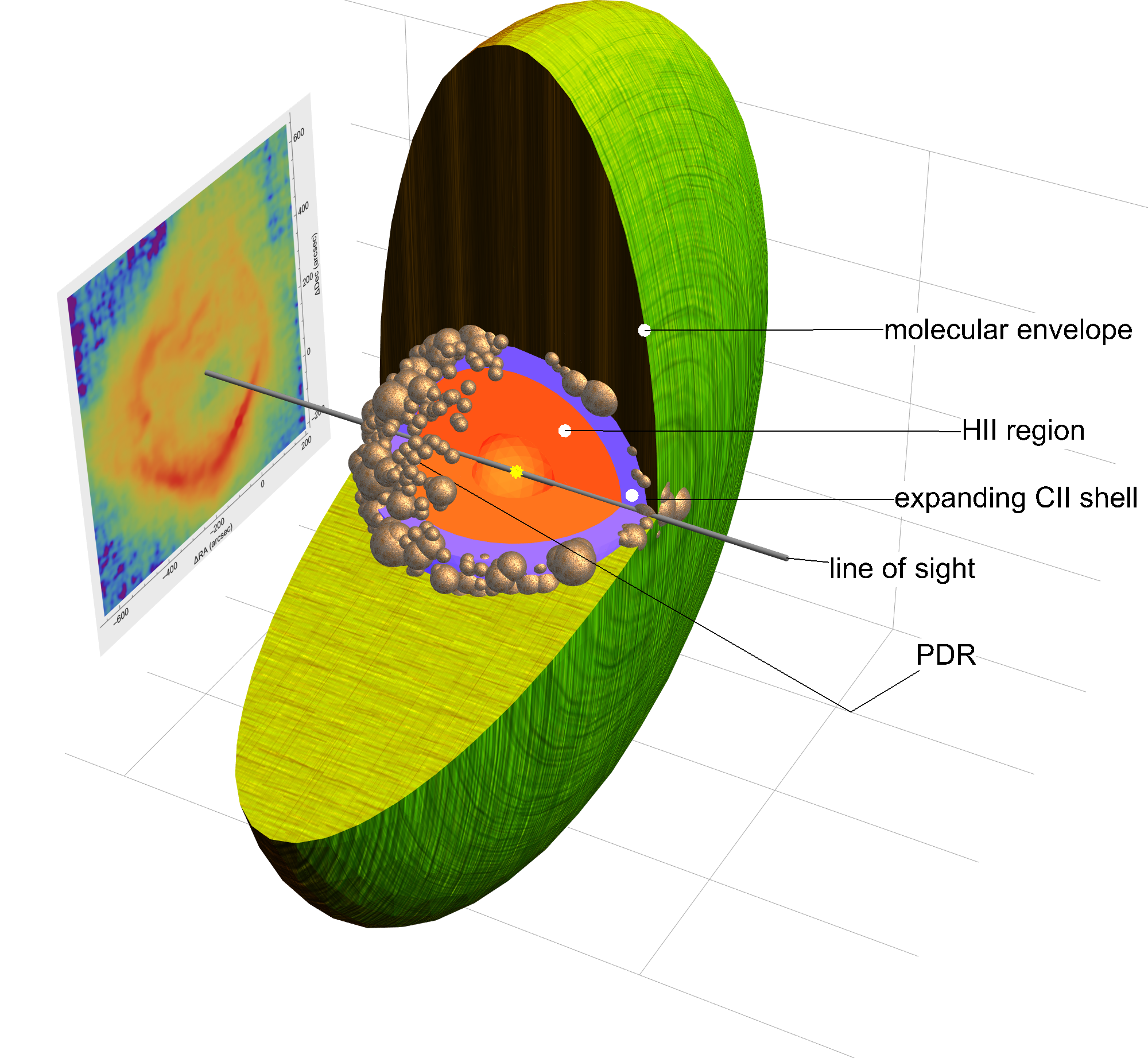}
        	\caption{3D visualization of the RCW120 region to show the oblate geometry of the molecular cloud. The observers perspective is from the right and the line integrated [\CII] map is projected on the left side. The spherical \HII\ region (red) is surrounded by the flat, low-density cloud (green). The clumpy PDR (golden clumps) and the expanding C$^+$ bubble (purple) are found in the interface region between \HII\ region and molecular cloud.} 
        	 \label{fig:cartoon2}
        \end{figure}

        Channel maps of the velocity resolved HISA column density are shown in Fig.~\ref{fig:HISA_channel_maps}. Similar to the cold molecular material most of the observed HISA mass is found at blue-shifted velocities or at the systemic velocity $\sim -7.5$ km s$^{-1}$. We emphasize again that though the filamentary structures are real, they are artificially enhanced because of the high gridding. Interestingly, we observe a weak ring-like structure, similar to the CO- and [\CII] ring emission around the bubble, starting at velocities around $-7$ km s$^{-1}$ and then continuing to red-shifted velocities.  
        This may be an indication of a possible inflow of atomic gas on the molecular torus, as already tentatively seen in CO (Sect.~\ref{CO-cube}).  

\section{Discussion}\label{sec:discussion}

    \subsection{The nature of the [CII] absorbing foreground layer} \label{discuss:foreground} 
	
        In Sect.~\ref{sec:result_cii} we found a layer of gas that is absorbing a significant amount of emission in [\CII]. Similar to the findings of \cite{Guevara2020}, who reported for various sources that there is a substantial column density of gas located in a cold [\CII] foreground layer which is atomic or molecular or a mixture of both. 
        
        We can now quantitatively investigate the hypothesis that the layer of cold C$^+$ is the same material that produces the HISA, this means that the [\CII{}] absorption occurs in atomic gas with HISA temperature. If C$^+$ is mixed with atomic hydrogen, we can compute the relation between HISA temperature, the [\CII] excitation temperature and the density of the gas. Together with the known column density of the foreground layer this turns into a length of the absorbing column.
        Figure~\ref{fig:density} shows the excitation temperature $T_{{\rm ex, [\CII]}}$ of [\CII], as a function of the density in the atomic gas, assuming that C$^+$ is only excited through collisions with atomic hydrogen \citep[collisional cross sections from][]{Barinovs2005}. The black dotted line indicates $T_{{\rm ex, [\CII]}}=30$~K, our upper limit, the curves were calculated for $T_{{\rm HISA}}$ between 60\,K and 90\,K. 
        The HISA temperature has a small influence on the density for the temperature regime of low  $T_{{\rm ex, [\CII]}}$. The relation allows us to discuss reasonable limits for $T_{{\rm ex, [\CII]}}$ . Using an average of HISA temperature of 80\,K and  $T_{{\rm ex, [\CII]}}=30$~K, we obtain a density of 500\,cm$^{-3}$ for the HISA layer and thus a LOS extent of $\sim$5\,pc. A somewhat smaller [\CII] excitation temperature of $\sim$20\,K results in a density of $\sim120$\,cm$^{-3}$ and a larger extent of $\sim$12\,pc for the HISA cloud. Even lower excitation temperatures can be excluded because they would result in a required length of the absorbing column that exceeds the size of the configuration by a large factor being highly unlikely for associated gas.

        Figure~\ref{fig:cartoon}, which will be discussed in more detail in the next subsection, qualitatively shows the layering of the \HII\ region bubble, the PDR ring, the molecular cloud and the enveloping \HI\ layer. 
        Though due to the uncertainties we can not further narrow down the values for the extent of the \HI\ layer, we demonstrate that the derived LOS extent of the \HI\ cloud is reasonable with respect to the diameter of the \HII\ region bubble (4.5\,pc) and the larger molecular cloud ($\sim$10\,pc). We note that \citet{Kirsanova2019} also concluded that there is a foreground absorbing cloud. However, they estimated a gas density $n = n_{\rm {HI}} + 2n_{\rm {H_2}}$ of 50 cm$^{-3}$ and a LOS extend of 1\,pc of the cloud, much lower than what we obtain.
        
        Summarizing, we think that in the case of RCW120 a large fraction of the C$^+$ absorbing foreground material is atomic. It is known since long that \HI\ halos exist around molecular clouds \citep{Burton1978,Williams1996,Lee2012,Motte2014,Imara2016,Wang2020} and RCW~120 is not an exception (see Fig.~\ref{fig:intensity_maps_co_j10} and \cite{Torii2015}). Detecting HISA features, however, is not always evident.  \citet{Jackson2002} stated that several criteria must be satisfied. First, the molecular cloud must be enveloped by a significant amount of cold \HI\ with a high enough opacity ($\tau_{\rm HI} \ge 1$), and second, there must be sufficient background emission from warm \HI. 
        These requirements are fulfilled in RCW~120. We assume that the carbon in the atomic cloud is ionized mostly by cosmic rays because even though UV radiation and X-rays "leak" through punctures in the PDR ring \citep{Luisi2021}, the UV field is fast attenuated with distance d by a factor 1/d$^2$. 
        This is an assumption, however, we can not investigate with the current data set. 

    \subsection{Flat molecular clouds around \HII\ region bubbles and a possible scenario of the RCW~120 region} 

        \cite{Beaumont2010} presented observations in the CO (3$\to$2) transition line of 43 \HII\ bubbles identified by the {\sl Spitzer} Space Telescope. They reported a deficit in CO emission towards the center of the observed rings, proposing that the associated molecular cloud must be "flat" and not spherical. We made the same observation for RCW~120 and showed quantitatively (Sects.~\ref{sec:result_co} and ~\ref{sec:simline}) that this CO deficit is not due to self-absorption effects, that could mimic missing emission, but a real shortfall of emission. 
        In addition, we proposed that CO self-absorption at red-shifted velocities could "hide" high-velocity gas that is falling on the dense molecular ring and similarly, at blue-shifted velocities, could prevent to detect expansion features in the molecular gas (no signs of expanding CO bubbles were found in the CO data that was not corrected for self-absorption). The density of this molecular gas is probably low, because high-velocity emission, i.e. infalling gas or an expanding molecular ring was not detected in $^{13}$CO, which is not affected by self-absorption. However, these conclusions are tentative and need further comparative studies of [\CII], CO, and \HI\ in other \HII\ region bubbles. 

        In Figs.~\ref{fig:cartoon} and \ref{fig:cartoon2}, we portray the RCW~120 star-forming region and speculate about its formation and evolution. We propose that we see RCW~120 in a stage of evolution in which a massive star formed inside a collapsing filament and the current \HII region now bursts out of the  remaining, nonfilamentary, but "flat" cloud structure. 
        The velocity resolved [\CII] spectra shown in \citet{Luisi2021} revealed an expanding C$^+$ bubble that swept up the surrounding material to a structure that is visible in the plane of the sky as a ring. The expanding bubble powered by stellar wind \citep{Luisi2021}, transfers momentum into the surrounding molecular cloud.
        The compression of the ambient molecular cloud was confirmed by observations of asymmetric column density profiles \citep{Zavagno2020} and double-peak features in the column density probability distribution functions \citep{Tremblin2014a}. This compression continues even after the bubble bursts out of the filament/sheet, in contrast to a pressure driven \HII region. The accumulation and compression of the ambient molecular clouds triggers additional star formation along the ring in RCW~120 \citep{Deharveng2009,Figueira2017}. Thus, missing young stellar objects (YSOs) along the line of sight towards the interior of the ring are an additional hint that the surrounding molecular cloud has a sheet-like geometry \citep{Anderson2015}. A longitudinally collapsing filament \citep{Schneider2010} additionally compresses the molecular gas and may have built up the prominent ridge seen in the south of RCW~120 (the north part is no longer visible because the \HII\ region created a hole in the structure). In general it is observed that the geometry of molecular clouds is flattened or filament-like \citep[e.g.,][]{Bally1989, deGeus1990, Heiles2003, Molinari2010, Andre2010}. Though there are different views of how dense structures are forming (quasi-static vs dynamic scenarios), our proposed scenarios fit in both cases.  
        In the quasi-static view, a collapsing cloud would first contract along its shortest axis towards a sheet-like geometry and then further collapse to a filament-like shape \citep{Lin1965}. Since the free fall time of an elongated cloud is much larger than the one of a spherical configuration \citep{Toala2012, Pon2012}, clumps inside a sheet or filament will collapse faster than the global molecular cloud, leading to star formation inside sheet-like molecular clouds. In turbulent, magneto-hydrodynamic simulations \citep[e.g.,][]{Audit2005,Banerjee2009,Vazquez2006}, sheets and filaments form naturally through gas compression in colliding atomic flows. 

        However, it is important to note that such a finding might be strongly biased towards sheet-like geometries. Ring-like structures are more likely to be observed originating from sheet-like molecular clouds due to the enhanced contrast between the emission from the surrounding torus and the inner ring. Potential bubbles embedded in molecular clouds sufficiently extended (the extension of the molecular cloud in all three dimensions is much larger than the \HII region) are more difficult to detect. The larger column density along the line of sight of an \HII region embedded in a molecular cloud increases the optical depth and thus reduces the contrast between the limb brightening shell and the interior of the \HII region. Moreover, magnetic pressure will act against the compression perpendicular to the magnetic field lines reducing the accumulation and compression at the ring and will distort the geometry of the \HII region \citep{Krumholz2007}. 

\section{Summary}\label{sec:summary}
	
    We studied the RCW~120 \HII\ region bubble and its associated molecular and atomic cloud, employing data from the SOFIA FEEDBACK [\CII] 158 $\mu$m legacy program, APEX $^{12}$CO and $^{13}$CO (3$\to$2) data, and the Southern Galactic Plane \HI\ 21cm survey. We analyzed self-absorption and optical depth in these lines and proposed a physical and geometrical model for the RCW~120 region. 

    \noindent $\bullet$ We determined optical depth effects along the ring of RCW~120 for [\CII] and CO (3$\to$2), employing information from [$^{13}$\CII] and $^{13}$CO (3$\to$2) data. 
    We observe large [\CII] optical depth and strong self-absorption in the [\CII] line in the velocity range $\sim-12$ to $\sim-4$ km s$^{-1}$ with a pronounced dip at $-$7.5 km s$^{-1}$, the systemic velocity of the cloud. To disentangle the weak [$^{13}$\CII] hyperfine-transition from the noise floor we averaged over square arcmin large regions along the ring using  dendrograms  \citep{Rosolowsky2008}. In a second step we solved the radiative transfer equations for multiple velocity components distributed between two layers. This analysis revealed a cold [\CII] layer, corresponding to a hydrogen column density of 3-7$\cdot$10$^{21}$ cm$^{-2}$, hiding a substantial amount of column number density, which would be otherwise undetected.  \\
    \noindent $\bullet$ The \HI\ spectral lines also show self-absorption (HISA) in the same velocity range as [\CII]. We performed a nonstandard HISA study and obtained similar \HI\ column densities as the ones from [\CII] for the absorbing foreground layer. We thus propose that the large column of cold C$^+$ in RCW~120 arises  from a cool, low-density ($<$500 cm$^{-3}$) atomic envelope around the region. The size of this HISA cloud is $\sim$5\,pc (the exact values depend on the HISA temperature and C$^+$ excitation temperature). 
    The carbon in this cool phase can only be ionized by cosmic rays, and UV-radiation and X-rays that leak through the punctured ring.\\     
    \noindent $\bullet$ We observe a deficit of CO (3$\to$2) emission along the LOS towards the \HII\ region bubble. $^{12}$CO has large optical depths and shows self-absorption at the bulk velocity ($-$7.5 km s$^{-1}$) of the molecular cloud. We solve the radiative transfer equation for two layers for the entire CO~(3$\to$2) spectral cube and demonstrate that the CO deficit is real and not mimicked by self-absorption. 
    To find initial model parameters for the entire data cube, we group the spectra of the cube using the Gaussian mixture model, an unsupervised machine learning approach to cluster multidimensional data. In a second step we solve the radiative transfer equations for the average spectra of each cluster. \\
    \noindent $\bullet$ The CO analysis revealed that red-shifted gas (with respect to the systemic velocity 
    of $\sim -7.5$ km s$^{-1}$) of RCW~120 is located close to the molecular ring, suggesting that the ring is still accreting material. Blue-shifted gas, that is not detectable without correction for self-absorption, indicates gas motion towards us and may thus signify an expanding molecular ring that was so far undetected. \\
    \noindent $\bullet$ We propose that RCW~120 could be a template for the evolution and geometry of \HII\ regions that expand from inside an initially filamentary or sheet-like structure. We speculate that the filament might 
    have experienced a longitudinal collapse while a massive star was forming inside the filament. The expanding \HII\ region and the collapse accumulated more mass in the \HII\ region/molecular cloud interface and formed a molecular shell that fragmented into clumps and star-forming cores. The expanding \HII\ region is currently bursting out of the sheet/filament and no filamentary structure remains. While the expansion of pressure driven \HII\ region eventually halts, the momentum transfer from the stellar wind will continue the expansion as observed in other sources \citep{Pabst2019,Tiwari2021,Luisi2021}, including RCW~120. This scenario is still highly speculative and needs further studies of bubbles in the key line tracers [\CII], CO, and \HI.  

\begin{acknowledgements}
    This study was based on observations made with the NASA/DLR Stratospheric Observatory for Infrared Astronomy (SOFIA). SOFIA is jointly operated by the Universities Space Research Association Inc. (USRA), under NASA contract NNA17BF53C, and the Deutsches SOFIA Institut (DSI), under DLR contract 50 OK 0901 to the University of Stuttgart. upGREAT is a development by the MPI f\"ur Radioastronomie and the KOSMA/Universit\"at zu K\"oln, in cooperation with the DLR Institut f\"ur Optische Sensorsysteme. \\
    Financial support for the SOFIA Legacy Program, FEEDBACK, at the University of Maryland was provided by NASA through award SOF070077 issued by USRA. \\
    The FEEDBACK project is supported by the BMWI via DLR, Projekt Number 50 OR 1916 (FEEDBACK) and Projekt Number 50 OR 1714 (MOBS - MOdellierung von Beobachtungsdaten SOFIA).\\
    N.S., R.S., and L.B. acknowledge support by the Agence National de Recherche (ANR/France) and the Deutsche Forschungsgemeinschaft (DFG/Germany) through the project "GENESIS" (ANR-16-CE92-0035-01/DFG1591/2-1). \\
    A.Z. thanks the support of the Institut Universitaire de France.\\
    The Australia Telescope Compact Array is part of the Australia Telescope National Facility (grid.421683.a) which is funded by the Australian Government for operation as a National Facility managed by CSIRO.\\
    The Parkes radio telescope is part of the Australia Telescope National Facility (grid.421683.a) which is funded by the Australian Government for operation as a National Facility managed by CSIRO.\\
    This work was supported by the Collaborative Research Centre 956,
    subprojects A4, C6 and C1, funded by the Deutsche Forschungsgemeinschaft
    (DFG), project ID 184018867.
\end{acknowledgements}

\bibliographystyle{aa} 
\bibliography{bibtex/bibliography} 

\newpage
\begin{appendix}

\section{Derivation of physical properties} \label{appendix-a}
		
    \subsection{Cold absorbing [\CII] excitation temperature range}
    
\begin{figure}[!h]
            \centering
        	\includegraphics[width=.45\textwidth]{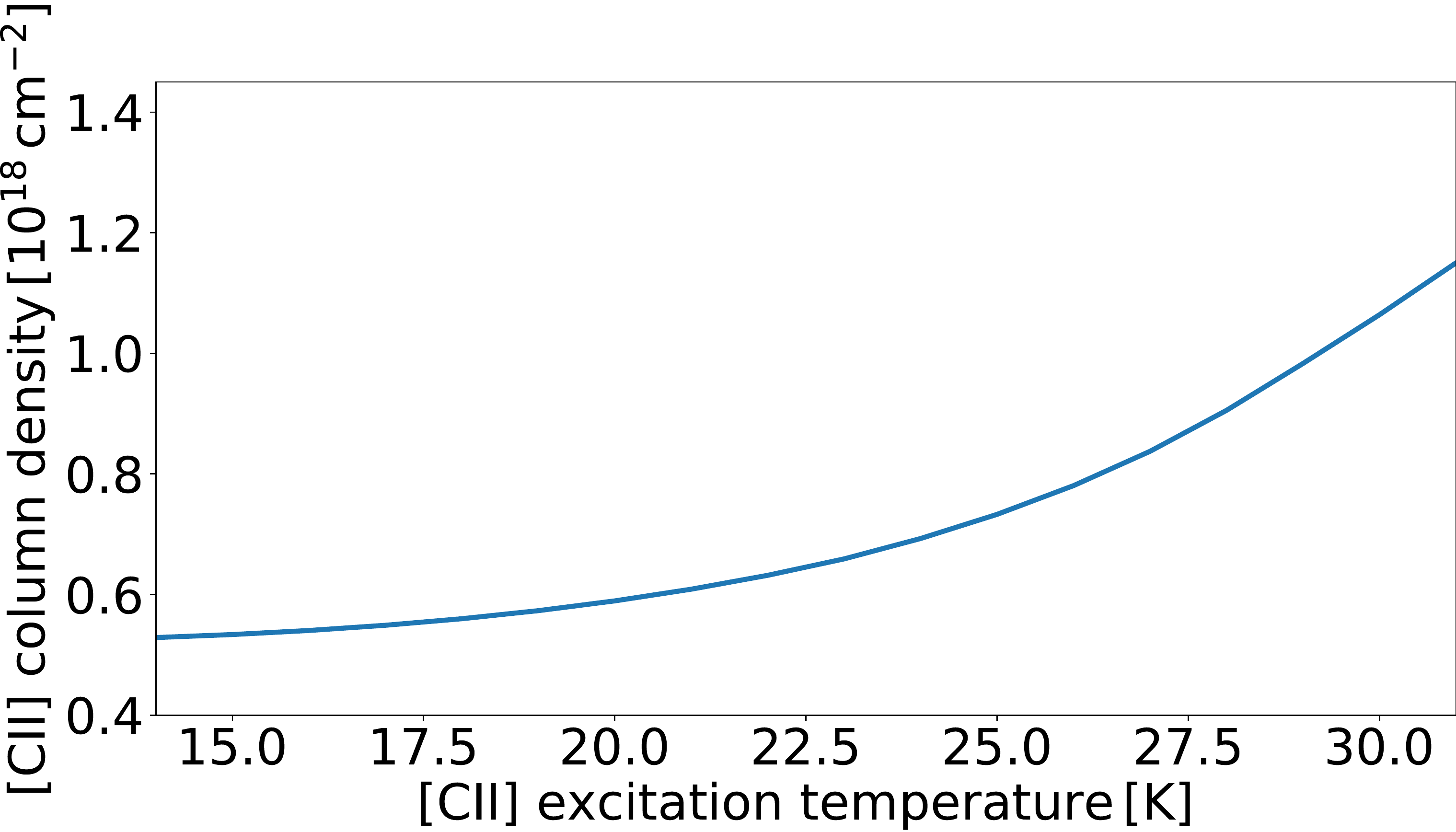}
        	\caption{The [\CII] column density of the cool foreground layer in the PDR ring as a function of the excitation temperature.} \label{fig:tex_colden_cii}
        \end{figure}

        We derived the physical properties of the cold absorbing [\CII] foreground in front of RCW~120 in Sec.~\ref{sec:result_cii} employing the two-layer multicomponent model \citep{Guevara2020}. The column number density is hereby a function of the optical depth and excitation temperature. Thus, for a chosen excitation temperature the optical depth is the remaining model parameter. To constrain the possible foreground excitation temperature we derived the lower limit from the [\CII] energy balance to $T_{\rm ex} \sim 15 $K. The upper temperature limit was calculated to $T_{\rm ex} \sim 30 $K from the hydrogen column density derived from the optical absorption measurements towards the ionized gas bubble of the \HII region in \citet{Zavagno2007}. We note that also the model fit goodness drastically decreases for temperatures above 30\,K. The resulting cold foreground [\CII] column density variation of  ${\rm N}_{\mathrm{ [\CII]}}$ as a function of excitation temperature is shown in Fig.~\ref{fig:tex_colden_cii}.
    
    \subsection{Mass and luminosity determinations}

        The two-layer model applied on the [\CII] and CO data gives us the synthetic background and foreground column density and corresponding line intensity. Using the output we can derive further physical parameter such as mass and luminosity.
        The [\CII] column density is converted into a mass of neutral gas by applying the hydrogen to carbon ratio $\mathrm{C/H}=1.6\cdot10^{-4}$ \citep{Sofia2004}:
        \begin{equation}
        	M_{\mathrm{H}}=\frac{N_{\mathrm{ [\CII]}}}{\mathrm{C/H}}m_{\mathrm{H}}A~,
        \end{equation}
        \noindent
        with the size $A$ of the target area and the hydrogen mass $m_{\mathrm{H}}$. In a similar fashion we derive the total molecular gas mass using the column density determined from the $^{13}$CO (3-2) transition. 
        \begin{equation}
        	M_{\mathrm{H_2}}=\frac{N_{\mathrm{CO}}}{\mathrm{CO/H_2}}m_{\mathrm{H_2}}A~,
        \end{equation}
        \noindent
        with the molecular hydrogen mass $m_{\mathrm{H_2}}$ and $\mathrm{CO/H_2} = 8.5\cdot10^{-5}$ as conversion factor \citep{Tielens2010}. 
        		
        The integrated line intensity $I$ [$\mathrm{K\,km\,s^{-1}}$] can be converted to the surface brightness $F$ [$\mathrm{W}\,\mathrm{m}^{-2}\,\mathrm{sr}^{-1}$]:
        	\begin{equation}
        		F = 2\kb I\left(\frac{\nu}{c}\right)^3
        	\end{equation}  
        	\noindent
        and multiplying this intensity with $4\pi A$ gives the luminosity $L$ of the region. 
		
\section{Optical depth error} \label{app:tau-error}

    In Sect.~\ref{sec:tau} we derived the velocity resolved optical depth along the torus in RCW~120 for [\CII] and CO~(3$\to$2) assuming a homogeneous medium. The optical depth itself is determined by comparing the observed intensities of the two isotopes. However, equation \ref{eq:optical_depth} is not analytically solvable, which makes the calculation of the corresponding error not trivial. To do so we first introduce a new variable $y$ which is simply:  

    \begin{equation}
    	y = \frac{T_{\mathrm{^{12}C}}}{\alpha T_{\mathrm{ ^{13}C}} }= \frac{1-e^{-\tau}}{\tau}~.
    \end{equation}
    \noindent
    The error of our new variable $y$ can now be derived from error propagation of the observed intensity errors $\sigma_{\mathrm{^{12}C}}$ and $\sigma_{\mathrm{^{13}C}}$ of both isotopes and $\Delta\alpha$ the uncertainty of the local $^{12}$C/$^{13}$C abundance ratio:

    \begin{equation}
    	\begin{aligned}
    	    \Delta y =& \sqrt{\left| \frac{\partial y}{\partial T_{\mathrm{^{12}C}}} \right|^2 \sigma^2_{\mathrm{^{12}C}} + \left| \frac{\partial y}{\partial T_{\mathrm{^{13}C}}} \right|^2 \sigma^2_{\mathrm{^{13}C}} + \left| \frac{\partial y}{\partial \alpha} \right|^2 \Delta\alpha }\\
    	    =& \sqrt{\frac{ \alpha^2 T_{\mathrm{^{13}C}}^2 \sigma^2_{\mathrm{^{12}C}} + \alpha^2T_{\mathrm{^{12}C}}^2 \sigma^2_{\mathrm{^{13}C}} +  \Delta\alpha^2 T_{\mathrm{^{13}C}}^2 T_{\mathrm{^{12}C}}^2  }{\alpha^4 T_{\mathrm{^{13}C}}^4}}
    		\label{eq:y_error_i}
    	\end{aligned}
    \end{equation}

    \noindent 
    on the other hand we can determine the error of $y$ by error propagation of the optical depth error:  

    \begin{equation}
    	\begin{aligned}
    	    \Delta y =& \sqrt{\left| \frac{\partial y}{\partial \tau} \right|^2 \Delta\tau^2} 
    	            = & \left| \frac{\partial y}{\partial \tau} \right| \Delta\tau 
    	            = &  \left| \frac{(1+\tau)e^{-\tau} -1}{\tau^2} \right| \Delta\tau~.
    	            \label{eq:y_error_ii}
    	\end{aligned}
    \end{equation}

    \noindent
    Combination of eq.~\ref{eq:y_error_i} and eq.~\ref{eq:y_error_ii} finally gives us an expression for the optical depth error:

    \begin{equation}
    	\Delta \tau = \left| \frac{\tau^2 }{(1+\tau)e^{-\tau}-1} \right| \sqrt{\frac{ \alpha^2 T_{\mathrm{^{13}C}}^2 \sigma^2_{\mathrm{^{12}C}} + \alpha^2T_{\mathrm{^{12}C}}^2 \sigma^2_{\mathrm{^{13}C}} +  \Delta\alpha^2 T_{\mathrm{^{13}C}}^2 T_{\mathrm{^{12}C}}^2  }{\alpha^4 T_{\mathrm{^{13}C}}^4}}~.
    \end{equation}

\section{Two-layer multicomponent fit parameter}

    \subsection{[\CII] fit parameter} \label{appendix-c1} 

        Tables C.1 to C.8 give the resulting fitting parameter of the two-layer multicomponent model fit for the [\CII] spectra along the ring/torus in RCW~120. We sort the results by the areas identified with Dendograms which are basically the brightest PDR regions of the ring.
        The tables are ordered by region and give the values for the background and foreground velocity components for the foreground [\CII] excitation temperatures 15\,K and 30\,K.

        \begin{table}[htb!]
            \caption{Two-layer multicomponent fit results of the RCW~120 PDR with $T_{\mathrm{ex, fg}}=15\,\mathrm{K}$}
            \label{tab:model_fit_cii_rcw_120_pdr_forg_15K}
            \begin{tabular}{l|cccc}
                \hline
                \hline
                Components & $\tau_{\mathrm{p, [\CII]}}$ & $T_{\mathrm{ex}}$  & $\vel_0$ & $w$ \\
                & & [K] & [km/s] & [km/s]\\
                \hline
                Bg. Comp. 1 & 2.3 & 50.0 & -7.6 & 3.9\\
                Bg. Comp. 2 & 0.08 & 50.0 & -14.8 & 9.4\\
                Bg. Comp. 3 & 0.5 & 50.0 & -7.3 & 7.6\\
                \hline
                Fg. Comp. 1 & 0.7 & 15.0 & -8.9 & 5.0\\
                \hline
            \end{tabular}
        \end{table}
        
        \begin{table}[htb!]
            \caption{Two-layer multicomponent fit results of the southwestern PDR with $T_{\mathrm{ex, fg}}=15\,\mathrm{K}$}
            \label{tab:model_fit_cii_south_western_pdr_forg_15K}
            \begin{tabular}{l|cccc}
                \hline
                \hline
                Components & $\tau_{\mathrm{p, [\CII]}}$ & $T_{\mathrm{ex}}$  & $\vel_0$ & $w$ \\
                & & [K] & [km/s] & [km/s]\\
                \hline
                Bg. Comp. 1 & 1.4 & 70.0 & -8.0 & 5.2\\
                Bg. Comp. 2 & 1.4 & 70.0 & -7.5 & 2.2\\
                Bg. Comp. 3 & 0.2 & 70.0 & -12.9 & 10.0\\
                \hline
                Fg. Comp. 1 & 0.3 & 15.0 & -8.1 & 2.2\\
                Fg. Comp. 2 & 0.8 & 15.0 & -11.4 & 3.0\\
                \hline
            \end{tabular}
        \end{table}
        
        \begin{table}[htb!]
            \caption{Two-layer multicomponent fit results of the southeastern PDR with $T_{\mathrm{ex, fg}}=15\,\mathrm{K}$}
            \label{tab:model_fit_cii_south_eastern_pdr_forg_15K}
            \begin{tabular}{l|cccc}
                \hline
                \hline
                Components & $\tau_{\mathrm{p, [\CII]}}$ & $T_{\mathrm{ex}}$  & $\vel_0$ & $w$ \\
                & & [K] & [km/s] & [km/s]\\
                \hline
                Bg. Comp. 1 & 3.0 & 60.0 & -6.9 & 3.8\\
                Bg. Comp. 2 & 0.8 & 60.0 & -7.1 & 8.1\\
                \hline
                Fg. Comp. 1 & 0.6 & 15.0 & -7.6 & 3.8\\
                \hline
            \end{tabular}
        \end{table}
        
        \begin{table}[htb!]
            \caption{Two-layer multicomponent fit results of the northeastern PDR with $T_{\mathrm{ex, fg}}=15\,\mathrm{K}$}
            \label{tab:model_fit_cii_north_eastern_pdr_forg_15K}
            \begin{tabular}{l|cccc}
                \hline
                \hline
                Components & $\tau_{\mathrm{p, [\CII]}}$ & $T_{\mathrm{ex}}$  & $\vel_0$ & $w$ \\
                & & [K] & [km/s] & [km/s]\\
                \hline
                Bg. Comp. 1 & 2.0 & 50.0 & -7.3 & 3.6\\
                Bg. Comp. 2 & 1.1 & 50.0 & -6.9 & 1.2\\
                Bg. Comp. 3 & 0.05 & 50.0 & -18.6 & 5.6\\
                Bg. Comp. 4 & 0.3 & 50.0 & -8.4 & 9.4\\
                \hline
                Fg. Comp. 1 & 0.5 & 15.0 & -9.3 & 2.2\\
                Fg. Comp. 2 & 0.4 & 15.0 & -7.3 & 1.8\\
                \hline
            \end{tabular}
        \end{table}
        
        \begin{table}[htb!]
            \caption{Two-layer multicomponent fit results of the RCW~120 PDR with $T_{\mathrm{ex, fg}}=30\,\mathrm{K}$}
            \label{tab:model_fit_cii_rcw_120_pdr_forg_30K}
            \begin{tabular}{l|cccc}
                \hline
                \hline
                Components & $\tau_{\mathrm{p, [\CII]}}$ & $T_{\mathrm{ex}}$  & $\vel_0$ & $w$ \\
                & & [K] & [km/s] & [km/s]\\
                \hline
                Bg. Comp. 1 & 2.3 & 50.0 & -7.6 & 3.8\\
                Bg. Comp. 2 & 0.1 & 50.0 & -13.8 & 9.9\\
                Bg. Comp. 3 & 0.3 & 50.0 & -5.7 & 6.2\\
                \hline
                Fg. Comp. 1 & 1.2 & 30.0 & -9.3 & 5.1\\
                \hline
            \end{tabular}
        \end{table}
        
        \begin{table}[htb!]
            \caption{Two-layer multicomponent fit results of the southwestern PDR with $T_{\mathrm{ex, fg}}=30\,\mathrm{K}$}
            \label{tab:model_fit_cii_south_western_pdr_forg_30K}
            \begin{tabular}{l|cccc}
                \hline
                \hline
                Components & $\tau_{\mathrm{p, [\CII]}}$ & $T_{\mathrm{ex}}$  & $\vel_0$ & $w$ \\
                & & [K] & [km/s] & [km/s]\\
                \hline
                Bg. Comp. 1 & 1.4 & 70.0 & -8.0 & 5.2\\
                Bg. Comp. 2 & 1.4 & 70.0 & -7.4 & 2.2\\
                Bg. Comp. 3 & 0.2 & 70.0 & -12.9 & 10.0\\
                \hline
                Fg. Comp. 1 & 0.4 & 30.0 & -8.1 & 2.3\\
                Fg. Comp. 2 & 1.4 & 30.0 & -11.8 & 3.1\\
                \hline
            \end{tabular}
        \end{table}
        
        \begin{table}[htb!]
            \caption{Two-layer multicomponent fit results of the southeastern PDR with $T_{\mathrm{ex, fg}}=30\,\mathrm{K}$}
            \label{tab:model_fit_cii_south_eastern_pdr_forg_30K}
            \begin{tabular}{l|cccc}
                \hline
                \hline
                Components & $\tau_{\mathrm{p, [\CII]}}$ & $T_{\mathrm{ex}}$  & $\vel_0$ & $w$ \\
                & & [K] & [km/s] & [km/s]\\
                \hline
                Bg. Comp. 1 & 2.7 & 60.0 & -6.9 & 3.8\\
                Bg. Comp. 2 & 0.8 & 60.0 & -7.1 & 8.1\\
                \hline
                Fg. Comp. 1 & 0.9 & 30.0 & -7.6 & 3.6\\
                \hline
            \end{tabular}
        \end{table}
        
        \begin{table}[htb!]
            \caption{Two-layer multicomponent fit results of the northeastern PDR with $T_{\mathrm{ex, fg}}=30\,\mathrm{K}$}
            \label{tab:model_fit_cii_north_eastern_pdr_forg_30K}
            \begin{tabular}{l|cccc}
                \hline
                \hline
                Components & $\tau_{\mathrm{p, [\CII]}}$ & $T_{\mathrm{ex}}$  & $\vel_0$ & $w$ \\
                & & [K] & [km/s] & [km/s]\\
                \hline
                Bg. Comp. 1 & 2.0 & 50.0 & -7.3 & 3.6\\
                Bg. Comp. 2 & 1.1 & 50.0 & -6.9 & 1.1\\
                Bg. Comp. 3 & 0.05 & 50.0 & -18.5 & 5.7\\
                Bg. Comp. 4 & 0.3 & 50.0 & -8.4 & 9.3\\
                \hline
                Fg. Comp. 1 & 0.5 & 30.0 & -7.3 & 1.7\\
                Fg. Comp. 2 & 1.1 & 30.0 & -9.6 & 2.7\\
                \hline
            \end{tabular}
        \end{table}


    \subsection{CO fit parameter}\label{appendix-c2} 

        Tables C.9 to C.13 give the fitting results of the two-layer multicomponent model fit for $^{12}$CO average spectra. The tables are ordered by the clusters identified with the GMM method and give the values for the background and foreground velocity components for the  CO excitation temperatures 40\,K for the background and 6 \,K for the foreground.

        \begin{table}[htb!]
            \caption{Two-layer multicomponent fit results of Cluster 1}
            \label{tab:model_fit_co_cluster_1}
            \begin{tabular}{l|cccc}
                \hline
                \hline
                Components & $\tau_{\mathrm{p, CO}}$ & $T_{\mathrm{ex}}$  & $\vel_0$ & $w$ \\
                & & [K] & [km/s] & [km/s]\\
                \hline
                Bg. Comp. 1 & 6.9 & 40.0 & -7.6 & 2.1\\
                Bg. Comp. 2 & 2.2 & 40.0 & -8.0 & 4.2\\
                Bg. Comp. 3 & 0.006 & 40.0 & 0.6 & 3.8\\
                Bg. Comp. 4 & 0.002 & 40.0 & -18.8 & 4.2\\
                \hline
                Fg. Comp. 1 & 1.7 & 6.0 & -5.4 & 4.9\\
                Fg. Comp. 2 & 1.3 & 6.0 & -11.0 & 4.7\\
                Fg. Comp. 3 & 0.3 & 6.0 & -8.5 & 2.4\\
                \hline
            \end{tabular}
        \end{table}
        
        \begin{table}[htb!]
            \caption{Two-layer multicomponent fit results of Cluster 2}
            \label{tab:model_fit_co_cluster_2}
            \begin{tabular}{l|cccc}
                \hline
                \hline
                Components & $\tau_{\mathrm{p, CO}}$ & $T_{\mathrm{ex}}$  & $\vel_0$ & $w$ \\
                & & [K] & [km/s] & [km/s]\\
                \hline
                Bg. Comp. 1 & 13.9 & 40.0 & -7.0 & 1.6\\
                Bg. Comp. 2 & 7.4 & 40.0 & -7.2 & 3.1\\
                Bg. Comp. 3 & 2.5 & 40.0 & -7.9 & 4.7\\
                Bg. Comp. 4 & 0.01 & 40.0 & 0.9 & 4.9\\
                Bg. Comp. 5 & 0.005 & 40.0 & -18.5 & 4.9\\
                \hline
                Fg. Comp. 1 & 1.7 & 6.0 & -3.8 & 4.2\\
                Fg. Comp. 2 & 1.4 & 6.0 & -11.3 & 4.9\\
                Fg. Comp. 3 & 0.7 & 6.0 & -8.0 & 2.2\\
                \hline
            \end{tabular}
        \end{table}
        
        \begin{table}[htb!]
            \caption{Two-layer multicomponent fit results of Cluster 3}
            \label{tab:model_fit_co_cluster_3}
            \begin{tabular}{l|cccc}
                \hline
                \hline
                Components & $\tau_{\mathrm{p, CO}}$ & $T_{\mathrm{ex}}$  & $\vel_0$ & $w$ \\
                & & [K] & [km/s] & [km/s]\\
                \hline
                Bg. Comp. 1 & 2.8 & 40.0 & -7.9 & 3.3\\
                Bg. Comp. 2 & 1.4 & 40.0 & -7.1 & 1.9\\
                Bg. Comp. 3 & 0.2 & 40.0 & -11.7 & 1.7\\
                Bg. Comp. 4 & 0.003 & 40.0 & 0.2 & 2.7\\
                \hline
                Fg. Comp. 1 & 1.9 & 6.0 & -8.7 & 4.9\\
                Fg. Comp. 2 & 1.4 & 6.0 & -12.4 & 2.0\\
                Fg. Comp. 3 & 1.4 & 6.0 & -5.0 & 3.3\\
                \hline
            \end{tabular}
        \end{table}
        
        \begin{table}[htb!]
            \caption{Two-layer multicomponent fit results of Cluster 4}
            \label{tab:model_fit_co_cluster_4}
            \begin{tabular}{l|cccc}
                \hline
                \hline
                Components & $\tau_{\mathrm{p, CO}}$ & $T_{\mathrm{ex}}$  & $\vel_0$ & $w$ \\
                & & [K] & [km/s] & [km/s]\\
                \hline
                Bg. Comp. 1 & 1.0 & 40.0 & -7.4 & 2.6\\
                Bg. Comp. 2 & 0.8 & 40.0 & -9.4 & 2.9\\
                Bg. Comp. 3 & 0.003 & 40.0 & 0.07 & 3.9\\
                \hline
                Fg. Comp. 1 & 1.6 & 6.0 & -6.6 & 4.5\\
                Fg. Comp. 2 & 1.4 & 6.0 & -10.5 & 4.4\\
                Fg. Comp. 3 & 1.2 & 6.0 & -5.1 & 1.1\\
                \hline
            \end{tabular}
        \end{table}
        
        \begin{table}[htb!]
            \caption{Two-layer multicomponent fit results of Cluster 5}
            \label{tab:model_fit_co_cluster_5}
            \begin{tabular}{l|cccc}
                \hline
                \hline
                Components & $\tau_{\mathrm{p, CO}}$ & $T_{\mathrm{ex}}$  & $\vel_0$ & $w$ \\
                & & [K] & [km/s] & [km/s]\\
                \hline
                Bg. Comp. 1 & 9.3 & 40.0 & -6.6 & 2.3\\
                Bg. Comp. 2 & 1.3 & 40.0 & -9.0 & 2.9\\
                Bg. Comp. 3 & 1.0 & 40.0 & -4.4 & 1.8\\
                Bg. Comp. 4 & 0.2 & 40.0 & -11.6 & 1.9\\
                Bg. Comp. 5 & 0.01 & 40.0 & -0.3 & 4.9\\
                \hline
                Fg. Comp. 1 & 1.7 & 6.0 & -3.7 & 3.4\\
                Fg. Comp. 2 & 1.7 & 6.0 & -7.9 & 3.4\\
                Fg. Comp. 3 & 1.5 & 6.0 & -11.8 & 3.7\\
                \hline
            \end{tabular}
        \end{table}

\section{\HI\ and [\CII] spectra} \label{HI-more} 
    Figure~D.1 displays an overlay between \HI\ and [\CII] spectra at the same angular resolution of $\sim$2$'$. 
    
    \begin{figure*}
        \centering
    	\includegraphics[width=11cm, angle=0]{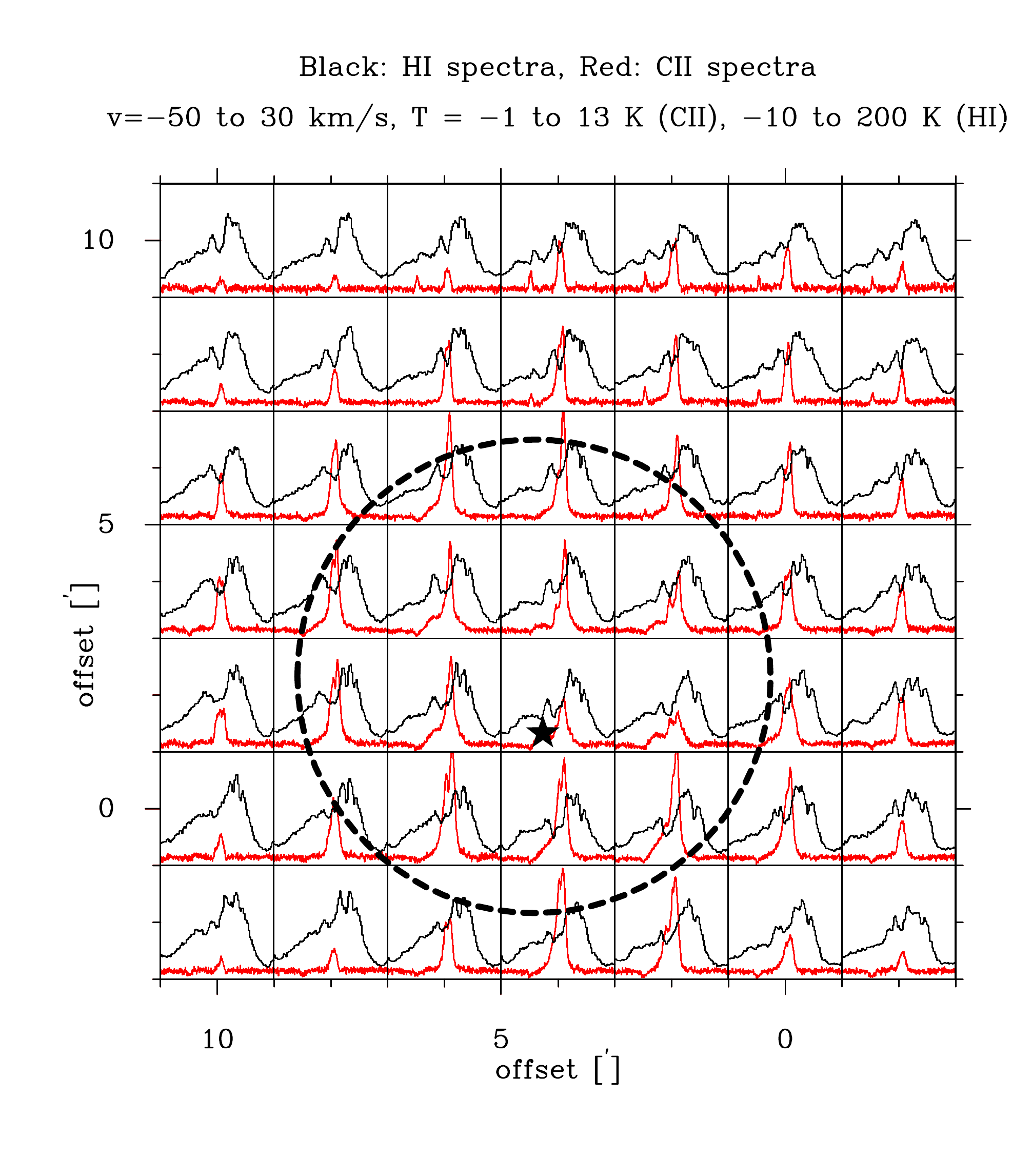}
    	\caption{Overlay between \HI\ (black) and [\CII] (red) spectra at an angular resolution of 2$'$ in a grid of 2$'$. The PDR ring of RCW~120 is indicated with a dashed black circle and the O star with a black star symbol. The velocity and temperature ranges are given in the upper caption. This plot is intended to show qualitatively that there is a systematic \HI\ depression at velocities around the bulk emission velocity of the cloud at $-8$ km s$^{-1}$. The [\CII] line shows self-absorption at that velocity (and flat-top spectra) which is not so prominent in this display because the spectra are smoothed to 2$'$ resolution. 
        }
        \label{fig:HI-map}
    \end{figure*}
\end{appendix}
	
\end{document}